%% file: main.tex
\documentclass[aoas]{imsart}

\usepackage[utf8]{inputenc}
\usepackage{preamble}
\usepackage{amsmath}
\usepackage{amsfonts}
\usepackage{amssymb}
\usepackage{tikz}
\usepackage{caption}
\usepackage{subcaption}
\usepackage{multirow}
\usepackage{graphicx}
\usepackage{upgreek}
\usepackage{booktabs}
\usepackage{siunitx}
\usepackage{xcolor}
\usepackage[authoryear]{natbib}
\usepackage{hyperref}
\usepackage{nicematrix}
\usepackage{lineno}
\usepackage{appendix}
\usepackage{makecell}
\usepackage{cleveref}
\usepackage{comment}
\DeclareCaptionTextFormat{tabletext}{\textup{#1}}% <================================================================

\newcounter{model}[section]
\newenvironment{model}[1][]{\refstepcounter{model}\par
   \noindent \textbf{Model~\themodel #1} \rmfamily}{}

\begin{document}

\begin{frontmatter}
\title{Poisson Cluster Process Models for Detecting Ultra-Diffuse Galaxies}
%\title{A sample article title with some additional note\thanksref{t1}}
\runtitle{Poisson Cluster Process for Detecting Ultra-Diffuse Galaxies}
%\thankstext{T1}{A sample additional note to the title.}

\begin{aug}
%%%%%%%%%%%%%%%%%%%%%%%%%%%%%%%%%%%%%%%%%%%%%%%
%% Only one address is permitted per author. %%
%% Only division, organization and e-mail is %%
%% included in the address.                  %%
%% Additional information can be included in %%
\author[A]{\fnms{Dayi}~\snm{Li}\orcid{0000-0002-5478-3966}\ead[label=e1]{dayi.li@mail.utoronto.ca}},
\author[B]{\fnms{Alex}~\snm{Stringer}\orcid{0000-0002-4133-6884}\ead[label=e2]{alex.stringer@uwaterloo.ca}},
\author[A]{\fnms{Patrick}~\snm{E. Brown}\orcid{0000-0003-2541-3744
}\ead[label=e3]{patrick.brown@utoronto.ca}},
\author[A, C]{\fnms{Gwendolyn}~\snm{M. Eadie}\orcid{0000-0003-3734-8177
}\ead[label=e4]{gwen.eadie@utoronto.ca}}
\and
\author[C]{\fnms{Roberto}~\snm{G. Abraham}\orcid{0000-0002-4542-921X}\ead[label=e5]{roberto.abraham@utoronto.ca}}
%%%%%%%%%%%%%%%%%%%%%%%%%%%%%%%%%%%%%%%%%%%%%%
%% Addresses                                %%
%%%%%%%%%%%%%%%%%%%%%%%%%%%%%%%%%%%%%%%%%%%%%%
\address[A]{Department of Statistical Sciences,
University of Toronto \printead[presep={,\ }]{e1} \printead[presep={,\ }]{e3} \printead[presep={,\ }]{e4}}

\address[B]{Department of Statistics and Actuarial Science,
University of Waterloo \printead[presep={,\ }]{e2}}

\address[C]{David A. Dunlap Department of Astronomy \& Astrophysics,
University of Toronto \printead[presep={,\ }]{e5}}

\end{aug}

\begin{abstract}

We propose a novel set of Poisson Cluster Process (PCP) models to detect Ultra-Diffuse Galaxies (UDGs), a class of extremely faint, enigmatic galaxies of substantial interest in modern astrophysics. We model the unobserved UDG locations as parent points in a PCP, and infer their positions based on the observed spatial point patterns of their old star cluster systems. Many UDGs have somewhere from a few to hundreds of these old star clusters, which we treat as offspring points in our models. We also present a new framework to construct a marked PCP model using the marks of star clusters. The marked PCP model may enhance the detection of UDGs and offers broad applicability to problems in other disciplines. To assess the overall model performance, we design an innovative assessment tool for spatial prediction problems where only point-referenced ground truth is available, overcoming the limitation of standard ROC analyses where spatial Boolean reference maps are required. We construct a bespoke blocked Gibbs adaptive spatial birth-death-move Markov chain Monte-Carlo algorithm to infer the locations of UDGs using real data from a \textit{Hubble Space Telescope} imaging survey. Based on our performance assessment tool, our novel models significantly outperform existing approaches using the Log-Gaussian Cox Process. We also obtained preliminary evidence that the marked PCP model may improve UDG detection performance compared to the model without marks. Furthermore, we find evidence of a potential new ``dark galaxy'' that was not detected by previous methods.
\end{abstract}


\end{frontmatter}

% Put the links to the files for each section here.
% Section files will be numbered by draft, like 01- is draft 1
% Different sections can be on different drafts, and this is where
% you change that
\input{sections/01-Introduction}
\input{sections/02-Data}

\input{sections/03-Method}

\input{sections/04-Inference}
\input{sections/05-Data-Analysis}
\input{sections/06-Discussion}
\input{sections/07-Acknowledgement}

\bibliographystyle{chicago}
\bibliography{bibliography}
% \appendix
% \input{sections/notation-and-assumptions}
\appendix
\input{sections/Notations-and-Symbols}
\input{sections/Supplementary-Materials}

\end{document}

%% file: sections/01-Introduction.tex
\section{Introduction}\label{sec:intro}

Ultra-diffuse galaxies (UDGs) are a class of extremely faint galaxies first found in abundance by \cite{VanDokkum2015}. UDGs have garnered significant attention due to their peculiar nature: despite their faintness, their sizes and masses are comparable to luminous galaxies like the Milky Way \citep{van_Dokkum_2017, Lim2018, Forbes_2020}. Observations indicate that UDGs contain unusual quantities (either too much, or too little, relative to most galaxies) of dark matter, a substance that is discernible only through its gravitational effects on visible matter. Because many UDGs have a much larger fraction of dark matter \citep{VanDokkum2015, Yagi2016, Wittmann2017, Janssens2019, Lim2020} compared to other known galaxy types, they are excellent places to test dark matter theories \citep{Hui2017, VanDokkum2019, Wasserman2019, Burkert2020}. As a result, investigation of UDGs has grown into one of the most active areas of galaxy formation research. 

\begin{figure}[t]
    \centering
    \includegraphics[width =0.6\textwidth]{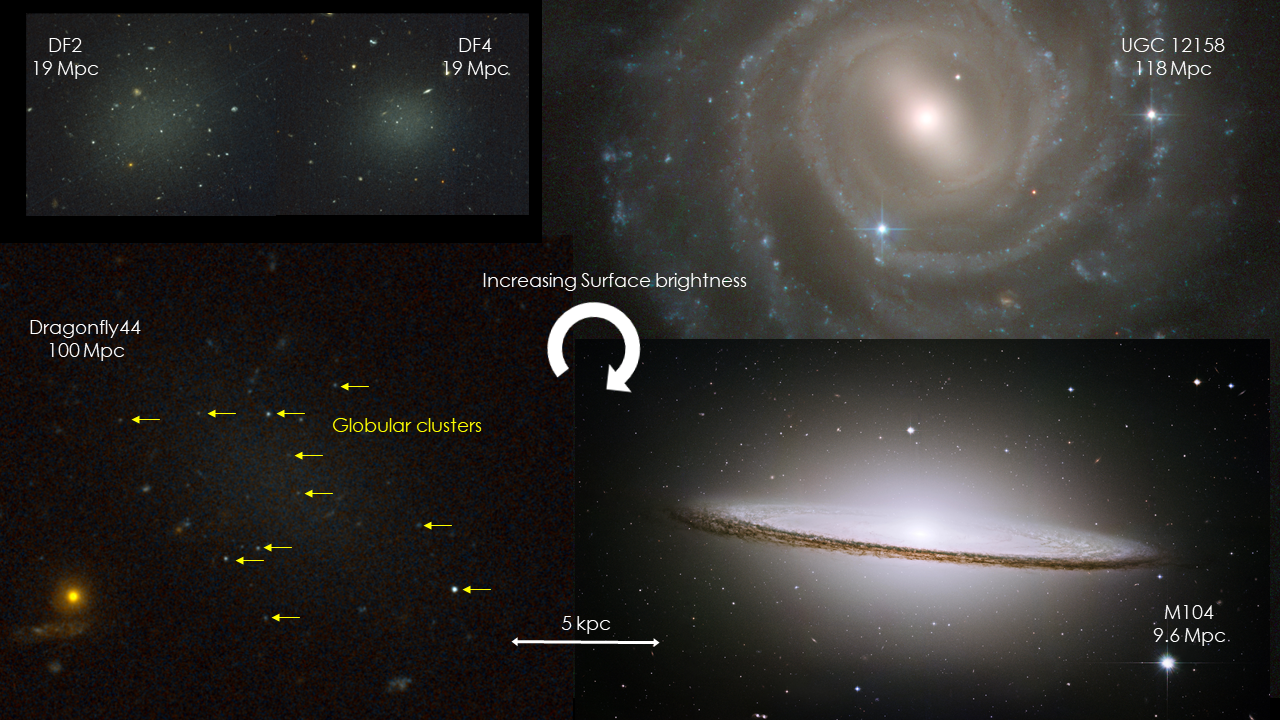}
    \caption{\footnotesize{Three UDGs (Dragonfly 44, NGC~1052-DF2, and NGC~1052-DF4) and two other typical luminous galaxies (UGC~12158 and M104). The galaxies are arranged in a clockwise order of increasing brightness on the same brightness scale, and their sizes are adjusted to be on the same spatial scales. The numbers below the galaxy IDs are the distances of the galaxies in mega-parsecs (Mpc, a distance unit). UGC~12158 is thought to resemble our own Milky Way Galaxy in appearance. M104 is a bright galaxy that has a spatial light profile most closely resembling that of a UDG, except for its brightness and the presence of a galactic disk. Arrows in the image of Dragonfly 44 indicate some of its globular clusters.}}
    \label{fig:gal_ex}
\end{figure}

UDGs typically appear as faint light patches in astronomical images. For example, three UDGs are shown on the left-hand side of \cref{fig:gal_ex} (labelled Dragonfly~44, DF~2, and DF~4), while two ``normal'' galaxies are shown on the right-hand side (UGC12158 and M104). In this figure, the spatial scales of these galaxies have been matched to highlight the substantial differences in the visibility of UDGs relative to ``normal" galaxies --- the UDGs have very low brightness, making them difficult to detect. UDGs are thus missing from many ``standard'' galaxy catalogs, and efforts have begun to adapt astronomical image analysis pipelines to enable efficient and accurate methods for detecting these enigmatic galaxies. 

%\gwen{You should mention the globular clusters in Figure 1 in the text, and to do that you need a short sentence or two about GCs around UDGs here, even if they are already mentioned in the abstract.}

Currently, UDGs are found using a computational search of their associated faint stellar light in images. The two major challenges with this method are: (1) the high computational cost of image segmentation needed to separate UDGs from the night sky background, and (2) contamination of samples of UDGs by imaging artifacts (such as scattered light) and light-reflecting diffuse gas clouds in the interstellar medium (so-called `Galactic cirrus'). Furthermore, this method could miss potential \textit{dark} UDGs --- ones with extremely faint stellar light that are of high astrophysical interest \citep[see][for more detailed discussions]{Li2022}.
% complements the shortcomings of finding UDGs by searching for their diffuse light patches.

Recently, \cite{Li2022} developed a log-Gaussian Cox process \citep[LGCP;][]{Moller1998} to detect UDGs via the spatial clustering signals of their associated globular clusters (GCs). Globular clusters are compact, spherical collections of hundreds of thousands of stars, and are easily detected, even at large distances. Numerous UDGs host GC populations \citep{Dokkum2016, Peng2016, Amorisco_2018, Lim2018, Lim2020, Danieli2021}, and these GCs typically appear on images as clusters of bright point sources on or near the body of the UDGs (e.g., Dragonfly 44 in \cref{fig:gal_ex}). 

\citeauthor{Li2022} treated the unexplained clustering of GCs in an LGCP as potential UDGs, and they successfully detected previously known UDGs in a \textit{Hubble Space Telescope} (\textit{HST}) imaging survey. Moreover, they detected a potential \textit{dark} galaxy, Candidate Dark Galaxy-1 (CDG-1).

Although an LGCP can detect UDGs, it suffers from several limitations:
\begin{enumerate}
    \item The LGCP cannot provide an inferential statement about the absence of a UDG in the data.  \citeauthor{Li2022} determined the existence of UDGs purely by visual inspection of the posterior distribution of the spatial random effect in an LGCP, which is only heuristic.
     \label{problem1}
     
    \item The underlying Gaussian process is problematic because it has inherent random fluctuations --- the LGCP may dismiss weaker GC clustering signals as noise. Moreover, the covariance structure of the Gaussian process has an implicit assumption  that clusters of GCs have the same physical sizes. Although a non-stationary covariance structure could alleviate the problem of varying cluster sizes, the exact structure required is unclear.
    \label{problem2}
       
    \item An LGCP lacks a physical interpretation. An LGCP assumes the GC point process intensity is generated through a Gaussian process, but this ignores the physical reality that galaxies are the progenitors of GCs. Moreover, the LGCP method only provides the locations of potential UDGs; it does not provide information about the physical properties of these GC systems, such as GC number and GC system size.
    \label{problem3}
 
    \item It is difficult to incorporate GC mark information in LGCP models. Physical theories and observations suggest that GCs in UDGs may have unique properties \citep{Lotz2004, Shen_2021b, van_Dokkum_2022, Saifollahi2022}. Thus, one can utilize these properties and consider a marked point process to potentially improve UDG detection. Although marked point process modeling under an LGCP framework has been studied extensively \citep[e.g.,][]{ho_modelling_2008, myllymaki_conditionally_2009, Diggle2010, waagepetersen_2016}, the mark distribution is linked to the Gaussian random field in these studies. This approach lacks the flexibility to incorporate physical theories and observations.
    \label{problem4}
\end{enumerate}

In this paper, we propose novel point process models to detect UDGs through the Poisson cluster process (PCP) framework \citep[e.g.,][]{Neyman1958, moller2003, moller2005} to tackle the issues of LGCP. The PCP is a doubly-stochastic process originating from the Neyman-Scott process \citep{Neyman1958}, which was initially developed to study galaxy clustering. A PCP creates point patterns by first generating unobserved cluster centers via a Poisson process. The observed points are then i.i.d. generated according to some probability density function around these centers. In our context, the unobserved cluster centers are the UDG centers, while GCs are the observed points associated with each UDG.

As noted by \citeauthor{Li2022}, the observed GCs originate from three distinct sub-populations: GCs in the intergalactic medium, GCs in luminous galaxies, and GCs in UDGs (see details in \cref{sec:GC suppop}). To detect UDGs, we must account for these GC sub-populations. We address this by modeling the GC point pattern in each sub-population using physically motivated parametric point process models, and we combine them using superposition, with the PCP specifically employed to model GCs from UDGs.

Our proposed models effectively address the issues of LGCP mentioned above. For issue \ref{problem1}, PCP can discern whether the data contain UDGs. For issue \ref{problem2}, PCP does not exhibit intrinsic intensity fluctuations as observed in a LGCP, ensuring the detection of weaker clustering signals. Moreover, PCP can seamlessly integrate physically motivated models, based on astronomical expertise, which allow each cluster of GCs to have varying physical sizes. This removes the restriction imposed by the covariance structure in a LGCP.\@ The usage of physical models also grants us the ability to deduce the physical characteristics of UDG GC systems, resolving issue \ref{problem3}. Lastly, for issue \ref{problem4}, incorporating GC mark information into the PCP framework is straightforward. 

Specifically, for the GC mark information, we introduce a novel marked PCP framework by modeling the GC marks as a mixture distribution based on the cluster assignment of observed points in the PCP.\@ Notably, parallel to the conception of this work, \cite{bu2023inferring} applied a similar idea to infer HIV transmission patterns. The application here demonstrates the versatility and applicability of the marked PCP framework in various disciplines.

To conduct inference, we design a bespoke blocked-Gibbs adaptive Metropolis birth-death-move Markov chain Monte-Carlo (MCMC) algorithm based on the adaptive Metropolis algorithm by \citep{Haario_2001, roberts2009} and the algorithm proposed by \cite{moller2005}. The use of this adaptive Metropolis scheme and our tailored proposal distributions for the birth-death-move update ensure good mixing of the chains.

We test our proposed models on data that include known UDGs; we fit our MCMC algorithm on the real GC data obtained from a \textit{HST} imaging survey. To compare our proposed models against LGCP, we develop a new overall performance assessment tool inspired by the ROC curve \citep{FAWCETT2006}. The traditional ROC curve is not suitable because we do not have the required spatial Boolean reference map \citep{Li2022}. Our adaptation, a $3$D-``ROC'' curve, overcomes this limitation by only requiring the known point locations of confirmed UDGs, while preserving the interpretability and functionality of the original ROC curve.

The paper is organized as follows: \cref{sec:data} and \cref{sec:method} describe the data and the modeling methodology respectively. \cref{sec:inf_est} introduces our MCMC algorithm as well as our performance assessment tool. \cref{sec:da} presents the data analysis results from fitting our models to the \textit{HST} data. \cref{sec:conclusion} provides conclusions and future directions of research.

%% file: sections/02-Data.tex
\section{Data \& Preliminaries}\label{sec:data}

\subsection{Data} The data considered here are from the Program for Imaging of the PERseus cluster (PIPER; \citealt{Harris2020}) survey, which are the same data used in \citeauthor{Li2022}. These data contain previously confirmed UDGs. The target region is the Perseus galaxy cluster at a distance of $75$ Megaparsec (Mpc) \footnote{One \textit{parsec} (pc) is approximately 3.2~light years or $3\times10^{16}$~km; One Mpc$=10^6$~pc and one kpc$=10^3$~pc.} \citep{Gudehus1995, Hudson1997}. The survey consists of 10 imaging visits. Each visit contains two images taken by the Advanced Camera for Survey (ACS) and the Wide Field Camera 3 (WFC3) on the
\textit{HST}\footnote{ACS and WFC3 are two different camera instruments abroad the \textit{HST}, with WFC3 offering broader wavelength coverage than ACS.}. Each image has an associated ID; for example, V11-ACS is the image taken by the ACS on the 11th visit. The GCs' locations are extracted from the images, and the GC brightnesses are measured by the magnitude in the F475W and F814W filters (\citealt{Harris2020}, each filter captures different wavelengths of light). Each image is roughly a square, with each side corresponding to $76$~kpc for the ACS camera and $62$~kpc for the WFC3. For simplicity, we scale all images and GC point patterns to $[0,1]^2$.

The color of a GC is defined as the difference between its magnitude in different wavelengths of light. Here it is the difference between the magnitude in $\mathrm{F475W}$ and $\mathrm{F814W}$:
\[
C = \mathrm{F475W} - \mathrm{F814W}. \nonumber
\]
The GC color $C$ is not the data we directly use in our model. We instead use the GC color variation $V$ which is derived from $C$. Essentially, $V$ is the difference between the color of a GC from the mean color of the GC sub-population it belongs to. However, mean GC color can vary between different GC sub-populations, e.g., the mean GC color can be different for different galaxies. This translates to a spatial variation in the mean of the GC color; furthermore, we do not know a priori which GC belong to which GC sub-population. We thus require the mean color variation across a spatial domain to obtain $V$. To this end, we consider a model-based approach to obtain $V$ (Section \ref{subsec: col_var}).

To validate our method, we use a confirmed UDG catalog from \cite{Wittmann2017}, and a UDG catalog compiled for the \textit{HST} proposal for the PIPER survey \citep{Harris2020}.

\subsection{GC Sub-populations}\label{sec:GC suppop}

Globular clusters are believed to originate from dense gas clouds within dark matter halos near infant galaxies in the early universe \citep{Forbes_2018, Madau_2020}. Over cosmic time, some GCs detach from their initial host galaxies. We assume the entire GC population in an image is comprised of the following three distinct, independent sub-populations:
\begin{itemize}[itemsep=0.5pt]
    \item \textbf{GCs in the Inter-Galactic Medium (IGM):} The Inter-galactic medium (IGM) is the material between galaxies. GCs in the IGM are believed to have been ejected from their host galaxies. The spatial distribution of GCs in the IGM is random and relatively uniform.
    
    \item \textbf{GCs in Luminous Galaxies}: Globular clusters are most abundant in luminous galaxies, such as spiral and elliptical galaxies. In our data, we only encounter elliptical galaxies, named after their ellipsoidal shape. The GC distribution in elliptical galaxies is inhomogeneous; the GC intensity is highest at the galactic center and drops off with distance away from the center.

    \item \textbf{GCs in UDGs}: Globular clusters in UDGs have a similar spatial distribution to that in elliptical galaxies, but the overall number of GCs in UDGs is generally much lower. 
\end{itemize}

\subsubsection{S\'{e}rsic Profile} To model the intensity functions of GCs in elliptical galaxies and UDGs, we adopt a widely used physical model called the S\'{e}rsic profile \citep{Sersic1963, Harris1991, Wang2013,Peng2016, van_Dokkum_2017, Dutta_Chowdhury_2019, Janssens_2022, Saifollahi2022}. This profile describes the intensity of GCs in a galaxy as a function of distance from the galactic center to a point $s=(s_x, s_y) \in \mathbb{R}^2$:
    \[
    \text{S\'{e}rsic}\{s; c, (\varlambda, R, n, \varphi,\rho)\} = \frac{\varlambda}{2\pi R^2n\Gamma(2n)\rho}\exp\left[-\left\{\frac{r(s; c, \varphi, \rho)}{R}\right\}^{1/n}\right],
    \]
    where
\begin{itemize}[itemsep=0.5pt]
    \item $c = (c_x, c_y) \in \mathbb{R}^2$ is the location of the galactic center,
    \item $\varlambda \in \mathbb{R}^+$ is rate parameter for the mean number of GCs in the host galaxy,
    \item $R\in \mathbb{R}^+$ is the characteristic size of the GC system,
    \item $n\in \mathbb{R}^+$ is the S\'{e}rsic index which describes the spatial concentration of the GCs,
    \item $\varphi\in (0, \pi)$ is the rotation angle of the host galaxy measured from the $x$-axis,
    \item $\rho \in \mathbb{R}^+$ is the ratio of the semi-axes of the GC system.
\end{itemize}
The function $r^2(s; c, \varphi,\rho)$ defines a 2D-quadratic function of $s$, and $\forall t>0$, $C_t = \{s \in \mathbb{R}^2: r(s; c, \varphi,\rho) = t\}$ is an ellipse centered at $c$ with rotation angle $\varphi$ and semi-axes ratio $\rho$: 
\[
r^2(s; c, \varphi, \rho) = \mathbf{r}'(s;c)\mathbf{H}^{-1}( \varphi,\rho)\mathbf{r}(s;c), \ \mathbf{r}(s;c) = (s_x - c_x, s_y - c_y)',
\]
and
\[
\mathbf{H}(\varphi,\rho) = \begin{bmatrix}
\cos^2\varphi +\rho^{2}\sin^2\varphi & \hspace{0.2cm} \sin\varphi\cos\varphi(\rho^{2}-1) \\[0.15cm]
            \sin\varphi\cos\varphi(\rho^{2}-1) & \hspace{0.15cm} \sin^2\varphi +\rho^{2}\cos^2\varphi \\
\end{bmatrix}.
\]

Note that the normalized S\'{e}rsic profile, $\text{S\'{e}rsic}\{s; c, (\varlambda, R, n, \rho, \varphi)\}/\varlambda$, is a 2D-Generalized Gaussian density \citep{dytso_analytical_2018}. If $n = 1/2$, it reduces to a 2D-Gaussian density.

\subsection{LGCP Model}\label{sec:lgcp}
For $n \in \mathbb{N}^+$, let $[n] = \{1,\dots,n\}$ and $\langle n\rangle =  \{0,\dots,n\}$. Suppose we observe GCs in a study region (an image) $S \subset \mathbb{R}^2$, with $s$ being a point in $S$. Denote $\mathbf{X} = \{\mathbf{x}_1, \dots, \mathbf{x}_n\} \subset S$ the point process for the GCs, which is an inhomogeneous Poisson process (IPP) with intensity $\Lambda(s) \geq 0$. \citeauthor{Li2022} proposed the following model based on the LGCP framework:
\begin{equation}
\begingroup
\allowdisplaybreaks
    \begin{aligned}\label{eqn:lgcp model}
    \mathbf{X} &\sim \text{IPP}\{\Lambda(s)\}, \\
    \log(\Lambda(s)) &= \beta_0 + \sum_{k = 1}^{N_G}f_{gal,k}(s; c_{k}^g, \bm{\beta}) + \mathcal{U}(s),\\
    \mathcal{U} &\sim \mathcal{GP}\{\mathbf{0}, \sigma^2\mathcal{M}_\nu(\cdot; h)\}.
\end{aligned}
\endgroup
\end{equation}
In \citeauthor{Li2022}'s strategy, they consider $\beta_0$ to capture the IGM GC intensity; $f_{gal,k}(s; c_{k}^g, \bm{\beta}), k \in [N_G]$ is a covariate term to account for GCs in elliptical galaxies, where $N_G$ is the number of elliptical galaxies in $S$ and $c_{k}^g$ is the center of the $k$-th elliptical galaxy. The term $\mathcal{U}(s)$ is a zero-mean Gaussian process with a Mat\'{e}rn covariance function $\sigma^2\mathcal{M}_\nu(\cdot; h)$ that captures unexplained clustering signals of GCs. Detection of UDGs is carried out by finding extreme value regions in the posterior distribution of $\exp\{\mathcal{U}(s)\}$.

\subsection{Generalized shot-noise Cox processes}

Our proposed Poisson cluster process (PCP) is a form of generalized shot-noise Cox process \citep{moller2005}, and is a three-level hierarchical process. The first level is a Poisson point process that generates the locations of parent points (cluster centers), the second level assigns a random number of offspring to each parent point, and the third level generates i.i.d. offspring locations around the parent points based on some p.d.f. that may be different for each parent point. Only the offspring locations are observed (the parent of each offspring is unknown).

Denote $\mathbf{X}_{\mathrm{off}} = \{\mathbf{x}_1, \dots, \mathbf{x}_{N_{\mathrm{off}}}\} \subset \mathbb{R}^2$ as the offspring point process, which is an IPP with intensity $\lambda(s) > 0$. The parent point process $\mathbf{X}_c = \{\mathbf{c}_1, \dots, \mathbf{c}_J\}\subset \mathbb{R}^2$ is an IPP with intensity $\lambda_c(s) > 0$. A PCP is written as:
\[\label{eq:pcp}
    \mathbf{X}_c &\sim \text{IPP}\{\lambda_c(s)\}, \\
    \lambda(s) \mid \mathbf{X}_c &= \sum_{j = 1}^J \varlambda_j h_j(s; \mathbf{c}_j), \\
    \mathbf{X}_{\mathrm{off}} \mid \lambda &\sim \text{IPP}\{\lambda(s)\}.
\]
In the above, $\varlambda_j, \ j \in [J]$ is the mean number of offspring points from the $j$-th parent and $h_j(\cdot; \mathbf{c}_j)$ is a p.d.f. on $\mathbb{R}^2$ centered at $\mathbf{c}_j$ with parameters that vary with $j$ but that are from the same parametric family. In our application, the parents $\mathbf{c}_j, \ j \in [J]$ represent the locations of UDGs, and the offspring points are GCs. Note that $J$ is a random variable.

The \cite{Neyman1958} process \citep[or NSP, see][]{moller2003} has a long history in astrophysics and has a number of similarities with the PCP presented here.   The original specification of the NSP has $\varlambda_j \equiv \varlambda$ and $h_j \equiv h, \ j \in [J]$,  although more recently the term ``Neyman-Scott process" has been used to refer to generalized Cox processes where these quantities  vary with $j$ \citep[see e.g.][]{hong2022, Wang2023}.  Allowing for cluster-specific $\varlambda_j$ and $h_j$ should not only improve cluster detection if the physical clusters have different sizes and shapes, but also provide scientific information about the cluster-to-cluster variation in the distribution of offspring.

%% file: sections/03-Method.tex
\section{Methods}\label{sec:method}

In this section, we present our models for GC point process. We treat the observed GC point pattern as an independent superposition of GCs from three distinct GC sub-populations mentioned in \cref{sec:GC suppop}, aiming to separate GCs in UDGs from others. The primary objective is inferring the unknown locations of UDGs from known GC point patterns, and we model the GCs from UDGs using a PCP model. As we are proposing several models with highly complex structures and numerous (hyper-) parameters, we provide a list of notation and symbols used (\cref{tab:notation}, \cref{appendix:notation}).

\subsection{Our Models}

We consider the three GC sub-populations mentioned in \cref{sec:GC suppop}
as three independent point processes:  $\mathbf{X}_{\mathrm{IGM}}$ with intensity $\Lambda_{\mathrm{IGM}}(s)$ for GCs in the IGM; $\mathbf{X}_G$ with intensity $\Lambda_G(s)$ for GCs in elliptical galaxies; and $\mathbf{X}_U$ with intensity $\Lambda_U(s)$ for GCs in UDGs. The set of observed GC locations is the union of the three processes, with $\mathbf{X} = \mathbf{X}_{\mathrm{IGM}} \cup \mathbf{X}_G \cup \mathbf{X}_U$, and the independence assumption gives
\[\label{eqn: superpose}
    \mathbf{X} \sim \text{IPP}\{\Lambda(s)\}, \qquad 
 \Lambda(s) = \Lambda_{\mathrm{IGM}}(s) + \Lambda_G(s) + \Lambda_U(s).
\]
Note that we only observe $\mathbf{X}$ and do not have membership information on a GC belonging to any one of $\mathbf{X}_{\mathrm{IGM}}$, $\mathbf{X}_G$ or $\mathbf{X}_U$. 

Motivated by the physical properties of the phenomena described in \cref{sec:GC suppop}, the three GC sub-populations are modelled as follows.
\begin{itemize}
    \item[\textbf{Inter-Galactic Medium}:] We model $\mathbf{X}_{\mathrm{IGM}}$ as a homogeneous Poisson process (HPP) with $\Lambda_{\mathrm{IGM}}(s) \equiv \beta_0 > 0$. Although the intensity of IGM GCs may exhibit radial inhomogeneity associated with the distance to the galaxy cluster center, we do not consider it here since such radial variations are negligible within the scope of each image \citep{Harris2020}.
    \item[\textbf{Luminous Galaxies}:] Assume $N_G$ elliptical galaxies are in $S$, with $\{c_k^g\}_{k=1}^{N_G}$ being their centers. Write the other S\'{e}rsic profile parameters as $\boldsymbol{G}_k = (\varlambda_k^g, R_k^g, n_k^g, \varphi_k^g, \rho_k^g), k \in [N_G]$. The superscript $g$ denotes the association with an elliptical galaxy. The $k$-th elliptical galaxy then has GC intensity function $\text{S\'{e}rsic}(s; c_k^g, \boldsymbol{G}_k)$. We assume that $\mathbf{X}_G$ is the union of the GC point processes from all $N_G$ elliptical galaxies and that each elliptical galaxy is independent, thus
    \[
    \Lambda_G(s) = \sum_{k=1}^{N_G}\text{S\'{e}rsic}(s; c_k^g, \boldsymbol{G}_k).
    \]
    For elliptical galaxies, the parameters $c, \varphi$, and $\rho$ are measured with high accuracy based on their spatial light distributions (e.g., using software such as SExtractor, \citealt{Bertin1996}). It has already been shown that the GC system of an elliptical galaxy generally shares these parameters with the spatial light distribution \citep{Harris1991, Wang2013}. Hence, we treat these three parameters as known, with other parameters being unknown since accurate estimates are unavailable.
    \item[\textbf{UDGs}:] We model GCs from UDGs using a PCP based on \cref{eq:pcp}. We assume that the parent point process is a homogeneous Poisson process with intensity $\lambda_c > 0$. We replace $\varlambda_jh_j(s;\mathbf{c}_j)$ in \cref{eq:pcp} by a S\'{e}rsic profile. Denote $N_U$ as the random variable for the number of UDGs and write $\boldsymbol{U}_j = (\varlambda_j^u, R_j^u, n_j^u, \varphi_j^u, \rho_j^u), \ j \in [N_U]$. The superscript $u$ denotes the association with a UDG. We have
    \[\label{eqn: PCP_UDG}
    \mathbf{X}_c &\sim \text{HPP}(\lambda_c), \\
    \Lambda_U(s) \mid \mathbf{X}_c &= \sum_{j = 1}^{N_U} \text{S\'{e}rsic}(s;  \mathbf{c}_j^u, \boldsymbol{U}_j), \\
     \mathbf{X}_U \mid \Lambda_U(s) &\sim \text{IPP}\{\Lambda_U(s)\}.
    \]
    $\mathbf{X}_c = \{\mathbf{c}_1^u, \dots, \mathbf{c}_{N_U}^u\}$ are then the UDGs centers and the goal of inference. Conditioning on $\mathbf{X}_c$ and $N_U$, the GC intensity in the $j$-th UDG follows a Sérsic profile with parameters $\boldsymbol{U}_j$ centered at $\mathbf{c}_j^u$. However, different from elliptical galaxies, we treat $(\mathbf{c}_j^u,\boldsymbol{U}_j),\ j \in [N_U]$ as unknown since we have no information about UDGs. 
\end{itemize}

\subsubsection{Model 1: Unmarked Process}\label{sec:model1}
Let $\boldsymbol{\Xi} = (\beta_0, \boldsymbol{\Gamma}, \boldsymbol{\Upsilon})$, $\boldsymbol{\Gamma} = \{\boldsymbol{G}_k\}_{k=1}^{N_G}$, and $\boldsymbol{\Upsilon} = \{\boldsymbol{U}_j\}_{j = 1}^{N_U}$ where $\boldsymbol{G}_k$ and $\boldsymbol{U}_j$ are as before. Based on \cref{eqn: superpose} and the model structures outlined previously, we present below our first model for detecting UDGs using only GC point patterns:
\begin{model}\label{model1}
\begingroup
\allowdisplaybreaks
\begin{align*}
    \mathbf{X}_c &\sim \text{HPP}(\lambda_c), \\
    \Lambda(s)\mid \mathbf{X}_c, \boldsymbol{\Xi} &= \beta_0 + \sum_{k=1}^{N_G}\text{S\'{e}rsic}(s;  c_k^g, \boldsymbol{G}_k) + \sum_{j=1}^{N_U}\text{S\'{e}rsic}(s; \mathbf{c}_j^u, \boldsymbol{U}_j), \\
    \mathbf{X} \mid \Lambda &\sim \text{IPP}\{\Lambda(s)\}.
\end{align*}
\endgroup
\end{model}
Our model, compared to the LGCP model, offers several key advantages. Firstly, the PCP can accommodate scenarios with no clusters, enabling direct inferential statements about the presence of UDGs in an image through $\mathbb{P}(N_U = 0 \mid \mathbf{X})$. Secondly, unlike the LGCP, our model consistently searches for GC clustering signals without inherent fluctuations in GC intensity caused by a Gaussian random field, thus reducing the likelihood of missing any signals. Lastly, our method incorporates a physically motivated model to explicitly describe the UDG GC intensity function. This approach not only detects UDGs but also derives their physical properties, embodied by the S\'{e}rsic parameters, offering insights valuable to astrophysicists.

\subsubsection{Model 2: Marked Process}

Physical theories and observations may imply special properties (marks) of GCs in UDGs, which we can utilize to potentially enhance detection.Thus, we incorporate GC marks into Model \ref{model1} to introduce a marked point process model.

Recall the GC point process $\mathbf{X} = \{\mathbf{x}_1, \dots, \mathbf{x}_n\}$. Denote by $\mathbf{M} = \{\mathbf{M}(\mathbf{x}_1), \dots,\mathbf{M}(\mathbf{x}_n)\}$ the mark of the GCs. We model the conditional distribution of $\mathbf{M}$ given all other information using a mixture model with $M = N_U + 1$ components. The $m$-th mixture component is characterized by a random variable $\mathcal{M}_{\Sigma_m}, m \in \langle N_U\rangle$, with parameters $\Sigma_m$. We write the corresponding probability distribution as $\pi_{\Sigma_m}\{\mathbf{M}(\mathbf{x}_i)\}$. A value of $m=0$ denotes the environment of IGM or luminous galaxies, and $m > 0$ is one of the $N_U$ UDGs. We introduce an indicator variable $Z_i \in \langle N_U\rangle$ to represent the environment of the $i$-th GC. Writing $\boldsymbol{\Sigma} = \{\Sigma_m\}_{m=0}^{N_U}$
we have,
by the law of total probability:
\[
    \mathbf{M}(\mathbf{x}_i) \mid \mathbf{x}_i, \mathbf{X}_c, \boldsymbol{\Xi}, \boldsymbol{\Sigma} \sim \sum_{m=0}^{N_U} p_{im}\mathcal{M}_{\Sigma_m}.
\]
Here $p_{im}$ is the full conditional probability that the $i$-th GC belongs to the $m$-th environment,
\[
p_{im} = \mathbb{P}\left(Z_i = m \mid \mathbf{x}_i = x_i, \mathbf{X}_c, \boldsymbol{\Xi}, \boldsymbol{\Sigma}\right) = \frac{\lambda_m(x_i)}{\sum_{k=0}^{N_U}\lambda_k(x_i)} = \frac{\lambda_m(x_i)}{\Lambda(x_i)},
\]
where $\lambda_m$ is the intensity of the $m$-th environment: $\lambda_0(s) = \beta_0 + \Lambda_G(s)$, and $\lambda_m(s) = \text{S\'{e}rsic}(s; \mathbf{c}_m^u, \boldsymbol{U}_m)$, $ m \in [N_U]$. Derivation of $p_{im}$ is in Section 2 of the Supplementary Material. 

With the above model structure, we present below a marked point process model for detecting UDGs using both GC location and mark information:
\vspace{0.25cm}
\begin{model}\label{model2}
\begingroup
\allowdisplaybreaks
\begin{align*}
    \mathbf{X}_c &\sim \text{HPP}(\lambda_c), \\
    \Lambda(s)\mid \mathbf{X}_c, \boldsymbol{\Xi} &= \beta_0 + \sum_{k=1}^{N_G}\text{S\'{e}rsic}(s;  c_k^g, \boldsymbol{G}_k) + \sum_{j=1}^{N_U}\text{S\'{e}rsic}(s; \mathbf{c}_j^u, \boldsymbol{U}_j), \\
    \mathbf{M}(\mathbf{x}_i) \mid \mathbf{x}_i, \mathbf{X}_c, \boldsymbol{\Xi}, \boldsymbol{\Sigma} &\sim \sum_{m=0}^{N_U} p_{im}\mathcal{M}_{\Sigma_m}, \\
    p_{i0} = 1-\Lambda_U(\mathbf{x}_i)/\Lambda(\mathbf{x}_i) &, \ p_{im} = \text{S\'{e}rsic}(\mathbf{x}_i; \mathbf{c}_m^u, \boldsymbol{U}_m)/\Lambda(\mathbf{x}_i), \ m \in [N_U] \\
    \mathbf{X} \mid \Lambda &\sim \text{IPP}\{\Lambda(s)\}.
\end{align*}
\endgroup    
\end{model}

Model \ref{model2} offers another notable advantage over the LGCP by allowing the explicit incorporation of physical theories and observations about the mark distribution of GCs in UDGs. This is achieved by embedding these assumptions into the prior distribution of $\boldsymbol{\Sigma}$. Such embedding is possible under PCP since we can construct direct physical models for GC intensity functions and explicitly assign each GC to a GC sub-population.

Specifically, denote a physical theory/observation about a mark of GCs in UDGs by $\mathcal{P}_M$. The prior distribution of $\boldsymbol{\Sigma}$ when $\mathcal{P}_M$ is not considered is $\pi(\boldsymbol{\Sigma}) = \prod_{m=0}^{N_U}\pi(\Sigma_m)$. Once $\mathcal{P}_M$ is imposed, it constrains the parameter space of $\Sigma_m, \ m \in [N_U]$ based on $\Sigma_0$, since $\mathcal{P}_M$ leverages $\Sigma_0$ as a reference to differentiate GCs in UDGs from GCs in other sub-populations. Hence, the prior distribution of $\boldsymbol{\Sigma}$ is now $\pi(\boldsymbol{\Sigma}) = \prod_{m=1}^{N_U}\pi(\Sigma_m \mid \Sigma_0)\pi(\Sigma_0)$ where the prior of $\Sigma_m, \ m \in [N_U]$ are conditionally independent given $\Sigma_0$. The exact formulation of $\pi(\Sigma_m \mid \Sigma_0)$ depends on $\mathcal{P}_M$. 

%In contrast, in the LGCP model, where the intensity functions of UDGs are entirely driven by a Gaussian process, it's impossible to provide direct physical models for GC intensity in UDGs. Thus, LGCP lacks a mechanism for GC membership assignment, and the introduction of the mark parameters $\Sigma_m$ and $\mathcal{P}_M$ is not possible. 

We now illustrate how to incorporate physical assumptions regarding GC marks into our model by considering the brightness and the color variation of GCs. 

\paragraph{GC Brightness}

The GC brightness (magnitude) distribution is called the globular cluster luminosity function (GCLF), and it is well-characterized by a Gaussian distribution \citep{harris_globular_1979, Harris1991, Rejkuba_2012}. However, due to telescope detection limits, fainter GCs are not observed, and thus the observed GCLF is right-truncated. \footnote{Due to historical reasons, brighter objects are defined to have more negative values, while fainter objects have more positive values.} 

Denote by $\text{Mag}(\mathbf{x}_i)$ the magnitude of the $i$-th GC. Let the known detection limit be $\text{Mag}_t$, which for our dataset is $\text{Mag}_t = 25.5$~mag. We set $\text{Mag}(\mathbf{x}_i) \mid Z_i = m \sim \mathcal{M}_{\Sigma_m} = \mathcal{N}_t(\mu_m, \sigma_{m,B}^2; \text{Mag}_t)$ with $\Sigma_m = (\mu_m, \sigma_{m,B}^2)$, which is a right-truncated normal distribution with truncation value $\text{Mag}_t$. As mentioned, some UDGs have abnormally brighter GCs on average than GCs from other environments \citep{Shen_2021b}. Therefore, we can impose the prior condition that $\mu_m > \mu_0, \forall m > 0$. Note that we only consider the GC magnitude in F814W. The magnitude in F475W generally does not provide additional information since the GCLFs in F475W and F814W are typically shifted by a constant amount with the standard deviation unchanged. 

\paragraph{Color Variations}\label{subsec: col_var}

Physical observations and theories indicated that GCs in UDGs may have smaller color variations than GCs from other environments \citep{Lotz2004, van_Dokkum_2022, Saifollahi2022}. Thus, GC color variations may be used as mark. As mentioned in \cref{sec:data}, we need to first acquire the GC color variations $V$ through GC colors. Suppose $C(\mathbf{x}_i)$ is the color of the $i$-th GC and $V(\mathbf{x}_i)$ is the color variation of the $i$-th GC. We derive $V(\mathbf{x}_i)$ from $C(\mathbf{x}_i)$ by treating $C(\mathbf{x}_i)$ as a geostatistical process governed by a Gaussian random field due to the spatial variation in $C(\mathbf{x}_i)$. Specifically, we consider
\[\label{eqn:col_var}
    C(\mathbf{x}_i) &\sim \mathcal{N}(\mu_C + \mathcal{W}(\mathbf{x}_i), \sigma_C^2), \ i = 1, \dots, n \\
    \mathcal{W}(s) &\sim \mathcal{GP}(0, \sigma_{\mathcal{W}}^2\mathbf{C}_{\mathcal{W}}(\cdot; h_w)).
\]
Here, $\mu_C$ is the population mean GC color in an image $S$, $\mathcal{W}(s)$ is the spatial variation in the mean GC color at $s \in S$, $\sigma_C$ is the observational-level noise of GC color, and $\mathbf{C}_{\mathcal{W}}(\cdot; h_w)$ is a Mat\'{e}rn covariance function with $\nu = 1$ and length scale $h_w$ and variance $\sigma_{\mathcal{W}}^2$. The color variation is then obtained by
\[
V(\mathbf{x}_i) = C(\mathbf{x}_i) - \hat{\mu}_C - \hat{\mathcal{W}}(\mathbf{x}_i),
\]
where $(\hat{\mu}_C, \hat{\mathcal{W}}(\mathbf{x}_i))$ are the posterior mean of $(\mu_C, \mathcal{W}(\mathbf{x}_i))$. 

We fit the above model using the integrated-nested Laplace approximation \citep[INLA;][]{Rue2009}. INLA is a fast approximate Bayesian inference method through numerical integration that provides an accurate approximation of the posterior distribution \citep[see][for theoretical guarantee on closely related methods]{Bilodeau_2024}. Further details such as the prior setting for parameters in \cref{eqn:col_var} are contained in Section 1 of the Supplementary Materials.

Given $V(\mathbf{x}_i)$, we model $V(\mathbf{x}_i) \mid Z_i = m \sim \mathcal{M}_{\Sigma_m} = \mathcal{N}(0, \sigma_{m,C}^2)$, where $\Sigma_m = \sigma_{m,C}$. We then impose a condition on the prior distributions with $\sigma_{0,C} > \sigma_{m,C}$, $\forall m > 0$, which represents the assumption that color variation of GCs in UDGs is smaller on average than other GC sub-populations. We also consider the scenario where $V$ is affected by error since the error can be quite high relative to the measured values. The error includes the uncertainty of $V$ through \cref{eqn:col_var}, and the measurement error of the color. We then have $V(\mathbf{x}_i) \mid Z_i = m \sim \mathcal{N}(0, \sigma_{m,C}^2 + e_i^2)$ where $e_i^2 = e_{i, P}^2 + e_{i, M}^2$. Both $e_{i, P}$ and $e_{i, M}$ are known, and they are respectively the error from model prediction and the measurement for the $i$-th GC.

\subsection{Priors}

The priors and hyper-priors for our model parameters and hyperparameters are listed in \cref{tab:prior}. We also provide the physically meaningful parameter space for these parameters. The choice of priors and hyper-prior is based on extensive previous astrophysical research \citep[e.g.][]{van_Dokkum_2017, Burkert2017, Forbes2017, Amorisco_2018, Lim2018}. However, the detailed motivation and explanation for the choice of the priors and hyper-prior require substantial astrophysical background, which is out of the scope of the main paper. We thus defer these details to Section 4 of the Supplementary Material. 
\begin{table}[t]
\begin{center}
\resizebox{\columnwidth}{!}{%
\begin{tabular}{llll}
    \toprule
    Components & (Hyer-)Parameters & (Hyer-)Prior & Motivation \\
      \midrule
    Mean UDG Count & $\lambda_c \in (0,\infty)$ & $\mathrm{Unif}(0,5)$ & No Information \\
    \midrule
    \multirow{5}{*}{UDGs, \hspace{0.05cm} $j \in [N_U]$} & $\varlambda_j^u\in (0,\infty)$ & $\mathrm{LN}(\log(7.6), 0.87^2)$ & Empirical \\
    & $R^u_j\in (0,\infty)$ & $\mathrm{LN}(\log(0.03), 0.5^2)$ &  Empirical \\
    & $n^u_j\in (0,\infty)$ & $\mathrm{LN}(\log(1), 0.75^2)$ & Empirical \\
    & $\varphi^u_j\in [0,\pi)$ & $\mathrm{Unif}(0, \pi)$ & No Information \\
    & $\rho^u_j\in (0,\infty)$ & $\mathrm{LN}(\log(1), 0.3^2)$ & Empirical \\
    \midrule
    \multirow{3}{*}{Elliptical Galaxies,\hspace{0.05cm} $k \in [N_G]$} & $\varlambda_k^g\in (0,\infty)$ & $\mathrm{LN}(\log(\mathcal{N}_k^g), 0.25^2)$ & Empirical ($\mathcal{N}_k^g$ varies between galaxies) \\
    & $R^g_k\in (0,\infty)$ & $\mathrm{LN}(\log(\mathcal{R}_k^g), 0.25^2)$ & Empirical ($\mathcal{R}_k^g$ varies between galaxies) \\
    & $n^g_k\in (0,\infty)$ & $\mathrm{LN}(\log(0.5), 0.5^2)$ & Empirical\\
    \midrule
    IGM & $\beta_0\in (0,\infty)$ & $\mathrm{LN}(\log(b_0), 0.5^2)$ & Empirical ($b_0$ varies between images) \\
     \midrule
    \multirow{3}{*}{Mark: Magnitude} & $\mu_0\in \mathbb{R}$ & $\mathrm{Unif}(23, 27)$ & Empirical \\
    & $\mu_m \mid \mu_0 \in \mathbb{R}$ & $\mathrm{Unif}(23, \mu_0), \ m \in [N_U]$ & Empirical and Assumption \\
    & $\sigma_{m,B}\in (0,\infty)$ & $\mathrm{Unif}(0.5, 1.9), \ m \in \langle  N_U \rangle$ & Empirical \\
    \midrule
    \multirow{2}{*}{Mark: Color} & $\sigma_{0, C}\in (0,\infty)$ & $\mathrm{Gamma}(0.2, 0.05)$ & Empirical \\
    & $ \sigma_{m,C} \mid \sigma_{0,C}\in (0,\infty)$ & $\mathrm{Unif}(0, \sigma_{0,C}), \ m \in [N_U]$ & Empirical and Assumption \\
 \bottomrule
\end{tabular}%
}
\end{center}
\caption{Prior and hyper-prior for our model components. The last column provides the reasoning for choosing such a prior. LN stands for log-normal. See the Section 4 of the Supplementary Material for detailed explanation and references.}
\label{tab:prior}
\end{table}

%% file: sections/04-Inference.tex
\section{Inference \& Computational Methods}\label{sec:inf_est}

\subsection{Inference} The primary inference goal is the locations of UDGs, $\mathbf{X}_c$, with the secondary inference goal being the physical parameters of the UDGs. We denote the data by $\mathbf{D} = \mathbf{X}$ for a model without marks or $\mathbf{D} = (\mathbf{X}, \mathbf{M})$ 
for a model with marks. Due to the complex structure of our models, conducting inference for the UDG locations and their physical parameters requires full posterior distributions through Bayesian inference. Traditionally, PCP models are fitted using methods such as moment-matching, minimum-contrast, or maximum likelihood \citep[e.g.][]{waagepetersen_2007, diggle_2013, moller2014, BADDELEY2022} for parameter estimation not concerned with cluster parameters. However, for inference of cluster locations, the task is much more difficult. A limited number of works \citep[e.g.][]{castelloe1998, moller2014} have previously managed to conduct full Bayesian inference on real-world data using the reversible jump MCMC (RJMCMC; \citealt{Green1995}) or its variants. Our problem is substantially more complicated, however, and our novel models require innovative inference techniques.
% However, the models considered in previous works are much simpler than the ones we propose, thus the inference algorithm is sub-optimal for our purpose. 
More recently, the collapsed Gibbs sampling under mixture of finite mixture models has been used \citep{Wang2023} to facilitate fast inference for clustering problems. Such methods require conjugacy to integrate out the physical parameters of UDGs so that the collapsed Gibbs sampling can be leveraged \citep{Wang2023}, and thus is not suitable for our problem. \cite{hong2023variational} considered variational inference for prediction problems under PCP models, but their method has no guarantee of posterior convergence. Thus, to conduct inference, we design a bespoke MCMC algorithm tailored to our models.

\subsection{Blocked-Gibbs Adaptive-Metropolis Birth-Death-Move Algorithm}\label{sec:GB_ABDM_MH}

We group our model parameters as $\boldsymbol{\Theta} = (\beta_0, \lambda_c, \boldsymbol{\Gamma}, \Sigma_0)$, $\boldsymbol{\Phi} = \{(\mathbf{c}_j^u, \boldsymbol{U}_j, \Sigma_j)\}_{j=1}^{N_U}$. $\boldsymbol{\Theta}$ is the parameter block that is always present in our models while $\boldsymbol{\Phi}$ is the trans-dimensional part. $\boldsymbol{\Phi}$ has varying cardinality as $N_U$ is random. We construct a blocked-Gibbs algorithm by first sampling $\boldsymbol{\Theta}$ using an adaptive Metropolis update \citep{Haario_2001, roberts2009} and then sampling $\boldsymbol{\Phi}$ through an improved birth-death-move update based on \cite{Geyer1994, moller2005}. The adaptive Metropolis ensures $\boldsymbol{\Theta}$ does not jeopardize the mixing performance of the entire algorithm. We also construct specialized proposal distributions for $\boldsymbol{\Phi}$ to improve upon the mixing performance of the algorithm based on \cite{Geyer1994, moller2005}.

\subsubsection{Algorithm Details}
We illustrate our algorithm based on Model \ref{model2}. The algorithm for Model \ref{model1} is similar. Assume $N_U$ is temporarily fixed. Denote $|\cdot|$ the Lebesgue measure. The posterior distribution with fixed $N_U$ is
\[\label{eqn:likelihood}
    \pi(\boldsymbol{\Theta}, \boldsymbol{\Phi} \mid \mathbf{D}) &\propto \exp\left(-\int_S \Lambda(s)ds\right)\prod_{i = 1}^n\left\{\Lambda(\mathbf{x}_i)\left[\sum_{m = 0}^{N_U}p_{im}\pi_{\Sigma_m}(\mathbf{M}(\mathbf{x}_i))\right]\right\}\pi(\boldsymbol{\Theta}, \boldsymbol{\Phi}).
\]
The prior $\pi(\boldsymbol{\Theta}, \boldsymbol{\Phi})$ is
\begin{multline}\label{eqn:prior}
    \pi(\boldsymbol{\Theta}, \boldsymbol{\Phi}) \propto  
    \\
    \exp(-\lambda_c|S|)\left[\prod_{m = 1}^{N_U}\lambda_c\pi(\Sigma_m \mid \Sigma_0)\pi(\boldsymbol{U}_j)\right]
\left[\prod_{k=1}^{N_G}\pi(\boldsymbol{G}_k)\right]\pi(\beta_0)\pi(\lambda_c)\pi(\Sigma_0).
\end{multline}
The integration of $\Lambda(s)$ on $S$ in \cref{eqn:likelihood} is computed by a fine-grid approximation, discretizing $S$ into equidistant cells and assuming $\Lambda(s)$ in each cell is constant.

Suppose the state of the chain is $(\boldsymbol{\Theta}, \boldsymbol{\Phi})$ at an update step. The algorithm is as follows.
\begin{enumerate}
    \item \textbf{Adaptive Metropolis (AM)} For $\boldsymbol{\Theta}$, we sample a new state $\boldsymbol{\Theta}'$ from the conditional posterior $\pi(\boldsymbol{\Theta} \mid \boldsymbol{\Phi}, \mathbf{D}) \propto \pi(\boldsymbol{\Theta}, \boldsymbol{\Phi} \mid \mathbf{D})$ according to a proposal density $q_\theta(\cdot \mid \cdot)$ based on an adaptive Metropolis update \citep{Haario_2001}:
\begin{equation}
  \boldsymbol{\Theta}_{i+1}\mid \boldsymbol{\Theta}_i \sim
    \begin{cases}
      \mathcal{N}(\boldsymbol{\Theta}_i, C_\theta) & \text{$i \leq 1000$},\\
      \mathcal{N}(\boldsymbol{\Theta}_i, \gamma\mathrm{Cov}(\{\boldsymbol{\Theta}_j\}_{j=1}^i)  + \gamma\varepsilon \boldsymbol{I}) & \text{$i > 1000$},
    \end{cases}       
\end{equation}
$C_\theta$ is a diagonal matrix, which is user-defined. $\gamma > 0$ is a scaling factor with $\gamma = 2.38^2/\mathrm{dim}(\boldsymbol{\Theta})$ based on the optimal scaling result from \cite{Haario_2001, Roberts_2001}. $\varepsilon > 0$ is user-defined to ensure the invertibility of the covariance matrix of the adaptive kernel. The proposed state $\boldsymbol{\Theta}^*$ is then accepted with probability
\[
A_{\text{AM}} = \min\left\{1, \frac{\pi(\boldsymbol{\Theta}^*, \boldsymbol{\Phi} \mid \mathbf{D})}{\pi(\boldsymbol{\Theta}, \boldsymbol{\Phi} \mid \mathbf{D})}\right\}.
\]
Components of $\boldsymbol{\Theta}$ are either transformed into log-scale or logit-scale based on their priors.

\begin{table}[t]
\begin{center}
\begin{tabular}{lll}
    \toprule
      Component & Parameters & Proposals \\
      \midrule
\multirow{3}{*}{UDGs} & $\mathbf{c}$ & $\mathbf{c} \mid \mathbf{c}_j^u, \boldsymbol{U}_j, \Sigma_j \sim \mathcal{N}(\mathbf{c}_j^u, (\delta_cR^u_j)^2)$  \\
     & $\boldsymbol{U}_{-\varphi}$ & $\log(\boldsymbol{U}_{-\varphi}) \mid \boldsymbol{U}_{j, -\varphi} \sim \mathcal{N}(\log(\boldsymbol{U}_{j, -\varphi}), \Delta_U)$ \\
     & $\varphi^u$ & $\varphi \mid \varphi_j^u \sim \mathcal{N}(\varphi_j^u, \delta_\varphi^2)$ \\
     \midrule
    \multirow{2}{*}{Mark: Magnitude} & $\mu$ & $\mu \mid \mu_j \sim \mathcal{N}(\mu_j, \delta_\mu^2)$ \\
    & $\sigma_{B}$ & $\sigma_{B} \mid \sigma_{j,B} \sim \mathcal{N}(\sigma_{j,B}, \delta_{\sigma,B}^2)$ \\
    \midrule
    \multirow{1}{*}{Mark: Color}& $\sigma_{C}$ & $\log(\sigma_{C}) \mid \sigma_{j,C} \sim \mathcal{N}(\log(\sigma_{j,C}), \delta_{\sigma,C}^2)$ \\
 \bottomrule
\end{tabular}
\end{center}
\caption{Proposal densities of parameters when a move-step is considered in the spatial birth-death-move transition kernel.}
\label{tab:proposal}
\end{table}

\item \textbf{Birth-Death-Move Metropolis-Hasting (BDM-MH)} For $\boldsymbol{\Phi}$, given the previously sampled $\boldsymbol{\Theta}'$, we sample a new state $\boldsymbol{\Phi}'$ from the conditional posterior $\pi(\boldsymbol{\Phi} \mid \boldsymbol{\Theta}', \mathbf{D}) \propto \pi(\boldsymbol{\Theta}', \boldsymbol{\Phi} \mid \mathbf{D})$ using a birth-death-move Metropolis-Hasting \citep[BDM-MH;][]{Geyer1994} scheme. At each update step, one can propose to add a new UDG (birth), remove an existing UDG (death), or tweak the configuration of an existing UDG (move).
We set the respective probabilities for birth, death, and move proposal to be $p_b = 1 - p_d = 1/2$ and $p_m = 1/3$ \footnote{Interpretation of $p_b$, $p_d$, and $p_m$ follows that in \cite{moller2005} where we first determine whether the update is a move step based on $p_m$; If it is not a move-step, then birth or death is chosen based on $p_b$ and $p_d$.}. If a birth step is chosen, we uniformly generate a new UDG at $\mathbf{c} \in S$. We generate its associated S\'{e}rsic profile parameters $\boldsymbol{U}$ and the GC mark distribution parameter $\Sigma$ according to proposal densities $q_b^u(\cdot)$ and $q_b^s(\cdot)$, which are their respective prior distributions. If a death step is chosen, an existing UDG is then randomly chosen to be removed. The birth-death Metropolis-Hasting ratio is then \citep{Geyer1994, moller2005}:
\begin{align}\label{eqn:BDMH_ratio}
    r[(\boldsymbol{\Theta}, \boldsymbol{\Phi}), (\boldsymbol{\Theta}, (\mathbf{c}, \boldsymbol{U}, \Sigma))] &= \frac{p_d\pi(\boldsymbol{\Theta}, \boldsymbol{\Phi} \cup \{(\mathbf{c}, \boldsymbol{U}, \Sigma)\} \mid \mathbf{D})|S|}{p_b\pi(\boldsymbol{\Theta}, \boldsymbol{\Phi} \mid \mathbf{D})(|\boldsymbol{\Phi}|+1)q_b^u(\boldsymbol{U})q_b^s(\Sigma)}.
\end{align}
The move step, on the other hand, follows a standard Metropolis-Hasting update scheme. For the proposal distribution of $\boldsymbol{\Phi}$, we set the move step proposals as follows: suppose the $j$-th existing cluster $(\mathbf{c}_j^u, \boldsymbol{U}_j, \Sigma_j)$ is chosen, then write $\boldsymbol{U}_{-\varphi} = (\varlambda^u, R^u, n^u, \rho^u)$ and $\boldsymbol{U}_{j, -\varphi} = (\varlambda_j^u, R_j^u, n_j^u, \rho_j^u)$. We propose a new cluster $(\mathbf{c}, \boldsymbol{U}, \Sigma)$ based on the densities listed in \cref{tab:proposal}. 

\end{enumerate}

In \cref{tab:proposal}, $(\delta_c,\Delta_U,\delta_\varphi, 
\delta_\mu,\delta_{\sigma,B},\delta_{\sigma,C})$ are all user-defined scale parameters. We provide the details of these choices in Section 5 of the Supplementary Material. Our proposal for the UDG center $\mathbf{c}$ adopts an ``adaptive" approach, where $\mathbf{c}$ depends on the radius $R^u_j$ of the UDG GC system. A smaller $R^u_j$ leads to less uncertainty in the central location, while a larger radius increases this uncertainty. Our choice of proposal addresses inefficiencies in the original algorithm by \cite{moller2005}, who used a uniform distribution to propose moves for $(\mathbf{c},\boldsymbol{U},\Sigma)$. Our method enhances the mixing of the chain by using the move step proposal for precision adjustments and the birth-death steps for exploration.

Given the above two update schemes, we sample the entire posterior distribution by iteratively updating $\boldsymbol{\Theta}$ and $\boldsymbol{\Phi}$ using a blocked-Gibbs sampling scheme:
\begin{itemize}
    \item Sample $\boldsymbol{\Theta}' \sim \pi(\boldsymbol{\Theta} \mid \mathbf{D}, \mathbf{\Phi}) \propto \pi(\boldsymbol{\Theta},\mathbf{\Phi} \mid \mathbf{D})$ through AM.
    \item Sample $\boldsymbol{\Phi}' \sim \pi(\boldsymbol{\Phi} \mid \mathbf{D}, \mathbf{\Theta}') \propto \pi(\boldsymbol{\Theta}', \mathbf{\Phi} \mid \mathbf{D})$ through BDM-MH.
\end{itemize}
We provide a preliminary convergence analysis of our algorithm in Section 5 of the Supplementary Material. The conditions that guarantee the convergence of our algorithm require novel theoretical results that are not yet available, and are out of the scope of this paper.

\subsection{Existence and Locations of UDGs}

\subsubsection{UDG Existence}\label{subsec: udg_existence}

After a posterior sample is obtained using the above MCMC algorithm, we first infer whether there are UDGs in an image. Let $\{\lambda_{c,i}^{pos}\}_{i=1}^{n_{pos}}$ be the posterior samples of $\lambda_c$ obtained using the method in \cref{sec:GB_ABDM_MH}. The probability that there is at least one UDG is $p_0^{pos} \equiv \mathbb{P}(N_U > 0\mid \mathbf{D})$ and is estimated from the posterior predictive distribution:
\[
\mathbb{P}(N_U > 0 \mid \mathbf{D}) = \int \mathbb{P}(N_U > 0 \mid \lambda_c)\pi(\lambda_c \mid \mathbf{D})d\lambda_c\approx 1-\frac{1}{n_{pos}}\sum_{i=1}^{n_{pos}}\exp(-\lambda_{c,i}^{pos}|S|).
\]
We determine the existence of UDGs by comparing $p_0^{pos}$ and the corresponding prior probability $p_0^{pri}\equiv \mathbf{P}(N_U > 0)$. If a UDG likely exists, the data will shift the posterior of $\lambda_c$ away from $0$ compared to the prior, resulting in $p_0^{pos} > p_0^{pri}$ and vice versa. 

Note that we recommend the above approach only as a reference instead of a gold standard. If there are very few UDGs with weak GC clustering signals, the likelihood may not be highly informative. Thus, in cases where $p_0^{pos} $ and $ p_0^{pri}$ are similar (within $5\%$ difference), we suggest to also visually inspect the posterior distribution of the UDG locations $\mathbf{X}_c$.

%We compare (i) $\mathbb{P}(N_U = 0 \mid \mathbf{D})$ and (ii) the posterior probability an UDG exists at a certain location. Comparing $\mathbb{P}(N_U = 0 \mid \mathbf{D})$ to $\mathbb{P}(N_U = k \mid \mathbf{D})$ for some $k > 0$ is not a suitable rule to determine the existence of UDGs due to the well-known label-switching problem \citep{Jasra_2005} of RJMCMC-type algorithms. For example, the posterior samples where $N_U = 1$ can represent multiple spurious clusters of GCs at different locations, thus $N_U = 1$ does not mean an actual UDG exists. We estimate (i) using the proportion of $N_U = 0$ in the posterior sample of $N_U$. For (ii), we estimate
%\[\mathbb{P}(\mathrm{UDG \ in } \ \omega \mid \mathbf{D}) = \mathbb{P}(\mathrm{UDG \ in } \ \omega \mid \mathbf{D}, N_U > 0)(1-\mathbb{P}(N_U = 0 \mid \mathbf{D})),
%\] 
%for any fixed region $\omega \subset S$ where $\omega$ has a size that can only contain no more than one UDG. We first discretise $S$ into equal-sized cells $\omega_i$. The first term in the RHS above is estimated by the ratio of the number of sampled UDG centers in $\omega_i$ to the total number of sampled UDG centers. If the maximum of the estimated quantity for all $\omega_i$ is smaller than the estimate of $\mathbb{P}(N_U = 0 \mid \mathbf{D})$, we conclude there is no UDG in the pointing and vice versa. For the cell size, we set its side length to $10$ kpc, which ensures that all the sampled UDG centers in a cell represent the same UDG, but not too small that the estimated probability is arbitrarily smaller than $\mathbb{P}(N_U = 0 \mid \mathbf{D})$.

\subsubsection{Inferring UDG Locations}\label{subsec: udg_detection}
We infer the locations of UDGs using the posterior samples of $\mathbf{X}_c \mid \mathbf{D}$ from our MCMC algorithm. For $\alpha \in (0,1)$, we define the $\alpha$-confidence level detection regions of UDGs as
\[\label{eqn: detection region}
B_{\mathbf{D}}(\alpha) = \argmin_{B\subset S}\{|B|: \mathbb{P}(\mathbf{X}_c \subset B \mid \mathbf{D}) \geq \alpha\}.
\]
$B_{\mathbf{D}}(\alpha)$ is the smallest region that contains UDG locations with at least probability $\alpha$. $B_{\mathbf{D}}(\alpha)$ is computationally costly because it requires a brute force optimization over $B \subset S$. We consider instead the dual problem of obtaining $B_{\mathbf{D}}(\alpha)$, which is finding the maximum probability of having UDGs in a region $B \subset S$ with $|B| \leq \vartheta|S|, \ \vartheta \in (0,1)$. We define this probability as the $\vartheta$-restricted detection probability:
\[\label{eqn: detection probability}
p_{\mathbf{D}}(\vartheta) = \sup\{\mathbb{P}(\mathbf{X}_c \subset B \mid \mathbf{D}): |B| \leq \vartheta|S|, \ B \subset S\}.
\]
In Section 8 of the Supplementary Materials, we demonstrate that computation of $p_{\mathbf{D}}(\vartheta)$ is straightforward compared to that of $B_{\mathbf{D}}(\alpha)$.

In contrast, \citeauthor{Li2022} detected UDGs by computing the excursion sets \citep{Bolin2015} of the posterior $\mathcal{U}(s)$ in \cref{eqn:lgcp model}. For an excursion level $u$, \citeauthor{Li2022} detected UDGs using the positive level $u$ excursion set with a $1-\alpha$ confidence level \citep{Bolin2015}:
\[\label{eqn: excursion sets}
E_{\alpha, u}(\mathcal{U}) = \argmax_{B \subset S}\{|B|: \mathbb{P}(B \subset A_u(\mathcal{U})\mid \mathbf{X}) \geq 1-\alpha\},
\]
where
\[
A_u(\mathcal{U}) = \{s \in S: \mathcal{U}(s) \geq u\}
\]
is the positive level $u$ excursion set \citep{Bolin2015}.
The choice of $u$, however, is user-defined, and it can impact the detection performance. Here, we instead directly obtain the posterior distribution of UDG locations, and do not require $u$ in \cref{eqn: excursion sets}. 
%LGCP detects UDGs by finding extreme value regions in the posterior of $\mathcal{U}(s)$, so $u$ is required. 
%: one can completely miss an UDG if $u$ is not well-chosen.

\subsection{Performance of UDG Detection}\label{subsec:performance tool}

%\textbf{[you need an intro paragraph here saying something along the lines of} "In order to test our method, we apply it to images for which there are XX known UDGs. The GCs of these GCs are used as the data ... etc. and we test the ability of our model to ``discover'' the known UDGs... etc... To quantify the performance of our model, we must device a quantative assessment, which we describe next...."]

\subsubsection{Performance Metric} Denote the chosen model as $\mathbf{G}$ (e.g., either one of the models presented here or the LGCP model in \citeauthor{Li2022}). We write the set of all possible detection regions $B$ produced by $\mathbf{G}$, defined by either \cref{eqn: detection region} or \cref{eqn: excursion sets}, as $\mathcal{B}_{\mathbf{G}}$. We define three performance metrics based on the principles that $B$ needs to be (i) reliable, (ii) precise, and (iii) of high certainty. Based on these metrics, we combine them to introduce a novel overall performance assessment tool for $\mathbf{G}$. 

Suppose an image contains $n_U$ number of known UDGs. Define $C_i, i = 1, \dots, n_U$ as
\[
C_i = \{s \in S: d(s,u_i) \leq R_{\mathrm{eff}}^i\},
\]
where $d$ is the Euclidean metric, $u_i$ and $R_{\mathrm{eff}}^i$ are the known location and effective radius \footnote{Effective radius here is defined as the radius that contains half of the light of a galaxy.} of the $i$-th UDG, respectively. We define the \textit{reliability metric} of $B \in \mathcal{B}_{\mathbf{G}}$ as
\[
\mathscr{R}(B) = \frac{1}{n_U}\sum_{i=1}^{n_U}\boldsymbol{1}\{B \cap C_i \neq \emptyset\}, 
\]
which checks the proportion of known UDGs detected by $B$. We define a detection of the $i$-th UDG if its neighborhood of $C_i$ intersects with $B$. To measure the precision of $B$ (i.e., $B$ needs to have small area), we define the \textit{precision metric} as $\nu(B) = 1 - |B|/|S|$. An uncertainty metric is also necessary; two models could produce two detection regions with the same reliability and precision metrics, so the one with higher confidence should be preferred. The uncertainty of $B$ can be measured by $\alpha$ in both \cref{eqn: detection region} and \cref{eqn: excursion sets}. We here define the \textit{uncertainty metric} as $\varpi(B) = 1 - \alpha$;
we elaborate on the motivation for this choice in \cref{subsec:performance compare}. 

\subsubsection{The Reliability-Uncertainty-Precision Curve}\label{sec: RUP curve} We now propose an overall performance assessment tool based on the idea of the ROC curve \citep{FAWCETT2006}. The traditional ROC curve is unsuitable for our context, as it requires a spatial boolean reference map --- something which is absent in our case of point-referenced UDGs \citep[see][for more details]{Li2022}. Our method, while not reliant on such a map, retains the ROC curve's functional benefits. We illustrate our idea based on $B$ defined by \cref{eqn: detection region}, and, for brevity, we omit the dependence of the metrics on $B$ in what follows.

For any $B \in \mathcal{B}_{\mathbf{G}}$, the previous three metrics are represented by a unique point $(\varpi, \nu, \mathscr{R})\in[0,1]^3$. Mapping all possible $(\varpi, \nu, \mathscr{R})$ of $B \in \mathcal{B}_{\mathbf{G}}$ yields a 3D performance curve $\mathcal{C}(\mathbf{G}) \subset [0,1]^3$ connecting $(0,0,1)$ and $(1,1,0)$. We term this curve the Reliability-Uncertainty-Precision (RUP) curve. A better model has $\mathcal{C}(\mathbf{G})$ closer to $\mathcal{C}^*(\mathbf{G}) = \{(\varpi, 1, 1): \varpi \in [0,1]\}\cup\{(0,\nu, 1): \nu \in [0,1]\}\cup\{(1,1,\mathscr{R}): \mathscr{R} \in [0,1]\}$. $\mathcal{C}^*(\mathbf{G})$ represents a theoretically perfect model with detection region containing only points at the locations of UDGs with $100\%$ certainty. 
\begin{figure}
    \centering
    \includegraphics[width = 0.8\textwidth]{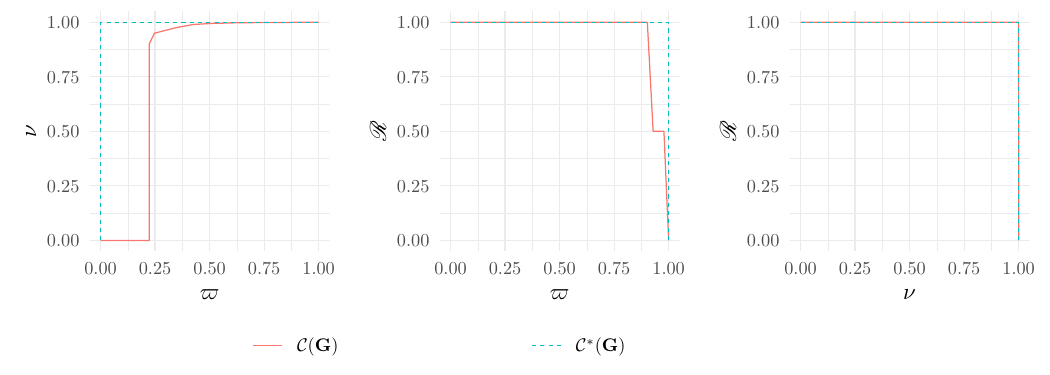}
    \caption{\footnotesize{Example of projected performance curves of $\mathcal{C}(\mathbf{G})$ (red solid lines) obtained from our Model \ref{model1} fitted to the GC data in the image V11-ACS. The projected perfect performance curves of $\mathcal{C}^*(\mathbf{G})$ (blue dashed lines) are also plotted for comparison. Both performance curves are projected onto the $\varpi$-$\nu$, $\varpi$-$\mathscr{R}$, and $\nu$-$\mathscr{R}$ planes.}}
    \label{fig:demo_SG}
\end{figure}

To evaluate model performance, we introduce a quantity akin to the area under the curve (AUC) in ROC analysis. We first project $\mathcal{C}(\mathbf{G})$ onto $\varpi$-$\nu$, $\varpi$-$\mathscr{R}$, and $\nu$-$\mathscr{R}$ planes, creating 2D curves $C_\varpi^{\nu}$, $C_\varpi^{\mathscr{R}}$ and $C_{\nu}^{\mathscr{R}}$ respectively. For each, we then obtain an AUC value, respectively denoted by $A_\varpi^{\nu}, A_\varpi^ {\mathscr{R}}, A_{\nu}^{\mathscr{R}} \in [0,1]$. We define our performance statistics as
\[\label{eqn: performance stats}
\mathcal{A}(\mathbf{G}) =  (A_\varpi^{\nu}A_\varpi^ {\mathscr{R}} A_{\nu}^{\mathscr{R}})^{1/3}.
\]
We use the geometric mean for its robustness against large values, granting us more distinguishing power. In \cref{eqn: performance stats}, $\mathcal{A}(\mathbf{G}) = 1$ if and only if $\mathcal{C}(\mathbf{G}) = \mathcal{C}^*(\mathbf{G})$. A model with $\mathcal{A}(\mathbf{G})$ nearing $1$ is preferable. Similar to ROC curves, $C_\varpi^{\nu}$ from a random model aligns with $\varpi=\nu$, since the certainty of the detection region grows proportionally with the area. However, $C_\varpi^ {\mathscr{R}}$ and $C_{\nu}^{\mathscr{R}}$ under a random model are step functions and the exact shapes depend on the locations and the number of known UDGs. \cref{fig:demo_SG} shows an example of projected performance curves of $\mathcal{C}(\mathbf{G})$ and $\mathcal{C}^*(\mathbf{G})$. Note that $\mathcal{C}(\mathbf{G})$ and $\mathcal{A}(\mathbf{G})$ can also be constructed under \cref{eqn: excursion sets} (with $u$ chosen) for LGCP. Under \cref{eqn: excursion sets}, a theoretically perfect model yields excursion sets pinpointing UDG locations for any $1-\alpha$, and its performance curve matches $\mathcal{C}^*(\mathbf{G})$.

%% file: sections/05-Data-Analysis.tex
\section{Results}\label{sec:da}

In this section, we apply the methods from \cref{sec:inf_est} to fit the models described in \cref{sec:method} to a selected dataset (\cref{sec:data}). We fit four models (no-mark, magnitude-marked, color-marked, and color-marked with error model) to 12 of the 20 images in the data set. The specifics of image selection are discussed later. Our focus is on detecting GC-rich UDGs $(N_{\mathrm{GC}} \geq 3)$, as those with $N_{\mathrm{GC}} \leq 2$ are indistinguishable from randomness. We also fit all our models to a set of simulated GCs as a sanity check. The results of the simulation are given in Section 11 of the Supplementary Material. 

To fit LGCP models, we used INLA \citep{Rue2009}. We ran four independent MCMC chains with 300k-500k iterations for each of our model, tailoring the chain length to the image complexity: more iterations for images with more luminous elliptical galaxies. Due to memory constraints, we thin our samples every $10$ iterations after burn-in. The details on the initialization and implementation of the MCMC algorithm are given in Section 5 of the Supplementary Material. MCMC convergence diagnostics, prior sensitivity analysis, and posterior predictive checks are contained in Sections 6, 7, and 10 of the Supplementary Material respectively. 

Our findings are: (i) our models successfully identify the existence of GC-rich UDGs; (ii) our models reveal a previously unnoticed potential UDG with three GCs; (iii) most of our models are superior to LGCP in terms of overall performance measured by the RUP curve; and (iv) including the GC marks improves the detection performance for the majority of images considered.

\subsection{UDG Existence}

Our analysis covers $12$ selected images, with seven lacking previously confirmed GC-rich UDGs and five containing them. Other images, both with and without GC-rich UDGs, were excluded due to having too few GCs ($\leq 20$). We opted for cases that are interesting and likely to be challenging. The selection of images with known GC-rich UDGs is explained later in \cref{subsec:performance compare}, where we compare model performance.

\begin{figure}[t]
    \centering
         \includegraphics[width=
         \textwidth]{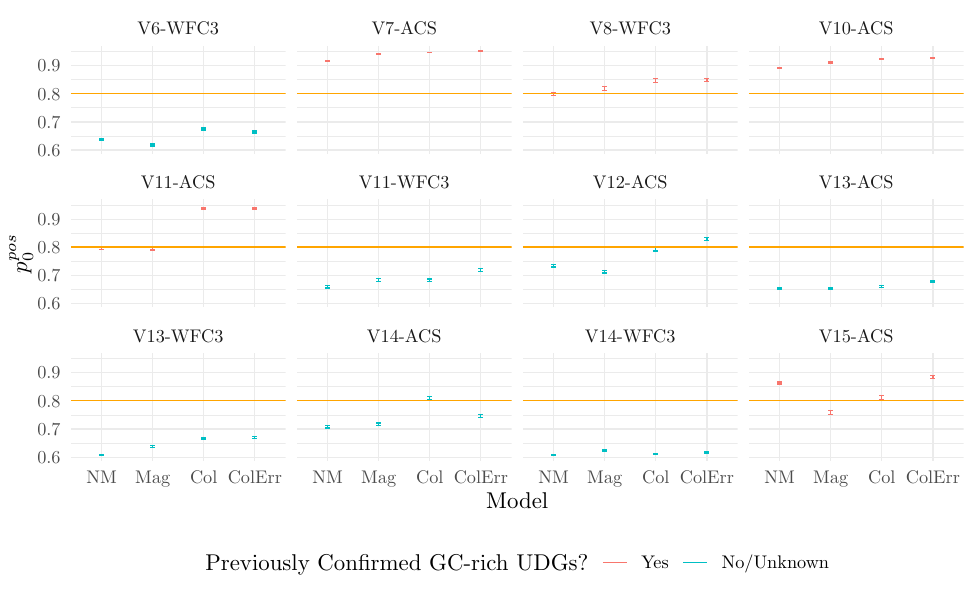}
     \caption{\footnotesize{Posterior probabilities $p_0^{pos}$ as in \cref{subsec: udg_existence} compared to the prior probability $p_0^{pri}$ (orange line). Error bars are the $95\%$ Monte Carlo confidence intervals. NM is the no-mark model; Mag is the magnitude-marked model; Col is the color-marked model; and ColErr is the color-marked model with error considered. All our models strongly suggest that there are no UDGs in five of the images with no previously confirmed GC-rich UDGs. Our models also provide confirmations that the five images with confirmed GC-rich UDGs indeed have UDGs. Our models also suggest that V12-ACS and V14-ACS seem to have previously undiscovered UDGs indicated by the comparison between $p_0^{pos}$ and $p_0^{pri}$ based on the color-marked models.}}
     \label{fig:no_UDG_prob}
\end{figure}

\cref{fig:no_UDG_prob} shows $p_0^{pos}$ obtained by fitting all four of our models to the aforementioned images and applying the procedures described in \cref{subsec: udg_existence}. $p_0^{pri}$ is shown as the orange horizontal line. For five images without previously confirmed GC-rich UDGs, all four of our models provide strong evidence that they do not have UDGs. For images that do contain known GC-rich UDGs, our models made the correct identification with $p_0^{pos} \gtrsim p_0^{pri}$. The only exception is V15-ACS under the magnitude-marked model, where $p_0^{pos} < p_0^{pri}$. This exception is explained by the rather faint GCs in the UDG in this image --- the model assigned a higher probability of it having no UDG. However, the difference between $p_0^{pos}$ and $p_0^{pri}$ is rather small and it warrants a visual check of $\mathbf{X}_c\mid \mathbf{D}$. The visual result suggests that the known UDG is still detected with a very strong signal (see Section 9 of the Supplementary Material for figures).

Interestingly, in \cref{fig:no_UDG_prob}, the color-marked models suggest there are two images, V12-ACS and V14-ACS, that seem to contain GC-rich UDGs not previously found. We investigate this in the next section.

\subsection{A Potential Dark Galaxy}

\begin{figure}[t]
    \centering
     \subfigure[Posterior intensity from our models for V12-ACS]{\includegraphics[width = 0.5\linewidth]{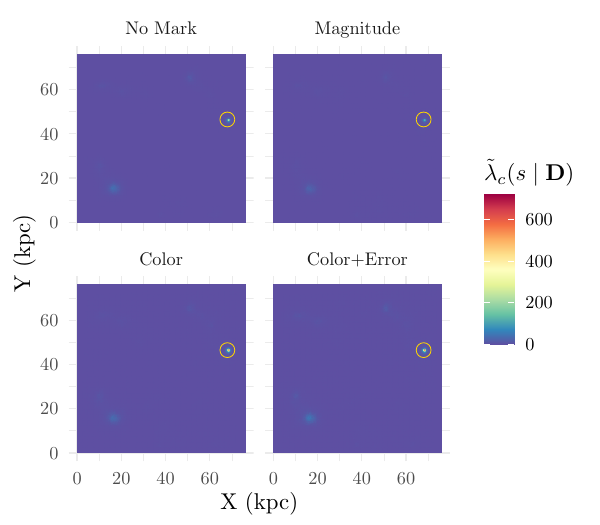}\label{fig:a}}%
     \subfigure[Posterior intensity from our models for V14-ACS]{\includegraphics[width = 0.5\linewidth]{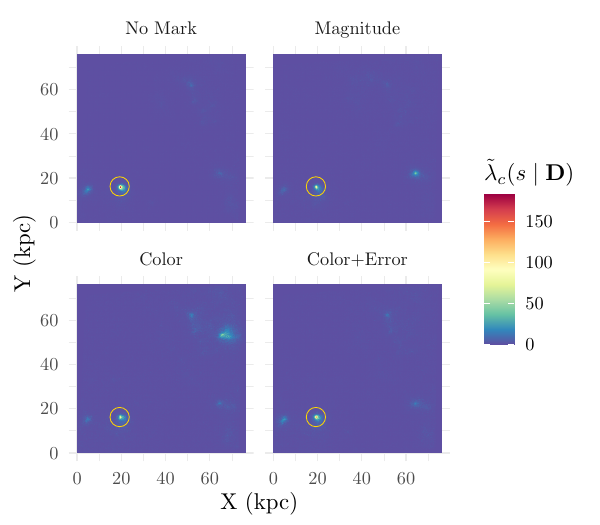}\label{fig:b}}\\
     \subfigure[$\mathbb{E}(\exp(\mathcal{U}(s)) \mid \mathbf{X})$ from LGCP for V12-ACS]{\includegraphics[width = 0.4\linewidth]{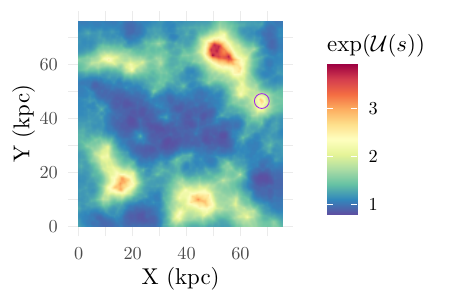}\label{fig:c}}%
     \hspace{14mm}
     \subfigure[$\mathbb{E}(\exp(\mathcal{U}(s)) \mid \mathbf{X})$ from LGCP for V14-ACS]{\includegraphics[width = 0.4\linewidth]{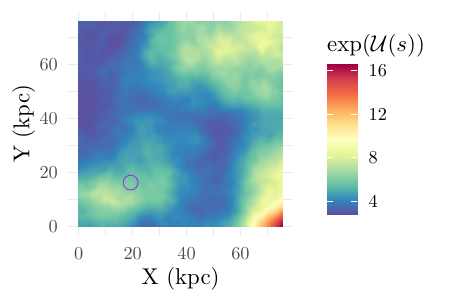}\label{fig:d}}
    \caption{\footnotesize{Posterior results for (a) our models in V12-ACS; (b) our models in V14-ACS; (c) LGCP in V12-ACS; (d) LGCP in V14-ACS. The posterior intensity from our models is scaled by the prior mean of $\lambda_c$ for comparison: $\Tilde{\lambda}_c(s\mid \mathbf{D}) \equiv \mathbb{E}(\lambda_c(s)\mid \mathbf{D})/\mathbb{E}(\lambda_c)$. The golden circles in (a) and (b) indicate the location of CDG-2. We see that LGCP fails to detect it in both V12 and V14-ACS as shown by the purple circles in (c) and (d).}}
    \label{fig:CDG-2}
\end{figure}

\begin{figure}[t]
    \centering
    \includegraphics[width = 0.5\textwidth]{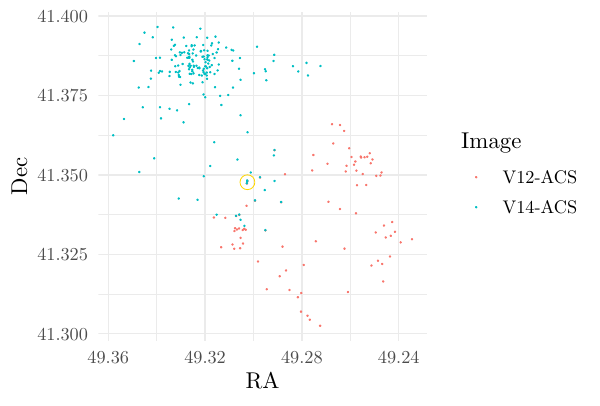}
    \caption{\footnotesize{GC locations from V12 (red points) and V14-ACS (blue points) in celestial coordinates right ascension (RA) and declination (Dec). The clump of three GCs, CDG-2, is marked by the golden circle.}}
    \label{fig:v12-v14}
\end{figure}

We visually inspected the posterior distribution of $\mathbf{X}_c$ for V12 and V14-ACS. These two images share an overlapping region, and a rather strong detection signal for a cluster of three GCs in the overlapping region is present in both images. \cref{fig:CDG-2}(a) and (b) provide the scaled posterior mean intensity $\Tilde{\lambda}_c(s\mid \mathbf{D}) \equiv \mathbb{E}(\lambda_c(s)\mid \mathbf{D})/\mathbb{E}(\lambda_c)$\footnote{Scaled intensity is used here for better comparison.}, where $\mathbb{E}(\lambda_c(s)\mid \mathbf{D})$ and $\mathbb{E}(\lambda_c)$ are the posterior and prior mean intensity respectively. We see that all our models picked up a strong signal in both V12 and V14-ACS, indicated by the golden circles. This cluster contains three tightly clumped GCs with no detectable faint stellar light. We dub this object Candidate Dark Galaxy-2 (CDG-2) following the discovery of another potential dark galaxy --- CDG-1 by \citeauthor{Li2022}. \cref{fig:v12-v14} shows the GC data from both V12 and V14-ACS in celestial coordinates, with the location of CDG-2 marked by the golden circle.

\cref{fig:CDG-2}(c) and (d) contain the results from LGCP fitted to each image. The posterior mean of $\exp(\mathcal{U}(s))$ from LGCP appears as pure random noise for both images, and the locations (indicated by the purple circles) of CDG-2 exhibit no apparent clustering. LGCP fails at detecting CDG-2 mostly due to the Gaussian random field structure. LGCP ignores weaker clustering signals such as CDG-2, which are taken as inevitable fluctuations caused by the underlying Gaussian random field. Our models, on the other hand, search for clumps of GCs, and will detect them even if their signals are relatively weak.

\begin{figure}[t]
     \centering
     \subfigure[Posterior Intensity from our models.]{
      \includegraphics[width=0.58\textwidth]{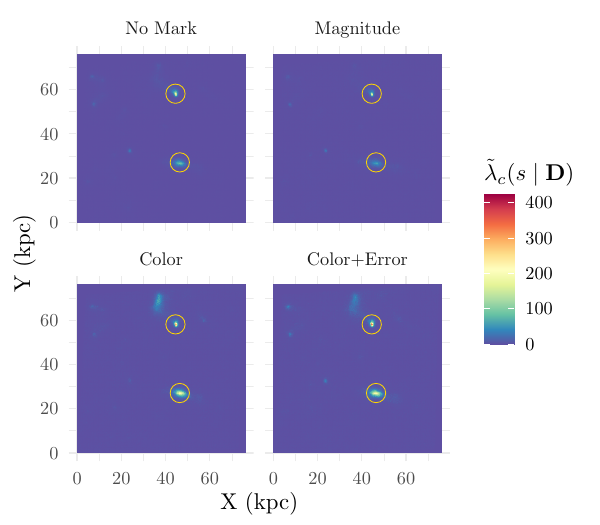}
    }
     \subfigure[Posterior mean of $\exp(\mathcal{U}(s))$ from LGCP]{\raisebox{16mm}{
      \includegraphics[width=0.37\textwidth]{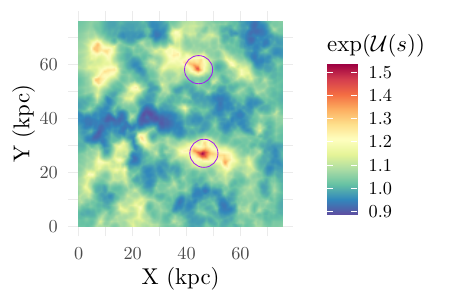}}
    }
          \caption{\footnotesize{(a) The scaled posterior intensity estimates $\Tilde{\lambda}_c(s \mid \mathbf{D})$ (same as in \cref{fig:CDG-2}) of $\mathbf{X}_c$ for the field V11-ACS from all four of our models. (b) Result from the LGCP model by \citeauthor{Li2022}. Golden circles in (a) and purple circles in (b) indicate the locations of two confirmed UDGs in the image from \cite{Wittmann2017}. All of our models produce strong detection signals (high intensity) for the two UDGs, with the color-marked models producing the strongest detection signals. The relative signals for the two UDGs from LGCP are much weaker.}}
     \label{fig:v11acs_results}
\end{figure}

%Note that the posterior intensity at CDG-2 for V12-ACS is much higher than in V14-ACS. This is because the V14-ACS contains a giant elliptical galaxy that adds additional noise to the image, and the detection signal for CDG-2 becomes weaker. Moreover, the detection signal under color-marked models in V14-ACS decreased compared to the no-mark model while it is the reverse for V12-ACS. This is due to the data processing inconsistency, which is bound to happen occasionally in astronomy, rather than an issue of our models. Essentially, the photometric data are processed individually for each image which produces inconsistency in the color measurement of GCs, and this is what happened for one of the GCs in CDG-2. Nevertheless, the model where we consider the error of GC color also indicates that CDG-2 indeed has very strong signals in both images.

\subsection{Detection Performance for GC-Rich UDGs}\label{subsec:performance compare}

We now present the detection performance analysis for previously confirmed GC-rich UDGs from five images using the method described in \cref{subsec:performance tool}. The five images are selected based on two criteria: firstly, images with too few GCs ($\leq 20$) are excluded; secondly, images with giant elliptical galaxies or more than one known GC-rich UDG are selected. The second criterion is chosen because the existence of giant elliptical galaxies can introduce additional noise, while having more than one UDG can significantly affect detection performance due to an imbalance in their signal strengths.

\subsubsection{Example Detection Results with V11-ACS}

We first showcase the posterior results obtained from all our models and LGCP. \cref{fig:v11acs_results} displays these results for V11-ACS image, as an example. Results for other images are detailed in Section 9 of the Supplementary Material. \cref{fig:v11acs_results}(a) shows the scaled posterior mean intensity $\Tilde{\lambda}_c(s\mid \mathbf{D})$ from our models. \cref{fig:v11acs_results}(b) presents the posterior mean of $\exp(\mathcal{U}(s))$ from LGCP. All our models provide strong signals at two UDG locations, marked by golden circles. While, LGCP picks up the signals of these UDGs, the signal is much weaker.

\begin{figure}[t]
    \centering
     \subfigure[Detection regions from our models.]{\includegraphics[width=0.75\textwidth]{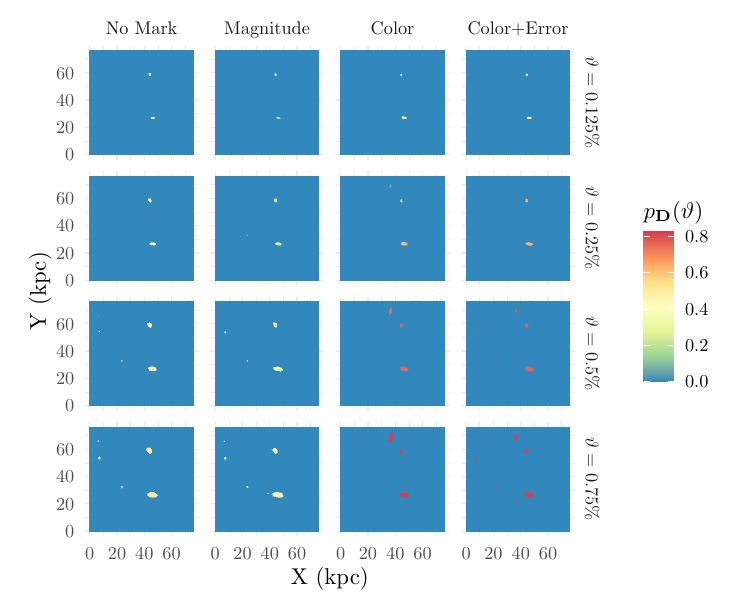}}
     \subfigure[Detection regions from LGCP.]{\includegraphics[width=0.7\textwidth]{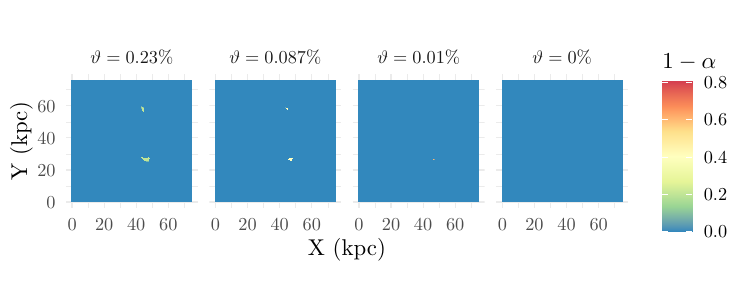}}
     \caption{\footnotesize{(a) Detection regions for V11-ACS with $\vartheta = |R|/|S| \in \{0.125\%, 0.25\%, \allowbreak 0.5\%, 0.75\%\}$ under our models and the corresponding $p_{\mathbf{D}}(\vartheta)$. (b) Excursion sets from LGCP with $u=0$ and $1-\alpha \in \{0.2, 0.4, 0.6, 0.8\}$ and the corresponding area ($\vartheta$) of the excursion sets. All our models managed to detect the two known UDGs in the image (see \cref{fig:v11acs_results}), with the color-marked models having the best performance where their detection regions have the highest confidence for any given level of $\vartheta$. In contrast, LGCP is only able to detect the UDGs for low level of confidence. As $1 - \alpha > 0.6$, LGCP fails to detect the two UDGs.}}
     \label{fig:v11acs_EX}
\end{figure}

We also provide an example of the detection regions defined by \cref{eqn: detection region} and \cref{eqn: excursion sets} using V11-ACS. For our models, we compute detection region $B$ with area $\vartheta = |B|/|S| \in \{0.125\%, 0.25\%, \allowbreak 0.5\%, 0.75\%\}$, which corresponds to typical UDG radii of $1.5-3.5$ kpc. We then compute $p_{\mathbf{D}}(\vartheta)$ defined in \cref{eqn: detection probability} for each model. For LGCP, following \citeauthor{Li2022}, we set the excursion level $u=0$ and confidence levels $1-\alpha \in \{0.2, 0.4, 0.6, 0.8\}$ to determine $E_{\alpha,0}(\mathcal{U})$ and $\vartheta$, using the \texttt{excursion} R package \citep{Bolin2018}. \cref{fig:v11acs_EX} contains the results of the detection regions.

\cref{fig:v11acs_EX}(a) demonstrates that all our models detected the two UDGs in V11-ACS shown in \cref{fig:v11acs_results}. Both color-marked models outperformed the no-mark and magnitude-marked models, exhibiting higher certainty across all $\vartheta$ values. For LGCP, UDG detection was effective only at lower confidence levels. Above a $0.6$ confidence level, LGCP's detection capability diminished, and at $0.8$, it failed entirely.

Note that as $\vartheta$ increases, our models appear to produce ``false-positives". However, when testing model performance, there is no ground truth for non-UDGs in the reference data, and thus the term ``false-positives" is not applicable nor well-defined. Moreover, since the impact of a false-positive detection is an expense of human labor to confirm the presence of a UDG, while the impact of a false-negative is missing the detection of a new UDG, false-positives can be regarded as the less severe of the two types of detection errors in this context.

Additionally, based on \cref{fig:v11acs_EX}, the relationship between the confidence level and the area of the detection region is reversed between our methods and the LGCP. Thus, we have defined the uncertainty measure $\varpi(B) = 1 - \alpha$ in \cref{subsec:performance tool} for both our models and LGCP. Although the physical interpretation of $\varpi(B)$ is reversed, the qualitative relationship between $\varpi(B)$ and the area is now the same. Moreover, the theoretically perfect model under the two frameworks now both have the same RUP curve, thus allowing for cross-model framework comparison.

\subsubsection{Overall Performance Assessment}
We now conduct overall performance assessment of all models fitted to the five selected images using the RUP curve in \cref{sec: RUP curve}. To reduce computational time and memory requirement, we thin our MCMC sample every $1000$ iterations. The large thinning interval effectively renders each sample independent, and we assume this is true when computing the uncertainty. For LGCP, we compute the excursion set by setting $u = 0$ for all five images as before.

\cref{tab:performance-tab} includes the overall performance assessment for each model based on the $\mathcal{A}(\mathbf{G})$ statistics introduced in \cref{sec: RUP curve}. For the image V7-ACS, INLA could not fit the LGCP to the GC data due to a computational error. Most likely, this image contains a UDG with very strong GC clustering signal. In principle, LGCP cannot produce such a strong clustering signal due to the thin-tailed nature of Gaussian distributions, and this led to the computational error by INLA.

\cref{tab:performance-tab} demonstrates that almost all of our models outperform LGCP in overall performance. Only magnitude and color-marked models perform worse than LGCP for the image V15-ACS. Furthermore, all the best performing models (indicated by purple numbers in \cref{tab:performance-tab}) are from our approach. This is strong evidence that our method is superior to LGCP.

\begin{table}[t]
    \centering
    \resizebox{0.8\columnwidth}{!}{%
    \begin{tabular}{cccccc}
    \toprule
    Image ID & No-Mark & Magnitude & Color & Color \& Error & LGCP \\
    \midrule
       \multirow{2}{*}{V7-ACS}  & $0.928$ & $\textcolor{purple}{0.951}$ & $0.921$ & $0.925$ & - \\
       & $(0.922, 0.933)$ & $(0.946, 0.956)$ & $(0.915, 0.926)$ & $(0.919, 0.930)$& - \\
       \midrule
       \multirow{2}{*}{V8-WFC3}  & $0.941$ & $ 0.929$ & $0.945$ & $\textcolor{purple}{0.949}$ & $0.871$ \\
       & $(0.936, 0.945)$ & $(0.923, 0.934)$ & $(0.941, 0.949)$ & $(0.945, 0.953)$& - \\
       \midrule
       \multirow{2}{*}{V10-ACS}  & $0.9931$ & $ 0.9923$ & $\textcolor{purple}{0.9937}$ & $0.9935$ & $0.972$ \\
       & $(0.992, 0.994)$ & $(0.9922, 0.9924)$ & $(0.992, 0.995)$ & $(0.992, 0.995)$& - \\
       \midrule
       \multirow{2}{*}{V11-ACS}  & $0.901$ & $0.885$ & $\textcolor{purple}{0.9821}$ & $0.973$ & $0.832$\\
       & $(0.896, 0.906)$ & $(0.879, 0.891)$ & $(0.9818, 0.9823)$ & $(0.972, 0.974)$& - \\
       \midrule
       \multirow{2}{*}{V15-ACS}  & $\textcolor{purple}{0.976}$ & $0.886$ & $0.929$ & $0.971$ & $0.963$\\
       & $(0.975, 0.977)$ & $(0.879, 0.893)$ & $(0.923, 0.933)$ & $(0.970, 0.973)$& - \\
       \bottomrule
    \end{tabular}%
    }
    \caption{$\mathcal{A}(\mathbf{G})$ statistics from \cref{subsec:performance tool} for all models fitted to the selected five images containing previously confirmed GC-rich UDGs. A value closer to $1$ is better. Brackets contain the $95\%$ confidence intervals of estimated $\mathcal{A}(\mathbf{G})$ from MCMC samples. For LGCP, there is no estimation error due to the deterministic nature of INLA. Purple number indicates the best performing model for each image. The best performances for each image all come from our models.}
    \label{tab:performance-tab}
\end{table}

\begin{figure}[h]
    \centering
    \includegraphics[width=\columnwidth]{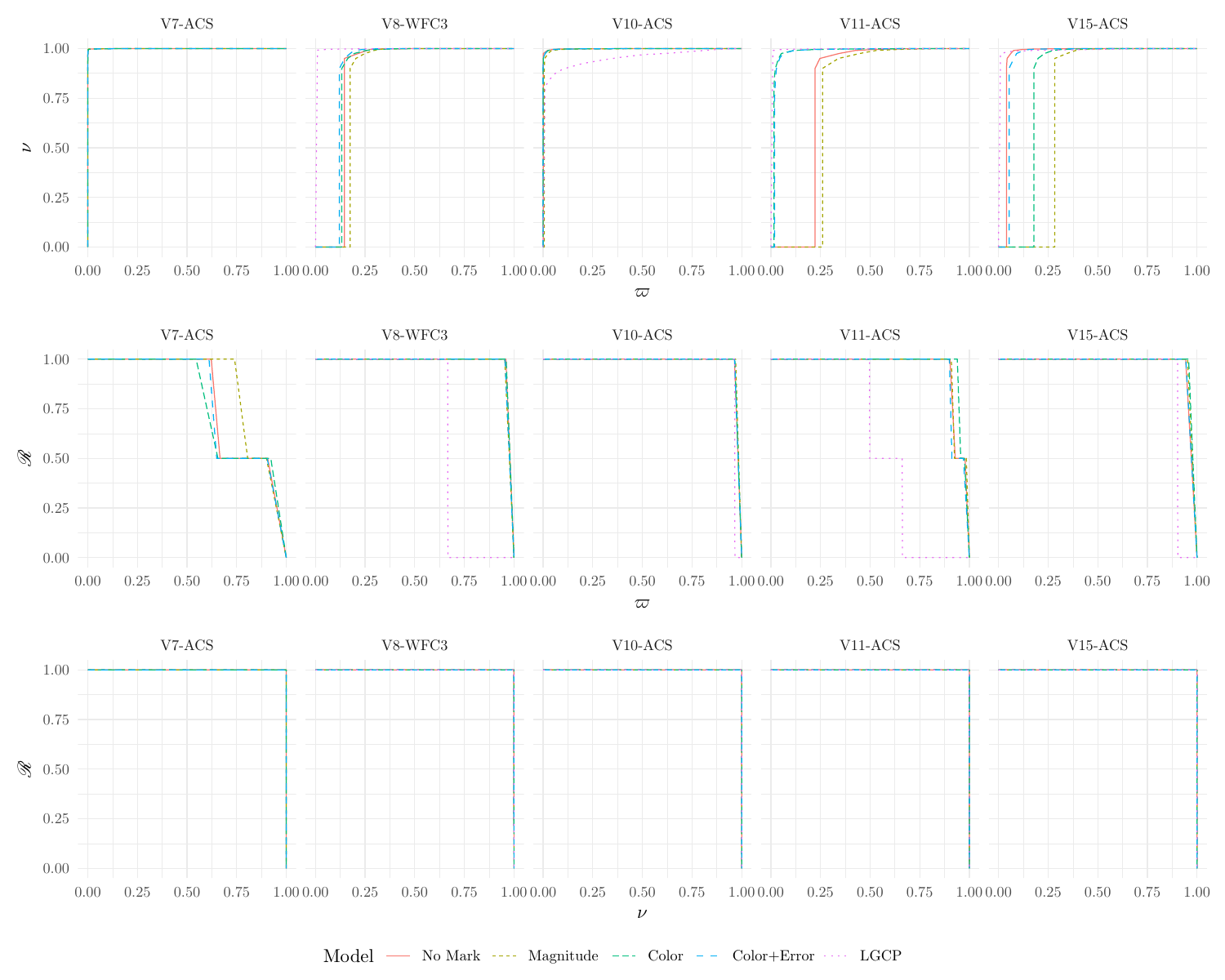}
    \caption{\footnotesize{Projected overall performance curves $C_\varpi^\nu$, $C_\varpi^{\mathscr{R}}$, $C_\nu^{\mathscr{R}}$ obtained for each model fitted to the five selected images.}}
    \label{fig:overall performance result}
\end{figure}

Within our own model framework, including the marks (either magnitude or color variations) improved performance over the no-mark model with the exception of V15-ACS. This result provides preliminary evidence that the marked model indeed may improve detection performance. %This suggests that the marked model can be safely used when marks are available.
% Such a result provides preliminary evidence that by including the GC mark information, the detection performance may indeed improve significantly.

We need to emphasize that including GC marks does not mean definitive improvement of UDG detection since physical properties of GCs in UDGs can exhibit variations and differ from existing physical theories and observations. The UDG in V15-ACS is an example. Such variations are mainly determined by the formation process and environments of UDGs, which are currently still in debate and an area of active research. Nevertheless, having the marked PCP model at our disposal enables the inclusion of GC mark information when available. In essence, astrophysicists can incorporate any GC marks into our marked PCP model to aid the detection of UDGs. For example, for UDG detection in the nearby Universe, including the spatial resolution of imaged point sources as marks may help distinguish bona-fide GCs from individual stars, thus decreasing the noise level and improving detection. Contrasting the detection results from different models may also improve decision power for discovering new UDGs.

To investigate exactly how the overall performance behaves across models, we now present the RUP curves. \cref{fig:overall performance result} shows the projected performance curves $C_\varpi^\nu$, $C_\varpi^{\mathscr{R}}$, and $C_\nu^{\mathscr{R}}$ obtained for each model fitted to the five images. For visual clarity, the uncertainty is not included.

Several key-points from \cref{fig:overall performance result} are the following: LGCP performs better than our models under the $C_\varpi^\nu$ curves for V8-ACS, V11-ACS, and V15-ACS. This is because these three images are quite contaminated, resulting in the UDG signals being relatively weak. Hence, when $\varpi$ is smaller ($\alpha$ is larger for $B_{\mathbf{D}}(\alpha)$), it starts to exhaust all the non-empty samples in $\mathbf{X}c$ from our models, and $B_{\mathbf{D}}(\alpha)$ reaches the entire study region $S$. Therefore, our models cannot reach any higher level of confidence, which is represented by the sudden drop of $\nu$ at small values of $\varpi$ in the $C_\varpi^\nu$ curves. Nevertheless, having empty $\mathbf{X}_c$ samples is an integral part of our models for identifying UDG existence, something the LGCP approach does not provide.

For the $C_\varpi^{\mathscr{R}}$ curves, the results show that our models outperform LGCP, especially for V8-WFC3 and V11-ACS. This is because the confidence level for the excursion sets from LGCP does not reach the full range of $(0,1)$, while our approach does. The difference in the potential confidence range is in part an inherent methodological difference. If we require a high confidence level for the excursion sets from LGCP, we may miss UDGs. Moreover, while LGCP \textit{has} the hypothetical potential to produce excursion sets that detect UDGs with a very high confidence level, it does not. The detection signals it produces are simply too weak for V8-WFC3 and V11-ACS. Granted, one can decrease the excursion level $u$ so that confidence level of the excursion sets can reach the range of $(0,1)$, but this comes at the cost of losing precision by significantly increasing the area of the excursion sets. 

Additionally, $C_\nu^{\mathscr{R}}$ for all models are effectively perfect, meaning that all models correctly identified the locations of known UDGs regardless of the area of the detection region. Such behavior is desired but also natural since the regions where known UDGs reside have the strongest signal strengths in all models. Thus, when we vary the area from smallest to largest, UDGs are always contained in the detection region. 

In terms of the effect of marks, the overall improvement for V8-WFC3 and V11-ACS observed in \cref{tab:performance-tab} is from the additional information provided by the GC colors, which significantly improved the performance under the $C_\varpi^\nu$. On the other hand, for V7-ACS, the image contains two UDGs with imbalanced GC clustering strengths. The introduction of the GC magnitude boosted the detection signal of the UDG with weaker GC clustering strength, and hence improved the performance under. In image V10-ACS, the signal strength of the UDG is so strong that the final performance results are nearly the same, regardless of inclusion of marks. 

%% file: sections/06-Discussion.tex
\section{Conclusion}\label{sec:conclusion}

We have introduced a set of novel point process models under the framework of a Poisson cluster process to detect ultra-diffuse galaxies through their globular clusters. Our models show clear advantages over the LGCP method by \cite{Li2022} in determining the non-existence of UDGs. Our models also managed to discover a potential new dark UDG that the LGCP method has previously missed. Moreover, we introduced a new overall performance assessment tool, the RUP curve, for spatial prediction problems that lack Boolean reference maps. Using the RUP curves, we showed that our proposed models outperformed the LGCP method. Moreover, based on the preliminary results obtained with the data at hand, we see promising signs that including GC mark information in the marked PCP model framework can potentially improve UDG detection performance.

While our focus here is on the introduction and preliminary detection analysis of our proposed PCP models, there are several directions of research in the future. Firstly, we can easily apply our model to data sets from other imaging surveys. Such an endeavor helps us further test and calibrate our models on data with different imaging quality that may come from different astrophysical environments. It may also reveal new UDGs that have previously eluded astrophysicists. Secondly, we can consider joint modeling of multiple marks of GCs to test the detection performance. We did not pursue this direction in the current paper due to the complexity of joint modelling of multivariate GC marks. Lastly, we can further improve the sampling performance of our MCMC algorithm by introducing adaptive proposals for the birth-death-move update of the MCMC. There are two potential avenues for this idea. The first is that the birth, death, and move probabilities of each update step may be adapted based on previous samples. The second is that the proposal distributions for the UDG locations and physical parameters within the birth and move step may be adapted as well.

%% file: sections/07-Acknowledgement.tex
\section*{Acknowledgement}

Authors D.L., P.E.B., and G.M.E. are also affiliated with the Data Sciences Institute at the University of Toronto. Author R.G.A. is also affiliated with the Dunlap Institute for Astronomy \& Astrophysics at the University of Toronto.

Part of the computation in this work is conducted using the computing clusters at the Digital Research Alliance of Canada, and the authors are grateful for the computational resources. The authors would like to thank the Associate Editor, the Editor, and the Referees for their helpful comments that significantly improved the paper. 

\section*{Funding}
D.L. is supported by the Data Sciences Institute Doctoral Student Fellowship at the University of Toronto and the Canadian Statistical Sciences Institute Multi-disciplinary Doctoral Trainee Program. G.M.E. acknowledges funding from Natural Sciences and Engineering Research Council (NSERC) through Discovery Grant RGPIN-2020-04554 and a Connaught New Researcher Award from the University of Toronto, both of which supported this research.
A.S. acknowledges funding from NSERC through Discovery Grant RGPIN-2023-03331.

%% file: sections/Notations-and-Symbols.tex
\section{Notations and Symbols}\label{appendix:notation}

\begingroup
\def\arraystretch{1.05}
\begin{table}[H]
\begin{center}
\resizebox{\columnwidth}{!}{%
\begin{tabular}{lllc}
    \toprule
    Notation/Symbol (Type) & Brief Description & Units & Parameter Groupings \\
    \midrule
    $[N]$ (const.) & $\{1, \dots, N\}$ & - & -\\
    $\langle N \rangle$ (const.) & $\{0, \dots, N\}$ & - & - \\
    \midrule
    $\mathbf{X}$ (PP)  & GC point process & - & - \\
    $\mathbf{x}$ (RV) & Unrealized GC location & (kpc, kpc)  & -\\
    $x$ (data) & Observed GC location & (kpc, kpc)  & -\\
    $S$ (const.) & Study region & -  & -\\
    $s$ (dummy) & A point in $S$ & (kpc, kpc)  & -\\
    $N_G$ (const.) & Number of elliptical galaxies (EG) in $S$ & -  & -\\
    $\text{Mag}$ (data) & GC brightness & magnitude & - \\
    $C$ (data) & GC color & magnitude & - \\
    $V$ (data) & GC color variations & magnitude & - \\
    $\mathbf{M}$ (data) & GC mark & -  & -\\
    $\mathcal{U}$ (RV) & Spatial random field under LGCP & - & - \\
    \midrule
    $\beta_0$ (RV) & GC intensity in the IGM & count/kpc$^2$ (normalized) & - \\
    \midrule
    $c_k^g$ (const.) & Center of the $k$-th EG & (kpc, kpc)  & - \\
    \midrule
    $\varlambda_k^g$ (RV) & Mean number of GCs in the $k$-th EG & count & \multirow{5}{*}{$\bm{G}_k, \ j \in [N_G]$}\\
    $R^g_k$ (RV) & Characteristic size of the GC system of the $k$-th EG & kpc & - \\
    $n^g_k$ (RV) & S\'{e}rsic index of the GC system of the $k$-th EG & unitless & - \\
    $\varphi^g_k$ (const.) & Orientation angle of the GC system of the $k$-th EG & radian & -\\
    $\rho^g_k$ (const.) & Semi-axis ratio of the GC system of the $k$-th EG & unitless & -\\
    \midrule
    $\lambda_c$ (RV) & Mean number of UDGs in $S$ & count & - \\
    $\mathbf{X}_c$ (PP) & Point process of UDG central locations & - & - \\
    $N_U$ (RV) & Number of UDGs in $S$ & count & - \\
    $\mathbf{c}_j^u$ (RV) & Center of the $j$-th UDG & (kpc, kpc) & $\mathbf{X}_c =\{\mathbf{c}_j^u\}_{j=1}^{N_U}$\\
    \midrule
    $\varlambda_j^u$ (RV) & Mean number of GCs in the $j$-th UDG & count & \multirow{5}{*}{$\bm{U}_j, \ j \in [N_U]$}\\
    $R^u_j$ (RV) & Characteristic size of the GC system of the $j$-th UDG & kpc &  \\
    $n^u_j$ (RV) & S\'{e}rsic index of the GC system of the $j$-th UDG & unitless & \\
    $\varphi^u_j$ (RV) & Orientation angle of the GC system of the $j$-th UDG & radian & \\
    $\rho^u_j$ (RV) & Semi-axis ratio of the GC system of the $j$-th UDG & unitless & \\
    \midrule
    $\mu_0$ (RV) & Mean magnitude for GCs not in UDGs & magnitude & \multirow{5}{*}{$\Sigma_m, \ m \in \langle N_U \rangle$} \\
    $\mu_m$ (RV) & Mean magnitude for GCs in the $m$-th UDGs, $m \in [N_U]$  & magnitude & \\
    $\sigma_{m,B}$ (RV) & \makecell[l]{Standard deviation of GC magnitude \\ in the $m$-th environment, $m \in \langle N_U \rangle$} &  magnitude & \\
    $\sigma_{0, C}$ (RV) & Color variation for GCs not in UDGs & magnitude &  \\
    $ \sigma_{m,C}$ (RV) &  Color variation for GCs in the $m$-th UDGs, $m \in [N_U]$ & magnitude & \\
    \midrule
    $\bm{\Gamma}$ & $\{\bm{G}_k\}_{k=1}^{N_G}$ & - & -  \\
    $\bm{\Upsilon}$ & $\{\bm{U}_j\}_{j=1}^{N_U}$ & - & - \\
    $\bm{\Xi}$ & $(\beta_0, \bm{\Gamma}, \bm{\Upsilon})$ & - & - \\
    $\bm{\Sigma}$ & $\{\Sigma_m\}_{m=0}^{N_U}$ & -& - \\
    \midrule
    $\bm{\Theta}$ & $(\beta_0, \lambda_c, \bm{\Gamma}, \Sigma_0)$ & - & - \\
    $\bm{\Phi}$ & $\{(\mathbf{c}_j^u, \bm{U}_j, \Sigma_j)\}_{j=1}^{N_U}$ & - & -  \\
    $\mathbf{D}$ & \makecell[l]{Data: either $(\mathbf{X})$ or $(\mathbf{X}, \mathbf{M})$ \\ depending on the model considered} & - & - \\
    \midrule
    $B$ & Detection region of UDGs & - & - \\
    $\mathscr{R}(B)$ & Reliability metric for $B$ & - & - \\
    $\nu(B)$ & Precision metric for $B$ & - & - \\
    $\varpi(B)$ & Uncertainty metric for $B$ & - & - \\
    $\mathcal{C}$ & A RUP curve & - & - \\
    $C_{\varpi}^{\nu}, C_{\varpi}^{\mathscr{R}}, C_{\nu}^{\mathscr{R}}$ & Projected RUP curves & - & - \\
    \bottomrule
\end{tabular}%
}
\end{center}
\caption{Major notation and symbols used in the paper and their meanings. const. = constant; PP = point process; RV = random variable; dummy = dummy variable. The notation and symbols are grouped roughly by topic (e.g., UDG model parameters, elliptical galaxy parameters, etc.) and order of appearance in the paper. For detailed explanations, see Section \ref{sec:method} and Section \ref{sec:inf_est}.}
\label{tab:notation}
\end{table}
\endgroup

%% file: sections/Supplementary-Materials.tex
\section{Supplementary Materials}
\subsection{Obtaining Globular Cluster Color Variations}

To obtain the GC color variations as described in Section 3 of the main text, we first consider the priors for $\mu_C$, $\sigma_C$, $\sigma_W$, and $h_w$. For $\mu_C$ and $\sigma_C$,  we  use the default independent prior in INLA where
\[
\mu_C \sim \mathcal{N}(0, 1000), \ \sigma_C \sim \text{Inv-Gamma}(1, 10^{-5}).
\]
For $\sigma_W$ and $h_w$, the priors are chosen based on the penalized-complexity priors from \cite{simpson_2017} and are informed by the type of environments in the images: if an image has no giant elliptical galaxies (i.e., only intergalactic-medium with potential UDGs), we set
\[P(\sigma_W > 0.1) = 0.1,\ P(h_W < 15 \text{ kpc}) = 0.9.\] 
For images with giant elliptical galaxies, the priors for $\sigma_W$ and $h_w$ are adjusted according to the number and the apparent size of the giant elliptical galaxies. Doing this accounts for additional GC color spatial variations introduced by these galaxies. The uncertainty of $V$ from model fitting is effectively the predictive error obtained by INLA.

\subsection{Derivation of $p_{im}$}
To derive the expression for $p_{im}$ in Model 2, let $\mathbf{x}_i$ be a random variable in $S$ that denotes the location of the $i$-th GC. We then have
\[
    p_{im} &= \mathbb{P}(Z_{i} = m \mid \mathbf{x}_i = x_i, \mathbf{X}_c, \beta_0, \boldsymbol{\Gamma}, \boldsymbol{\Upsilon}, \boldsymbol{\Sigma}) \\
    &= \frac{\pi(\mathbf{x}_i = x_i \mid Z_{i} = m, \mathbf{X}_c, \beta_0, \boldsymbol{\Gamma}, \boldsymbol{\Upsilon}, \boldsymbol{\Sigma})\mathbb{P}(Z_{i} = m \mid \mathbf{X}_c, \beta_0, \boldsymbol{\Gamma}, \boldsymbol{\Upsilon}, \boldsymbol{\Sigma})}{\pi(\mathbf{x}_i = x_i, \mathbf{X}_c, \beta_0, \boldsymbol{\Gamma}, \boldsymbol{\Upsilon}, \boldsymbol{\Sigma})} \\
    &= \frac{\pi(\mathbf{x}_i = x_i \mid Z_{i} = m, \mathbf{X}_c, \beta_0, \boldsymbol{\Gamma}, \boldsymbol{\Upsilon}, \boldsymbol{\Sigma})\mathbb{P}(Z_{i} = m \mid \mathbf{X}_c, \beta_0, \boldsymbol{\Gamma}, \boldsymbol{\Upsilon}, \boldsymbol{\Sigma})}{\sum_{k=0}^{N_U}\pi(\mathbf{x}_i = x_i \mid Z_{i} = k, \mathbf{X}_c, \beta_0, \boldsymbol{\Gamma}, \boldsymbol{\Upsilon}, \boldsymbol{\Sigma})\mathbb{P}(Z_{i} = k \mid \mathbf{X}_c, \beta_0, \boldsymbol{\Gamma}, \boldsymbol{\Upsilon}, \boldsymbol{\Sigma})}.
\]
The first term in the numerator above is the probability density that the $m$-th environment produces a GC at location $x_i$ given all other parameters. Denote the intensity of GCs from the $m$-th environment by $\lambda_m(s), m \in \langle N_U \rangle$. The required probability density is obtained by normalizing $\lambda_m(s)$ to a probability density on $S$ and evaluated at $x_i$. Thus, we have
\[
\pi(\mathbf{x}_i = x_i \mid Z_{i} = m, \mathbf{X}_c, \beta_0, \boldsymbol{\Gamma}, \boldsymbol{\Upsilon}, \boldsymbol{\Sigma}) = \frac{\lambda_m(x_i)}{\int_S\lambda_m(s)ds}.
\]
The second term in the numerator is the probability that the GC point process produces a GC from the $m$-th environment given the UDG centers and all parameters (notice that without conditioning on $\mathbf{x}_i$, the index $i$ does not hold any distinguishing property for different GCs). This is simply the ratio between the mean number of GCs produced by the $m$-th environment and the mean number of GCs produced by the entire point process. Hence,
\[
\mathbb{P}(Z_{i} = m \mid \mathbf{X}_c, \beta_0, \boldsymbol{\Gamma}, \boldsymbol{\Upsilon}, \boldsymbol{\Sigma}) = \frac{\int_S\lambda_m(s)ds}{\sum_{k = 0}^{N_U}\int_S\lambda_k(s)ds}.
\]
Thus, we arrive at
\[
p_{im} = \frac{\lambda_m(x_i)}{\sum_{k=0}^{N_U}\lambda_k(x_i)} = \frac{\lambda_m(x_i)}{\Lambda(x_i)}.
\]
Note that the dependence on $\boldsymbol{\Sigma}$ for the GC mark distribution was never utilized. In the data-generating process, the locations of GCs are generated first and the GC marks are generated second. The GC marks are then ``attached" to each GC based on environment. Thus, without conditioning on the GC marks $\mathbf{M}$, the membership probability of a GC does not depend on $\boldsymbol{\Sigma}$. 

\subsection{Derivation of Likelihood and Prior}

Following the notations in the main paper, assume that $N_U$ is fixed, write $\boldsymbol{\Theta} = (\beta_0, \lambda_c, \boldsymbol{\Gamma}, \Sigma_0)$, and $\boldsymbol{\Phi} = \{(c_j^u, \boldsymbol{U}_j, \Sigma_j)\}_{j=1}^{N_U}$. The likelihood is then
\begingroup
\allowdisplaybreaks
\begin{align*}
    \pi(\mathbf{X}, \mathbf{M} \mid \boldsymbol{\Theta}, \boldsymbol{\Phi}) &=\pi(\mathbf{X} \mid \boldsymbol{\Theta}, \boldsymbol{\Phi})\pi(\mathbf{M} \mid \mathbf{X}, \boldsymbol{\Theta}, \boldsymbol{\Phi})\\
    &=\pi(\mathbf{X} \mid \boldsymbol{\Theta}, \boldsymbol{\Phi})\pi(\mathbf{M} \mid \mathbf{X}, \boldsymbol{\Theta}, \boldsymbol{\Phi})\\
    &=\pi(\mathbf{X} \mid \boldsymbol{\Theta}, \boldsymbol{\Phi})\prod_{i=1}^{n}\pi(\mathbf{M}(\mathbf{x}_i) \mid \mathbf{X}, \boldsymbol{\Theta}, \boldsymbol{\Phi})\\
    &=\pi(\mathbf{X} \mid \boldsymbol{\Theta}, \boldsymbol{\Phi})\prod_{i=1}^{n}\left[\sum_{m=0}^{N_U} \mathbb{P}(Z_{i} = m \mid \mathbf{X}, \boldsymbol{\Theta}, \boldsymbol{\Phi}) \pi(\mathbf{M}(\mathbf{x}_i) \mid Z_i = m, \mathbf{X}, \boldsymbol{\Theta}, \boldsymbol{\Phi})\right] \\
    &=\pi(\mathbf{X} \mid \boldsymbol{\Theta}, \boldsymbol{\Phi})\prod_{i=1}^{n}\left[\sum_{m=0}^{N_U} p_{im} \pi(\mathbf{M}(\mathbf{x}_i) \mid Z_i = m, \mathbf{X}, \boldsymbol{\Theta}, \boldsymbol{\Phi})\right] \\
    &=\exp\left(|S|-\int_S \Lambda(s)ds\right)\prod_{i=1}^{n}\Lambda(x_n)\prod_{i = 1}^n\left[\sum_{m = 0}^{N_U}p_{im}\pi_{\Sigma_m}(\mathbf{M}(x_i))\right].
\end{align*}
\endgroup
For the prior distribution, we have
\begin{align*}
\pi(\boldsymbol{\Theta}, \boldsymbol{\Phi}) &= \pi(\beta_0, \lambda_c, \boldsymbol{\Gamma}, \mathbf{X}_c, \boldsymbol{\Upsilon}, \boldsymbol{\Sigma})\\
&= \pi(\mathbf{X}_c \mid \lambda_c)\pi(\beta_0)\pi(\lambda_c)\pi(\boldsymbol{\Gamma})\pi(\boldsymbol{\Upsilon})\pi(\boldsymbol{\Sigma}) \\
&= \pi(\mathbf{X}_c \mid \lambda_c)\left[\prod_{k=1}^{N_G}\pi(\boldsymbol{G}_k)\right]\left[\prod_{m = 1}^{N_U}\pi(\boldsymbol{U}_j)\right]\pi(\boldsymbol{\Sigma})\pi(\lambda_c)\pi(\beta_0) \\
&= \exp(|S|(1-\lambda_c))\lambda_c^{N_U}\left[\prod_{m = 1}^{N_U}\pi(\Sigma_m \mid \Sigma_0)\pi(\boldsymbol{U}_j)\right]\left[\prod_{k=1}^{N_G}\pi(\boldsymbol{G}_k)\right]\pi(\beta_0)\pi(\lambda_c)\pi(\Sigma_0).
\end{align*}

\subsection{Prior}

For $\lambda_c$, we set $\lambda_c \sim \mathrm{Unif}(0, 5)$. The specific value of $l_c$ is uncertain \citep{Lim2018, Venhola2017, Lim2020} so we provide it with a uniform distribution. The upper bound of $l_c = 5$ is a reasonable value for the maximum mean number of UDGs given the scope of each image in our dataset.

For $\varlambda^u_j, \ j \in [N_U]$, we set $\varlambda^u_j \sim \mathrm{LN}(\log(7.6), 0.87^2)$, based on results from \cite{Dokkum2016, Amorisco_2018, Lim2018, Lim2020}. Specifically, we obtained the said prior by combining the UDG GC count data from \citeauthor{Dokkum2016, Amorisco_2018, Lim2018, Lim2020} and fitted a log-normal distribution to the data. Note that we removed UDGs with no GCs when we are fitting the distribution, since our goal is to detect UDG through their GC population. 

For $R^u_j, \ j \in [N_U]$, we set $R^u_j \sim \mathrm{LN}(\log(0.03), 0.5^2)$ based on the typical size of an UDG of $1.5$~kpc \citep{VanDokkum2015, Forbes2017, Saifollahi2022}. The mode value of $0.03$ is a scaled value based on the scaling of images into $[0,1]^2$, this corresponds to a value of roughly $2$~kpc in physical scale. For the rest of the parameters, we set $n^u_j \sim \mathrm{LN}(\log(1), 0.75^2)$, $\rho^u_j \sim \mathrm{LN}(\log(1), 0.3^2)$ based on previous studies \citep{Burkert2017, Prole2019, Saifollahi2022} while $\varphi^u_j \sim \mathrm{Unif}(0, \pi)$. 

If we consider the brightness of GCs as marks, we assign
\begin{align*}
    \mu_0 &\sim \mathrm{Unif}(23, 27), \\
    \mu_m \mid \mu_0 &\sim \mathrm{Unif}(23, \mu_0), \ m > 0 \\
    \sigma_{m,B} &\sim \mathrm{Unif}(0.5, 1.9), \ m \in \{0\}\cup[N_U].
\end{align*}
The assignment of the above prior is based on known GCLF parameters \citep{Harris1991, Rejkuba_2012} and the distance to the Perseus cluster \citep{Harris2020}. Note that a more informative prior might be placed on $\mu_0$ and $\sigma_{m,B}$; the GCLF parameters are extensively well-studied, and for a long time were considered to be universal \citep{Harris1991, Rejkuba_2012}. However, the recent observation by \cite{Shen_2021b} has introduced significant doubt on the universality of the GCLF, especially for UDGs. Thus, we set a more uninformative, uniform prior on these parameters.

If the color variation is considered as marks, we set
\begin{align*}
    \sigma_{0, C} &\sim \mathrm{Gamma}(0.2, 0.05), \\
    \sigma_{m,C} \mid \sigma_{0,C} &\sim \mathrm{Unif}(0, \sigma_{0,C}), \ m > 0,
\end{align*}
based on the previous extensive studies on GC colors \citep[see][for example]{Gebhardt_1999} and the assumption that $\sigma_j$'s are smaller than $\sigma_0$.

For the elliptical galaxies, we set $\varlambda_k^g \sim \mathrm{LN}(\log(\mathcal{N}_g^k), 0.25^2)$ where $\mathcal{N}_g^k$ is determined based on the specific frequency relation \citep{Harris1991}. The specific frequency is an empirical relationship that connects the total brightness of a galaxy to its GC counts:
\[
S_N = \mathcal{N}_g10^{0.4(M_V + 15)}.
\]
Here $S_N$ is the specific frequency, $\mathcal{N}_g$ is the total GC count in a galaxy, and $M_V$ is the total brightness. This relationship is well-studied and used extensively for predicting the GC counts of giant elliptical galaxies. Since we can obtain accurate estimates of the total brightness of elliptical galaxies in our data, we use it to provide a crude estimate of the value of $\mathcal{N}_g^k$ by assuming $S_N = 3$, which is typical for elliptical galaxies \citep{Harris1991}.

For $R^g_k$, we set $R_k^g \sim \mathrm{LN}(\log(\mathcal{R}_g^k), 0.25^2)$ where $\mathcal{R}_g^k$ is obtained based on the results from \cite{Forbes2017}. Specifically, \citeauthor{Forbes2017} studied the relationship between the effective radius of an elliptical galaxy and the GC system radius. They found that on average, the GC system radius is roughly $3-4$ times that of the effective radius among elliptical galaxies. For $n^g_k$, we set $n^g_k \sim \mathrm{LN}(\log(0.5), 0.5^2)$.

Lastly, for the intensity parameter $\beta_0$ of the intergalactic medium, we provide it with a prior of $\beta_0 \sim \mathrm{LN}(\log(b_0), 0.5^2)$. Here, $b_0$ is different for different image pointings, and is chosen to conserve the total number of GCs in each point pattern (i.e., to ensure that the number of GCs in UDGs, normal galaxies, and the intergalactic medium equate to the total number of GCs in each point pattern).

\subsection{Inference Algorithm}
\begin{algorithm}[!ht]
\caption{Blocked Gibbs Spatial Birth-Death-Move MCMC}\label{alg:blocked_sbdm_MCMC}
\begin{algorithmic}
\REQUIRE Data $\mathbf{D}$; Starting value $(\boldsymbol{\Theta}_1, \boldsymbol{\Phi}_1)$; number of iteration $M$; proposal density $q_\theta(\cdot \mid \cdot)$; birth proposal density $q_b^u(\cdot), q_b^s(\cdot)$; move proposal density $q_m^{\mathbf{c}}(\cdot \mid \cdot), q_m^u(\cdot \mid \cdot), q_m^s(\cdot \mid \cdot)$; $p_b$, $p_d$, $p_m$; study region $S$.
\FOR{$i = 1$ to $M$}
    \STATE Current state $(\boldsymbol{\Theta},\boldsymbol{\Phi}) \gets (\boldsymbol{\Theta}_i,\boldsymbol{\Phi}_i)$.
    \STATE Generate $\boldsymbol{\Theta}' \sim q_\theta(\boldsymbol{\Theta}' \mid \boldsymbol{\Theta})$. Set $\boldsymbol{\Theta}_{i+1} = \boldsymbol{\Theta}'$ with probability
    \[
    \min\left\{1, \frac{\pi(\boldsymbol{\Theta}', \boldsymbol{\Phi} \mid \mathbf{D})}{\pi(\boldsymbol{\Theta}, \boldsymbol{\Phi} \mid \mathbf{D})}\right\}.
    \]
    \STATE Otherwise $\boldsymbol{\Theta}_{i+1} = \boldsymbol{\Theta}$.
    \STATE Generate $U_m \sim \mathrm{Uniform}(0,1)$.
    \IF{$U_m < 1 - p_m$}
        \STATE Generate $U_b \sim \mathrm{Uniform}(0,1)$.
        \IF{$U_b < p_b$ or $|\boldsymbol{\Phi}| = 0$}
            \STATE Generate new cluster proposal $(\mathbf{c}, \boldsymbol{U}, \Sigma)$ with $\mathbf{c} \sim \mathrm{Uniform}(S)$, $\boldsymbol{U} \sim q_b^u(\boldsymbol{U})$, $\Sigma \sim q_b^s(\Sigma)$. Set $\boldsymbol{\Phi}_{i+1} = \boldsymbol{\Phi} \cup (\mathbf{c}, \boldsymbol{U}, \Sigma)$ with probability \[\min\{1, r[(\boldsymbol{\Theta}_{i+1}, \boldsymbol{\Phi}), (\boldsymbol{\Theta}_{i+1}, (\mathbf{c}, \boldsymbol{U}, \Sigma))]\}.\]
            \STATE Otherwise $\boldsymbol{\Phi}_{i+1} = \boldsymbol{\Phi}$.
            \ELSE
            \STATE Select $j$ uniformly from $\{1,\dots, N_U\}$. Set $\boldsymbol{\Phi}_{i+1} = \boldsymbol{\Phi} \backslash (\mathbf{c}_{j}^u, \boldsymbol{U}_j, \Sigma_j)$ with probability \[\min\left\{1, r[(\boldsymbol{\Theta}_{i+1}, \boldsymbol{\Phi}\backslash (\mathbf{c}_{j}^u, \boldsymbol{U}_j, \Sigma_j)), (\boldsymbol{\Theta}_{i+1}, (\mathbf{c}_{j}^u, \boldsymbol{U}_j, \Sigma_j))]^{-1}\right\}.\]
            \STATE Otherwise $\boldsymbol{\Phi}_{i+1} = \boldsymbol{\Phi}$.
        \ENDIF
    \ELSIF{$U_m > 1 - p_m$ and $|\boldsymbol{\Phi}| > 0$}
        \STATE Select $j$ uniformly from $\{1,\dots, N_U\}$. Generate move proposal $(\mathbf{c}, \boldsymbol{U}, \Sigma)$ with $\mathbf{c} \sim q_m^{\mathbf{c}}(\mathbf{c} \mid \mathbf{c}_{j}^u, R_j^{u})$, $\boldsymbol{U} \sim q_m^u(\boldsymbol{U} \mid \boldsymbol{U}_j)$, $\Sigma \sim q_m^s(\Sigma \mid \Sigma_j)$.  Set $\boldsymbol{\Phi}_{i+1} = \boldsymbol{\Phi} \backslash (\mathbf{c}_{j}^u, \boldsymbol{U}_j, \Sigma_j) \cup (\mathbf{c}, \boldsymbol{U}, \Sigma)$ with probability \[\label{eqn:move kernel}
        \min\left\{1, \frac{\pi(\boldsymbol{\Theta}_{i+1}, \boldsymbol{\Phi} \backslash (\mathbf{c}_{j}^u, \boldsymbol{U}_j, \Sigma_j) \cup (\mathbf{c}, \boldsymbol{U}, \Sigma) \mid \mathbf{D})q_m^{\mathbf{c}}(\mathbf{c}_{j}^u \mid \mathbf{c}, R)q_m^u(\boldsymbol{U}_j \mid \boldsymbol{U})q_m^s(\Sigma_j \mid \Sigma)}{\pi(\boldsymbol{\Theta}_{i+1}, \boldsymbol{\Phi} \mid \mathbf{D})q_m^{\mathbf{c}}(\mathbf{c} \mid \mathbf{c}_{j}^u,R_j^{u})q_m^u(\boldsymbol{U} \mid \boldsymbol{U}_j)q_m^s(\Sigma \mid \Sigma_j)}\right\}.
        \]
        \STATE Otherwise $\boldsymbol{\Phi}_{i+1} = \boldsymbol{\Phi}$.
    \ELSE
    \STATE $\boldsymbol{\Phi}_{i+1} = \boldsymbol{\Phi}$.
    \ENDIF
\ENDFOR
\STATE Output $\{(\boldsymbol{\Theta}_1, \boldsymbol{\Phi}_1), \dots, (\boldsymbol{\Theta}_M, \boldsymbol{\Phi}_M)\}$.
\end{algorithmic}
\end{algorithm}

Algorithm \ref{alg:blocked_sbdm_MCMC} illustrates the detailed implementation of our MCMC inference algorithm. Notations follow the ones from the main paper.

\subsubsection{Convergence} 

Our algorithm requires novel theoretical results to provide sufficient conditions of convergence. The adaptive Metropolis update for $\bm{\Theta}$ given $\bm{\Phi}$ destroys the Markovian property of the chain, and therefore, standard Markov chain theory does not apply. 

\cite{haario_componentwise_2005} gave sufficient conditions of adaptive (blocked) Metropolis-within-Gibbs algorithm to be ergodic for posterior distribution with fixed dimensions, which state that the non-adaptive (blocked) Metropolis-within-Gibbs algorithm with Gaussian proposal for each individual block needs to be uniformly ergodic. 

Firstly, it is unclear whether the uniform ergodicity condition stated in \cite{haario_componentwise_2005} also extends to Algorithm \ref{alg:blocked_sbdm_MCMC} even if the update scheme for the trans-dimensional block $\bm{\Phi}$ is non-adaptive. Secondly, if the non-adaptive version of Algorithm \ref{alg:blocked_sbdm_MCMC} needs to be uniformly ergodic, then this is yet to be proven. No existing theoretical results provide sufficient conditions for uniform ergodicity. Proposition 6 in \cite{moller2005} gave sufficient conditions of $V$-uniform ergodicity for their spatial-birth-death-move algorithm where $V(\bm{\Phi}) = \beta^{|\bm{\Phi}|}$ with $\beta > 1$ under the simple generalized shot noise Cox process, but no conditions were given for uniform ergodicity of their algorithm. Moreover,  even for $V$-uniform ergodicity, it is unclear how to extend \cite{moller2005}'s conditions to our case where we have additional fixed-dimensional parameters $\bm{\Theta}$ and models for the GC marks, let alone uniform ergodicity.

To summarize, the theoretical results required to show the ergodicity of Algorithm \ref{alg:blocked_sbdm_MCMC} do not yet exist, and they are out of the scope of this paper. Nevertheless, in what follows, we do provide sufficient conditions so that our algorithm converges when the adaptive Metropolis update for $\bm{\Theta}$ is changed to standard Metropolis (Gaussian) update. 

Recall the support of all (hyper-) parameters defined in Table 1 of the main paper and $S$ the study region, write the support of $\bm{\Theta}$ as $S_{\bm{\Theta}}$, the support of $\bm{U}$ as $S_{\bm{U}}$, and the support of $\Sigma$ as $S_{\Sigma}$. We assume $\bm{\Theta} \subseteq Q \subseteq S_{\bm{\Theta}}$ where $Q$ is a bounded measurable set, and $\bm{\Phi} = \{(\mathbf{c}^u_j, \bm{U}_j, \Sigma_j)\}_{j=1}^{N_U}$ is almost surely finite. Moreover, for any $(\mathbf{c}^u_j, \bm{U}_j, \Sigma_j)$, assume $(\mathbf{c}^u_j, \bm{U}_j, \Sigma_j) \in \mathscr{F} = S_{ext}\times \mathscr{U}\times\mathscr{S}$ where $S \subseteq S_{ext} \subseteq \mathbb{R}^2$, $\mathscr{U} \subseteq S_{\bm{U}}$, $\mathscr{S} \subseteq S_{\Sigma}$ with $S_{ext}, \mathscr{U}, \mathscr{S}$ being bounded and measurable. Denote $\Omega_{N_U} = \left\{\{(\mathbf{c}^u_j, \bm{U}_j, \Sigma_j)\}_{j=1}^{N_U} \subseteq \mathscr{F}\right\}$ and $\Omega_{0} = \emptyset$. Write $\Omega = \bigcup_{i=0}^\infty \Omega_{i}$. 

Given the above assumptions, we need to show that i) the standard Metropolis update with Gaussian proposal for $\bm{\Theta}$ given $\bm{\Phi}$ leaves $\pi(\bm{\Theta}, \bm{\Phi} \mid \mathbf{D})$ invariant; ii) the BDM-MH update for $\bm{\Phi}$ given $\bm{\Theta}$ leaves $\pi( \bm{\Phi}, \boldsymbol{\Theta} \mid \mathbf{D})$ invariant. Given i) and ii), the combined transition kernels of i) and ii) trivially leave $\pi(\bm{\Theta}, \bm{\Phi} \mid \mathbf{D})$ invariant since i) and ii) form a cyclic transition kernel through blocked Gibbs update. Lastly, we just need to show the entire chain is irreducible and aperiodic.

For i), given the assumptions of the support of $(\bm{\Theta}, \bm{\Phi})$, $\pi(\bm{\Theta}, \bm{\Phi} \mid \mathbf{D})$ is well-defined and bounded from above and away from zero. Thus, for any $\bm{\Theta}',  \bm{\Theta} \in Q$ and $\bm{\Phi} \in \Omega$,
\[
\frac{\pi(\bm{\Theta}' \mid \bm{\Phi}, \mathbf{D})}{\pi(\bm{\Theta} \mid \bm{\Phi}, \mathbf{D})} = \frac{\pi(\bm{\Theta}', \bm{\Phi} \mid \mathbf{D})}{\pi(\bm{\Theta}, \bm{\Phi} \mid \mathbf{D})},
\]
and the ratio is well defined. It is therefore trivial that the standard Metropolis update with Gaussian proposal for $\bm{\Theta}$ given $\bm{\Phi}$ indeed leaves $\pi(\bm{\Theta}, \bm{\Phi} \mid \mathbf{D})$ invariant.

For ii), we follow the proof of convergence from \cite{Das_2019}. We first check the detailed balance condition for the birth-death step. Same as in i), for any $\bm{\Theta} \in Q$, and any $\bm{\Phi}, \bm{\Phi}' \in \Omega$, $\pi(\bm{\Phi}, \boldsymbol{\Theta} \mid \mathbf{D})$ and $\pi(\bm{\Phi}', \boldsymbol{\Theta} \mid \mathbf{D})$ are well-defined (bounded from above and away from zero) given the support of $(\bm{\Theta}, \bm{\Phi})$ and the fact that $\bm{\Phi}$ is almost surely finite. Thus,
\[
\frac{\pi(\bm{\Phi}' \mid \bm{\Theta}, \mathbf{D})}{\pi(\bm{\Phi} \mid \bm{\Theta}, \mathbf{D})} = \frac{\pi(\bm{\Phi}', \bm{\Theta} \mid \mathbf{D})}{\pi(\bm{\Phi}, \bm{\Theta} \mid \mathbf{D})}
\]
is well-defined. Now for the birth step, suppose we move from $\bm{\Phi}$ to $\bm{\Phi} \cup \{(\mathbf{c}, \bm{U}, \Sigma)\}$ with $\bm{\Theta} \in Q$ fixed, the probability of transition is
\[\label{eqn:birth trans}
& \pi( \bm{\Phi}, \boldsymbol{\Theta} \mid \mathbf{D})\times p_b \frac{1}{|S|}  q_b^u(\bm{U})q_b^s(\Sigma) \times \min\left\{1, \frac{p_d\pi(\bm{\Phi} \cup \{(\mathbf{c}, \bm{U}, \Sigma)\}, \boldsymbol{\Theta}\mid \mathbf{D})|S|}{p_b\pi(\bm{\Phi}, \boldsymbol{\Theta} \mid \mathbf{D})(|\bm{\Phi}| + 1)q_b^u(\bm{U})q_b^s(\Sigma)}\right\} \\
&=\min\left\{\pi(\bm{\Phi}, \boldsymbol{\Theta} \mid \mathbf{D})p_b\frac{1}{|S|}q_b^u(\bm{U})q_b^s(\Sigma), \quad \frac{1}{|\bm{\Phi}| + 1}p_d\pi(\bm{\Phi} \cup \{(\mathbf{c}, \bm{U}, \Sigma)\}, \boldsymbol{\Theta} \mid \mathbf{D})\right\}.
\]
For detailed balance to hold, we need to return from $\bm{\Phi} \cup \{(\mathbf{c}, \bm{U}, \Sigma)\}$ to $\bm{\Phi}$ with $\bm{\Theta}$ fixed. The corresponding probability of transition is
\[\label{eqn:death trans}
& \pi(\bm{\Phi} \cup \{(\mathbf{c}, \bm{U}, \Sigma)\}, \boldsymbol{\Theta} \mid \mathbf{D}) \times p_d  \frac{1}{|\bm{\Phi}| + 1}\times \min\left\{1, \frac{p_b\pi(\bm{\Phi}, \boldsymbol{\Theta}\mid\mathbf{D})(|\bm{\Phi}| + 1)q_b^u(\bm{U})q_b^s(\Sigma)}{   p_d\pi(\bm{\Phi} \cup \{(\mathbf{c}, \bm{U}, \Sigma)\}, \boldsymbol{\Theta}, \mid\mathbf{D})|S|}\right\} \\
&= \min\left\{\frac{1}{|\bm{\Phi}| + 1}p_d\pi(\bm{\Phi} \cup \{(\mathbf{c}, \bm{U}, \Sigma)\}, \boldsymbol{\Theta} \mid \mathbf{D}),  \quad \frac{1}{|S|}p_b\pi(\bm{\Phi} , \boldsymbol{\Theta} \mid\mathbf{D})q_b^u(\bm{U})q_b^s(\Sigma)\right\}.
\]
Clearly, \ref{eqn:birth trans}$=$\ref{eqn:death trans}. Moreover, all proposal distributions defined on $\mathscr{F}$ are bounded from above and away from zero in the above. Therefore, detailed balance condition holds for the birth-death step.

For the move step in the BDM-MH update presented in Algorithm \ref{alg:blocked_sbdm_MCMC}, the transition kernel is determined by a standard Metropolis-Hasting accept-reject step (\cref{eqn:move kernel}). Since all proposal distributions given in Table 2 of the main paper are bounded from above and away from zero on $\mathscr{F}$, the detailed balance condition also holds for the move step.

The entire transition kernel in our BDM-MH algorithm is a mixture of the birth-death kernel and the move kernel with mixture probabilities given by $1-p_m$ and $p_m$ respectively. Since both transition kernels satisfy detailed balance conditions, the transition kernel formed by their mixture also satisfies detailed balance condition \citep[see e.g.][]{Geyer1994, roodaki2012note}.

For irreducibility and aperiodicity, suppose the chain is at $(\bm{\Theta}, \bm{\Phi}) = \left(\bm{\Theta}, \{(\mathbf{c}_j^u, \bm{U}_j, \Sigma_j)\}_{j = 1}^{K}\right)$. Let $(\bm{\Theta}', \bm{\Phi}') = \left(\bm{\Theta}',\{(\mathbf{c}_j^{u'}, \bm{U}_j', \Sigma_j')\}_{j = 1}^{K'}\right)$ be such that $\bm{\Theta}' \in Q$, $\bm{\Phi}' \subseteq \mathcal{F}$ with $K' \in \mathbb{N}$, and $\pi(\bm{\Theta}', \bm{\Phi}' \mid \mathbf{D}) > 0$. Note that if $K' = 0$, $\bm{\Phi}' = \emptyset$. Clearly, one can reach from $\bm{\Theta}$ to $\bm{\Theta}'$ in one step with positive probability. One can also reach from $K$ to $K'$ in $|K-K'|$ steps with positive probability through the birth-death steps. After $K'$ is reached, the configuration $\{(\mathbf{c}_j^{u'}, \bm{U}_j', \Sigma_j')\}_{j = 1}^{K'}$ can be reached in $K'$ move steps with positive probability. Thus, the chain is irreducible. Since $(\bm{\Theta}', \bm{\Phi}')$ is arbitrary, the chain is also aperiodic. Therefore, the entire Makov chain is irreducible, aperiodic, with $\pi(\bm{\Phi}, \boldsymbol{\Theta} \mid \mathbf{D})$ as its stationary distribution.

Note that in practical implementation, not having the requirement of bounded support $Q, \mathscr{F}$ for $(\bm{\Theta}, \bm{\Phi})$ rarely affects the convergence of the algorithm as majority of the posterior mass is contained in some sufficiently large bounded region. For the convergence performance of our Algorithm \ref{alg:blocked_sbdm_MCMC}, we illustrate it by the simulation example in Section \ref{supp_sec: simulation}.

\subsubsection{Initialization, Burn-in, and Tuning Parameters}

To initialize the chains, all (hyper-) parameters are given initial values drawn from their respective (hyper-) priors. The UDG locations are initialized uniformly across the study region $S$ with the initial UDG numbers being Poisson with parameter $\lambda_c$ being the starting value. The burn-in period is determined by inspecting the traceplots and the details given in Table \ref{tab:fixed-dim diagnostic}.

For $C_\theta$ in the initial run of the adaptive MCMC algorithm for the fixed-dimensional parameters $\boldsymbol{\Theta}$, we set the diagonal element for $\beta_0$ to $0.025$ and $\lambda_c$ to $0.1$. When there are elliptical galaxies in the images, all components of the S\'{e}rsic parameters have corresponding diagonal elements in $C_\theta$ being $0.025$. For $\varepsilon$, we have effectively set it to $0$, our experiments showed that $\varepsilon = 0$ has no impact on the final MCMC results especially for uni-modal posteriors as indicated by \cite{Haario_2001}, which are the case here.

For the the scale parameters of the proposal distributions in the move step (shown in Table 2 of the main text) of our algorithm, we set $\delta_c = 0.25$, $\Delta_U = \text{diag}(0.05)$, $\delta_\varphi = 0.05$, $\delta_\mu = 0.1$, $\delta_{\sigma,B} = 0.05$, and $\delta_{\sigma,C} = 0.05$.

\subsection{MCMC Diagnostics}

To assess MCMC convergence, we monitored the potential scale reduction factor \cite[$\hat{R}$;][]{gelman1992, vats_2019, vats2021} and the effective sample size using the \texttt{stable.GR} \citep{vats_2019, vats2021} and \texttt{mcmcse} \citep{mcmcse} R packages. The methods by \cite{vats_2019, vats2021} are adopted here as they provide convergence criteria for multivariate MCMC results. The key quantities we monitor are:
\begin{itemize}
    \item \textbf{Model 1}: For a sample $(\mathbf{X}_{c}^{(t)}, \boldsymbol{\Xi}^{(t)})$ from the MCMC chains, compute 
    \[\label{eqn: model1_L}
    \log\{\mathcal{L}(\mathbf{x}_i)\} = \log\left\{\Lambda\left(\mathbf{x}_i; \mathbf{X}_c^{(t)}, \boldsymbol{\Xi}^{(t)}\right)\right\}, \ i = 1, \dots, n.
    \]
    \item \textbf{Model 2}: For a sample $(\mathbf{X}_c^{(t)}, \boldsymbol{\Xi}^{(t)}, \boldsymbol{\Sigma}^{(t)})$ from the MCMC chains, compute
    \[\label{eqn: model2_L}
    \log\{\mathcal{L}(\mathbf{x}_i)\} &= \log\bigg[\left\{\beta_0^{(t)} + \Lambda_{\mathrm{G}}\left(\mathbf{x}_i; \boldsymbol{\Gamma}^{(t)}\right)\right\}\pi\left\{\mathbf{M}(\mathbf{x}_i); \Sigma_0^{(t)}\right\} \\
    &+ \sum_{j = 1}^{N_U^{(t)}}\text{S\'{e}rsic}\left\{\mathbf{x}_i; c_{j}^{u,(t)}, \boldsymbol{U}_j^{(t)}\right\}\pi\left\{\mathbf{M}(\mathbf{x}_i); \Sigma_j^{(t)}\right\}\bigg], \ i = 1, \dots, n. 
    \]
\end{itemize}

For Model 1, we focus on the log-intensity of the point process at each GC location. For Model 2, we consider the adjusted log-intensity with mark distribution included. These quantities contain all model parameters except the hyper-parameter $\lambda_c$, and are scalar with consistent definitions and interpretations across model transitions \citep{hastie_model_2012}. We also monitor the multivariate convergence of $\boldsymbol{\Theta}$. Additionally, we consider the convergence diagnostic for the number of UDGs $N_U$ based on the MCMC diagnostics for categorical variables proposed by \cite{deonovic2017convergence}. 

We acknowledge the limitations of our diagnostic approach, as convergence diagnostics for RJMCMC are particularly challenging \citep{Sisson_2005}. While there has been little dedicated work, notable efforts include \cite{sisson_distance-based_2007} and 
\cite{heck_quantifying_2019} . The latter use a distance-based diagnostic by treating the samples as a Point process realization, and the former conduct convergence for RJMCMC using a discrete Markov model. However, we do not consider these methods due to their complex implementation and the significant computational demands they pose for our models.

\subsubsection{Convergence of Fixed-Dimensional Parameters}

We present the convergence diagnostics for the fixed-dimensional parameters $\boldsymbol{\Theta}$ for each model we have fitted to the $12$ images in Table
~\ref{tab:fixed-dim diagnostic}. We compute the multivariate PSRF and multivariate effective sample size based on \cite{vats_2019, vats2021}. Their method also provides a convergence criteria for the multivariate MCMC output analysis based on a required accuracy to construct $1- \alpha$ confidence region for the parameters. We set the relative precision of the the CI to $\epsilon = 0.1$ and $\alpha = 0.1$. The results in Table~\ref{tab:fixed-dim diagnostic} indicate that all our models have most likely reached convergence for the fixed-dimensional parameters.

\subsubsection{Trans-Dimensional Diagnostics}

\begin{figure}[!ht]
    \centering
    \includegraphics[width = 0.7\columnwidth]{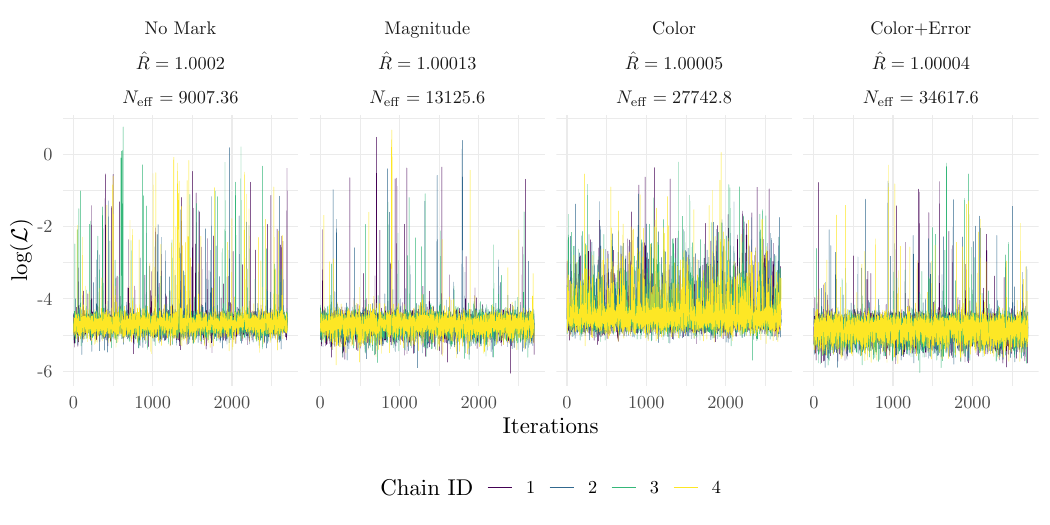}
    \caption{\footnotesize{Traceplots, PSRF, and effective sample size of quantities shown in Eq.~\ref{eqn: model1_L} and \ref{eqn: model2_L} for selected GCs in V6-WFC3.}}
    \label{fig:v6wfc3_trace}
\end{figure}

\begin{figure}[!ht]
    \centering
    \includegraphics[width = 0.7\columnwidth]{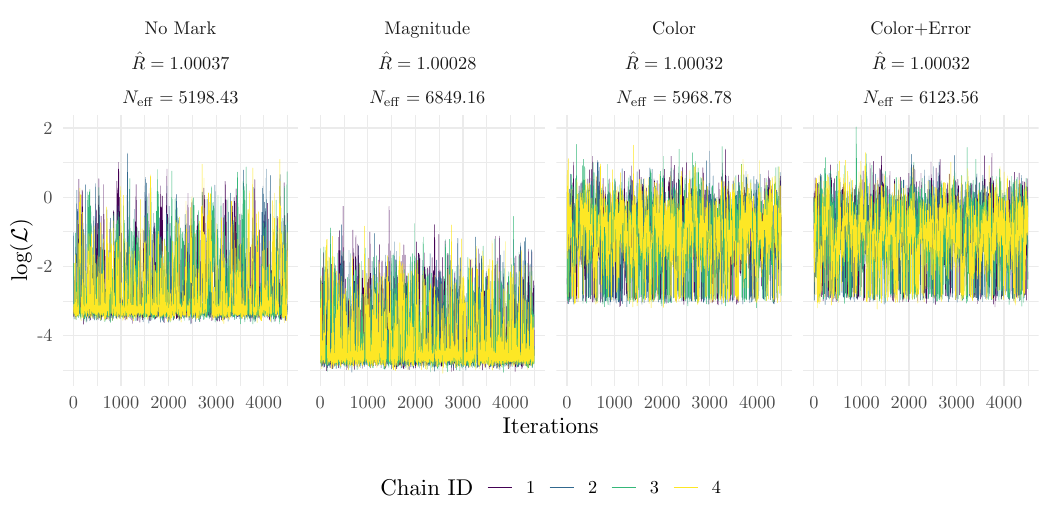}
    \caption{\footnotesize{Traceplots, PSRF, and effective sample size of quantities shown in Eq.~\ref{eqn: model1_L} and \ref{eqn: model2_L} for selected GCs in V11-ACS.}}
    \label{fig:v11acs_trace}
\end{figure}

For the trans-dimensional parameters (i.e., Eq.~\ref{eqn: model1_L} and \ref{eqn: model2_L}), we use the traceplots, PSRF, and effective sample sizes to assess convergence for each of the $12$ images. Because our data include a total of 12 images and thousands of GCs, we only present the results for selected GCs in the image V6-WFC3 and V11-ACS, shown in \cref{fig:v6wfc3_trace} and \cref{fig:v11acs_trace},  for succinctness. For each image, the GCs selected for diagnostics are either supposedly in a detected UDG or have the lowest effective sample size. The diagnostic results for other images are similar to the ones presented here, and are provided at \href{https://doi.org/10.5281/zenodo.10864105}{project repository}.

Due to memory constraints, we show the traceplots of chains thinned by every $100$ samples. Based on the PSRFs and visual inspection of the traceplots, all chains seem to have reached convergence.

\subsubsection{Convergence Diagnostic of $N_U$}

Another aspect of convergence we assess is the number of UDGs $N_U$ in an image obtained from the MCMC chains. As noted in \cite{deonovic2017convergence}, traditional diagnostics such as the PSRF is not suited for categorical variables such as $N_U$. \citeauthor{deonovic2017convergence} thus propose a convergence diagnostic for categorical variables based on a modified Pearson's Chi-Squared test for homogeneity that accounts for autocorrelation in MCMC samples. However, there is no available software that we can employ to conduct the test. The test also requires categorical time series from the MCMC chains to follow a certain model structure. Therefore, we opt to thin our MCMC chains for $N_U$ until the samples can be considered independent. We then use the thinned sample of $N_U$ to conduct the standard Pearson's Chi-Squared test for homogeneity.

\begin{table}[H]
    \centering
    \resizebox{\columnwidth}{!}{%
    \begin{tabular}{lclcccc}
    \toprule
       Image ID (Total Samples after Thinning) & Burn-in (per chain) & Models  &  $\mathrm{dim}(\boldsymbol{\Theta})$ & mPSRF & $N_{\mathrm{eff}}$ & Converged? \\
       \midrule
       \multirow{4}{*}{V6-WFC3 ($108$~K)} & $30$~K & No-Mark & $2$ & $< 1.0005$ & $36026.2$ & Yes \\
       & $30$~K & Magnitude & $4$ & $< 1.0005$ & $30022.2$ & Yes \\
       & $30$~K & Color & $3$ & $< 1.0005$ & $27486.6$ & Yes \\
       & $30$~K & Color+Error & $3$ & $< 1.0005$ & $19415.0$ & Yes \\
       \midrule
       \multirow{4}{*}{V7-ACS ($160$~K)} & $100$~K & No-Mark & $2$ & $< 1.0005$ & $44300.8$ & Yes \\
       & $100$~K & Magnitude & $4$ & $< 1.0005$ & $17923.6$ & Yes \\
       & $100$~K & Color & $3$ & $< 1.0005$ & $26029.5$ & Yes \\
       & $100$~K & Color+Error & $3$ & $< 1.0005$ & $24194.2$ & Yes \\
       \midrule
       \multirow{4}{*}{V8-WFC3 ($180$~K)} & $50$~K & No-Mark & $8$ & $< 1.0005$ & $22972.8$ & Yes \\
       & $50$~K & Magnitude & $10$ & $< 1.0005$ & $12300$ & Yes \\
       & $50$~K & Color & $9$ & $< 1.0005$ & $9432.1$ & Yes \\
       & $50$~K & Color+Error & $9$ & $< 1.0005$ & $12902.5$ & Yes \\
       \midrule
       \multirow{4}{*}{V10-ACS ($180$~K)} & $50$~K & No-Mark & $5$ & $< 1.0005$ & $37767.9$ & Yes \\
       & $50$~K & Magnitude & $7$ & $< 1.0005$ & $16028.1$ & Yes \\
       & $50$~K & Color & $6$ & $< 1.0005$ & $25877.9$ & Yes \\
       & $50$~K & Color+Error & $6$ & $< 1.0005$ & $22165.7$ & Yes \\
       \midrule
       \multirow{4}{*}{V11-ACS ($180$~K)} & $50$~K & No-Mark & $8$ & $< 1.0005$ & $29262.8$ & Yes \\
       & $50$~K & Magnitude & $10$ & $< 1.0005$ & $20823.6$ & Yes \\
       & $50$~K & Color & $9$ & $< 1.0005$ & $21653.5$ & Yes \\
       & $50$~K & Color+Error & $9$ & $< 1.0005$ & $20503.7$ & Yes \\
       \midrule
       \multirow{4}{*}{V11-WFC3 ($108$~K)} & $30$~K & No-Mark & $2$ & $< 1.0005$ & $34625.7$ & Yes \\
       & $30$~K & Magnitude & $4$ & $< 1.0005$ & $22859.9$ & Yes \\
       & $30$~K & Color & $3$ & $< 1.0005$ & $25438.7$ & Yes \\
       & $30$~K & Color+Error & $3$ & $< 1.0005$ & $21825.2$ & Yes \\
       \midrule
       \multirow{4}{*}{V12-ACS ($180$~K)} & $50$~K & No-Mark & $8$ & $< 1.0005$ & $34161.0$ & Yes \\
       & $50$~K & Magnitude & $10$ & $< 1.0005$ & $25374.8$ & Yes \\
       & $50$~K & Color & $9$ & $< 1.0005$ & $25660.5$ & Yes \\
       & $50$~K & Color+Error & $9$ & $< 1.0005$ & $24474.1$ & Yes \\
       \midrule
       \multirow{4}{*}{V13-ACS ($144$~K)} & $40$~K & No-Mark & $5$ & $< 1.0005$ & $42382.3$ & Yes \\
       & $40$~K & Magnitude & $7$ & $< 1.0005$ & $34907.8$ & Yes \\
       & $40$~K & Color & $6$ & $< 1.0005$ & $38790.3$ & Yes \\
       & $40$~K & Color+Error & $6$ & $< 1.0005$ & $35814.6$ & Yes \\
       \midrule
       \multirow{4}{*}{V13-WFC3 ($108$~K)} & $30$~K & No-Mark & $2$ & $< 1.0005$ & $71190.8$ & Yes \\
       & $30$~K & Magnitude & $4$ & $< 1.0005$ & $27684.1$ & Yes \\
       & $30$~K & Color & $3$ & $< 1.0005$ & $32568.6$ & Yes \\
       & $30$~K & Color+Error & $3$ & $< 1.0005$ & $33584.1$ & Yes \\
       \midrule
       \multirow{4}{*}{V14-ACS ($144$~K)} & $40$~K & No-Mark & $5$ & $< 1.0005$ & $39880.5$ & Yes \\
       & $40$~K & Magnitude & $7$ & $< 1.0005$ & $20764.2$ & Yes \\
       & $40$~K & Color & $6$ & $< 1.0005$ & $21759.3$ & Yes \\
       & $40$~K & Color+Error & $6$ & $< 1.0005$ & $28194.5$ & Yes \\
       \midrule
       \multirow{4}{*}{V14-WFC3 ($144$~K)} & $40$~K & No-Mark & $5$ & $< 1.0005$ & $63516.4$ & Yes \\
       & $40$~K & Magnitude & $7$ & $< 1.0005$ & $40729.3$ & Yes \\
       & $40$~K & Color & $6$ & $< 1.0005$ & $52589.9$ & Yes \\
       & $40$~K & Color+Error & $6$ & $< 1.0005$ & $53600.3$ & Yes \\
       \midrule
       \multirow{4}{*}{V15-ACS ($180$~K)} & $50$~K & No-Mark & $8$ & $< 1.0005$ & $23384.8$ & Yes \\
       & $50$~K & Magnitude & $10$ & $< 1.0005$ & $15575.0$ & Yes \\
       & $50$~K & Color & $9$ & $< 1.0005$ & $16069.5$ & Yes \\
       & $50$~K & Color+Error & $9$ & $< 1.0005$ & $7800.8$ & Yes \\
       \midrule
    \end{tabular}%
    }
    \caption{Multivariate PSRF  (mPSRF) and effective sample sizes for all our models fitted to the $12$ images. All models indicated convergence based on mPSRF. For the effective sample size, all models have reached the minimum number to obtain a $90\%$ confidence region with $0.1$ precision compared to the iid sample case.}
    \label{tab:fixed-dim diagnostic}
\end{table}

 %Although the effective sample sizes are on the low end for some of the GCs, especially the ones in images with known UDGs, it is not of concern since we are only using these quantities to check overall model convergence, rather than conducting accurate inference about, say, quantiles of these quantities.

%\begin{figure}[!ht]
%    \centering
%    \includegraphics[width = 0.7\columnwidth]{supp_figures/v7acs_traceplot.pdf}
%    \caption{\footnotesize{Traceplots, PSRF, and effective sample size of quantities shown in Eq.~\ref{eqn: model1_L} and \ref{eqn: model2_L} for selected GCs in V7-ACS.}}
%    \label{fig:v7acs_trace}
%\end{figure}

%\begin{figure}[!ht]
%    \centering
%    \includegraphics[width = 0.7\columnwidth]{supp_figures/v8wfc3_traceplot.pdf}
%    \caption{\footnotesize{Traceplots, PSRF, and effective sample size of quantities shown in Eq.~\ref{eqn: model1_L} and \ref{eqn: model2_L} for selected GCs in V8-WFC3.}}
%    \label{fig:v8wfc3_trace}
%\end{figure}

%\begin{figure}[!ht]
%    \centering
%    \includegraphics[width = 0.7\columnwidth]{supp_figures/v10acs_traceplot.pdf}
%    \caption{\footnotesize{Traceplots, PSRF, and effective sample size of quantities shown in Eq.~\ref{eqn: model1_L} and \ref{eqn: model2_L} for selected GCs in V10-ACS.}}
%    \label{fig:v10acs_trace}
%\end{figure}

Table \ref{tab:NU diagnostic} shows the homogeneity test results for $N_U$ based on the thinned MCMC chains from all models fitted to the $12$ images. The tests were conducted at $5\%$ significance level. All tests results have concluded that $N_U$ obtained from different chains are homogeneous, thus indicating convergence for $N_U$.

\begin{table}[!ht]
    \centering
    \resizebox{0.8\columnwidth}{!}{%
    \begin{tabular}{llcccc}
    \toprule
       Image ID &  & No-Mark & Magnitude & Color & Color+Error \\
       \midrule
       \multirow{3}{*}{V6-WFC3} & Thinning & 250 & 250 & 250 & 250 \\
       & $\chi^2$-test $p$-value & 0.533 & 0.506 & 0.167 & 0.878 \\
       & Test Conclusion & Homogeneous & Homogeneous & Homogeneous & Homogeneous \\
       \midrule
       \multirow{3}{*}{V7-ACS} & Thinning & 1000 & 1000 & 1000 & 1000 \\
       & $\chi^2$-test $p$-value & 0.409 & 0.361 & 0.443 & 0.564 \\
       & Test Conclusion & Homogeneous & Homogeneous & Homogeneous & Homogeneous \\
       \midrule
       \multirow{3}{*}{V8-WFC3} & Thinning & 1500 & 1500 & 1500 & 1500 \\
       & $\chi^2$-test $p$-value & 0.724 & 0.061 & 0.070 & 0.477 \\
       & Test Conclusion & Homogeneous & Homogeneous & Homogeneous & Homogeneous \\
       \midrule
       \multirow{3}{*}{V10-ACS} & Thinning & 1000 & 1000 & 1000 & 1000 \\
       & $\chi^2$-test $p$-value & 0.6205 & 0.333 & 0.423 & 0.814 \\
       & Test Conclusion & Homogeneous & Homogeneous & Homogeneous & Homogeneous \\
       \midrule
       \multirow{3}{*}{V11-ACS} & Thinning & 500 & 500 & 500 & 500 \\
       & $\chi^2$-test $p$-value & 0.865 & 0.125 & 0.651 & 0.253 \\
       & Test Conclusion & Homogeneous & Homogeneous & Homogeneous & Homogeneous \\
       \midrule
       \multirow{3}{*}{V11-WFC3} & Thinning & 500 & 500 & 500 & 500 \\
       & $\chi^2$-test $p$-value & 0.237 & 0.162 & 0.628 & 0.616 \\
       & Test Conclusion & Homogeneous & Homogeneous & Homogeneous & Homogeneous \\
       \midrule
       \multirow{3}{*}{V12-ACS} & Thinning & 1000 & 1000 & 1000 & 1000 \\
       & $\chi^2$-test $p$-value & 0.220 & 0.745 & 0.598 & 0.132 \\
       & Test Conclusion & Homogeneous & Homogeneous & Homogeneous & Homogeneous \\
       \midrule
       \multirow{3}{*}{V13-ACS} & Thinning & 500 & 500 & 500 & 500 \\
       & $\chi^2$-test $p$-value & 0.303 & 0.561 & 0.616 & 0.691 \\
       & Test Conclusion & Homogeneous & Homogeneous & Homogeneous & Homogeneous \\
       \midrule
       \multirow{3}{*}{V13-WFC3} & Thinning & 500 & 500 & 500 & 500 \\
       & $\chi^2$-test $p$-value & 0.999 & 0.647 & 0.737 & 0.517 \\
       & Test Conclusion & Homogeneous & Homogeneous & Homogeneous & Homogeneous \\
       \midrule
       \multirow{3}{*}{V14-ACS} & Thinning & 1000 & 1000 & 1000 & 1000 \\
       & $\chi^2$-test $p$-value & 0.222 & 0.576 & 0.478 & 0.918 \\
       & Test Conclusion & Homogeneous & Homogeneous & Homogeneous & Homogeneous \\
       \midrule
      \multirow{3}{*}{V14-WFC3} & Thinning & 1000 & 1000 & 1000 & 1000 \\
       & $\chi^2$-test $p$-value & 0.865 & 0.300 & 0.612 & 0.556 \\
       & Test Conclusion & Homogeneous & Homogeneous & Homogeneous & Homogeneous \\
       \midrule
       \multirow{3}{*}{V15-ACS} & Thinning & 1500 & 1500 & 1500 & 1500 \\
       & $\chi^2$-test $p$-value & 0.874 & 0.150 & 0.533 & 0.604 \\
       & Test Conclusion & Homogeneous & Homogeneous & Homogeneous & Homogeneous \\
       \midrule
    \end{tabular}%
    }
    \caption{Pearson's Chi-Squared test for homogeneity for thinned sample of $N_U$. Thinning is the interval we have thinned. The test is conducted at $5\%$ significance level. All models have indicated convergence based on the test results.}
    \label{tab:NU diagnostic}
\end{table}

\subsection{Prior Sensitivity Analysis}

Since our model is complex and computationally expensive, we opt for a simple method by \cite{kallioinen2023detecting} to conduct prior sensitivity analysis. \citeauthor{kallioinen2023detecting} proposed to conduct prior sensitivity analysis by first power-scaling the prior distribution, the power-scaling induces an importance weight that is then applied to the posterior samples to obtain a weighted posterior samples. The sensitivity analysis is then conducted by comparing the discrepancy between the original and the weighted posterior samples. If a parameter is insensitive to the prior, then there is little discrepancy between the original and weighted posterior samples.

Following the procedure in \citeauthor{kallioinen2023detecting}, we compute the power-scaled importance weight for each MCMC sample $\{(\boldsymbol{\Theta}^{(t)}, \boldsymbol{\Phi}^{(t)})\}_{t=1}^N$. Denote $\alpha$ the strength of the power-scaling applied to the prior distribution as in \citeauthor{kallioinen2023detecting}, the importance weight for each posterior sample is
\[
w_\alpha^{(t)}
= \left[\prod_{m = 1}^{N_U^{(t)}}\pi\left(\boldsymbol{U}_j^{(t)}\right)^{\alpha-1}\right]\left[\prod_{k=1}^{N_G}\pi\left(\boldsymbol{G}_k^{(t)}\right)^{\alpha-1}\right]\pi\left(\beta_0^{(t)}\right)^{\alpha-1}\pi\left(\Sigma_0^{(t)}\right)^{\alpha-1}.
\]
For no-mark model, the above becomes
\[
w_\alpha^{(t)}
= \left[\prod_{m = 1}^{N_U^{(t)}}\pi\left(\boldsymbol{U}_j^{(t)}\right)^{\alpha-1}\right]\left[\prod_{k=1}^{N_G}\pi\left(\boldsymbol{G}_k^{(t)}\right)^{\alpha-1}\right]\pi\left(\beta_0^{(t)}\right)^{\alpha-1}.
\]
The power-scaling for the prior of $\lambda_c$ is not required since there is no power-scaling for a uniform distribution. Moreover, for the magnitude-marked model, the power-scaling for the prior of $\Sigma_0$ does not exist due to the uniform prior assignment. The sensitivity diagnostic is based on the discrepancy between the empirical CDF of the quantity of interest and the corresponding $w_\alpha^{(t)}$-weighted empirical CDF measured by the Jensen-Shannon metric, relative to the strength of power-scaling. A diagnostic value greater than $0.05$ is served as a general rule for indicating sensitivity towards the prior \citep{kallioinen2023detecting}.

Since our primary goal is to infer the locations $\mathbf{X}_c$ of UDGs, and our secondary goal is to infer the associated physical parameters of detected UDGs, we focus on conducting prior sensitivity analysis for quantities associated with UDGs, i.e., $\lambda_c$ and $\boldsymbol{\Phi}$, where $\mathbf{X}_c$ being the most important. Other parameters such as $\beta_0$ and $\mathbf{G}_k$'s are not of concern.

Because $\boldsymbol{\Phi}$ is a marked point process ($\mathbf{X}_c$ is the point process while the physical parameters of the UDGs are the associated marks), directly applying the method from \citeauthor{kallioinen2023detecting} using the accompanying R package \texttt{priorsense} is infeasible. This is because $\boldsymbol{\Phi}$ does not have an associated CDF that is required for discrepancy analysis. We thus collapse all samples of $\{\boldsymbol{\Phi}^{(t)}\}_{t=1}^N$ into one sample of UDG locations and their associated physical parameters. We denote this collapsed sample by $\boldsymbol{\mathcal{X}} = \{(c_w^u, \boldsymbol{U}_w, \Sigma_w)\}_{w = 1}^{\mathcal{W}}$ with $\mathcal{W} = \sum_{t=1}^NN_U^{(t)}$, and we keep the information about the importance weight $w_\alpha^{(t)}$. Essentially, for any $\mathcal{X}_w \in \boldsymbol{\mathcal{X}}, \ w \in [\mathcal{W}]$, its corresponding importance weight is $w_\alpha^{(t)}$ if $\mathcal{X}_w$ comes from the $t$-th iteration of the chain. If $\boldsymbol{\Phi}^{(t)} = \emptyset$, then we do not need to consider the weight as it cannot affect an empty set.

To assess the prior effect on $\mathbf{X}_c$, we consider the prior effect on the marginal distributions of the $x$ and $y$ coordinates of $c_w^u, \ w \in \mathcal{W}$. We also consider the distribution of the distance $d(c_w^u, s)$ from $c_w^u$ to any fixed $s \in S$. For the physical parameters, we consider parameter values from $\mathcal{X}_w$'s that are located within a small radius of a detected UDG. This ensures that the considered $\mathcal{X}_w$'s all represent the same detected UDG.

Table \ref{tab:ps_no_UDG} shows the sensitivity diagnostic for $\lambda_c$ for images without detected UDGs, including V6-WFC3, V11-WFC3, V13-ACS, V13-WFC3, and V14-WFC3. Since these images do not contain detected UDGs, only $\lambda_c$ is required for analysis as it determines the existence of UDGs.

Table \ref{tab:ps_v11} and \ref{tab:ps_v12} show the prior sensitivity diagnostics for $\lambda_c$, the marginal $x$, $y$ coordinates of the UDG locations, $d(c_w^u, s)$ where $s=(0.1, 0.1), (0.3, 0.3), (0.5, 0.5),$ and $(0.8, 0.8)$, and the physical parameters of UDGs in the image V11-ACS and V12-ACS respectively. The results for other images with detected UDGs are similar to that in Table \ref{tab:ps_v11}, and are available in the provided project repository.

We only present the results of the mean number of GCs $\varlambda^u$, the GC system radius $R^u$, and the S\'{e}rsic index $n^u$ for an UDG since these are the most important parameters. The aspect ratio $\rho^u$ and orientation angle $\varphi^u$ are not inherent physical properties of a galaxy, but merely a result of our observational perspective from our location in the Universe.

\begin{table}[!ht]
    \centering
    \resizebox{0.65\columnwidth}{!}{%
    \begin{tabular}{lcccc}
    \toprule
       Image ID & No-Mark & Magnitude & Color & Color+Error \\
       \midrule
       V6-WFC3 & $0.013$ & $0.020$ & $0.033$ & $0.026$  \\
       V11-WFC3 & $0.014$ & $0.018$ & $0.035$ & $0.031$  \\
       V13-ACS & $0.012$ & $0.019$ & $0.023$ & $0.024$  \\
       V13-WFC3 & $0.016$ & $0.025$ & $0.037$ & $0.035$  \\
       V14-WFC3 & $0.019$ & $0.025$ & $0.024$ & $0.024$  \\
       \bottomrule
    \end{tabular}%
    }
    \caption{Prior sensitivity analysis of $\lambda_c$ for all four models fitted to the images that do not contain detected UDGs (V6-WFC3, V11-WFC3, V13-ACS, V13-WFC3, and V14-WFC3). Values greater than $0.05$ serves as a general rule for indicating the parameters/quantities are sensitive to the prior.}
    \label{tab:ps_no_UDG}
\end{table}

\begin{table}[!ht]
    \centering
    \resizebox{0.7\columnwidth}{!}{%
    \begin{tabular}{ccccc}
    \toprule
       Parameter/Quantities & No-Mark & Magnitude & Color & Color+Error \\
       \midrule
       $\lambda_c$ & $0.028$ & $0.008$ & $0.012$ & $0.016$  \\
       $x$-coordinate of $c_w^u$ & $0.037$ & $0.026$ & $0.025$ & $0.023$\\
       $y$-coordinate of $c_w^u$  & $0.033$ & $0.017$ & $0.018$ & $0.013$\\
       $d(c_w^u, s), s = (0.1, 0.1)$ & $0.040$ & $0.024$ & $0.028$ & $0.025$\\
       $d(c_w^u, s), s = (0.3, 0.3)$ & $0.021$ & $0.015$ & $0.018$ & $0.018$\\
       $d(c_w^u, s), s = (0.5, 0.5)$ & $0.030$ & $0.018$ & $0.011$ & $0.008$\\
       $d(c_w^u, s), s = (0.8, 0.8)$ & $0.046$ & $0.028$ & $0.029$ & $0.023$\\
       \midrule
       $\varlambda_1^u$ & $0.050$ & $\textbf{0.081}$ & $0.037$ & $\textbf{0.061}$  \\
       $R_1^u$ & $\textbf{0.124}$ & $\textbf{0.113}$ & $\textbf{0.137}$ & $\textbf{0.137}$  \\
       $n_1^u$ & $0.047$ & $\textbf{0.097}$ & $\textbf{0.079}$ & $\textbf{0.079}$  \\
       \midrule
       $\varlambda_2^u$ & $\textbf{0.092}$ & $\textbf{0.101}$ & $\textbf{0.155}$ & $\textbf{0.112}$  \\
       $R_2^u$ & $\textbf{0.122}$ & $\textbf{0.176}$ & $\textbf{0.148}$ & $\textbf{0.142}$  \\
       $n_2^u$ & $\textbf{0.073}$ & $\textbf{0.096}$ & $\textbf{0.129}$ & $\textbf{0.111}$  \\
       \bottomrule
    \end{tabular}%
    }
    \caption{Prior sensitivity analysis of parameters/quantities associated with UDGs for all four models fitted to the image V11-ACS. Values greater than $0.05$ serves as a general rule for indicating the parameters/quantities are sensitive to the prior. Bold numbers indicate diagnostic statistics being greater than $0.05$.}
    \label{tab:ps_v11}
\end{table}

\begin{table}[!ht]
    \centering
    \resizebox{0.7\columnwidth}{!}{%
    \begin{tabular}{ccccc}
    \toprule
       Parameter/Quantities & No-Mark & Magnitude & Color & Color+Error \\
       \midrule
       $\lambda_c$ & $0.013$ & $0.021$ & $0.041$ & $0.050$  \\
       $x$-coordinate of $c_w^u$ & $\textbf{0.085}$ & $\textbf{0.058}$ & $\textbf{0.056}$ & $\textbf{0.058}$\\
       $y$-coordinate of $c_w^u$  & $\textbf{0.082}$ & $\textbf{0.054}$ & $\textbf{0.057}$ & $\textbf{0.060}$\\
       $d(c_w^u, s), s = (0.1, 0.1)$ & $\textbf{0.098}$ & $\textbf{0.065}$ & $\textbf{0.067}$ & $\textbf{0.072}$\\
       $d(c_w^u, s), s = (0.3, 0.3)$ & $\textbf{0.094}$ & $\textbf{0.063}$ & $\textbf{0.055}$ & $\textbf{0.061}$\\
       $d(c_w^u, s), s = (0.5, 0.5)$ & $0.021$ & $0.021$ & $0.023$ & $0.023$\\
       $d(c_w^u, s), s = (0.8, 0.8)$ & $\textbf{0.106}$ & $\textbf{0.072}$ & $\textbf{0.066}$ & $\textbf{0.075}$\\
       \midrule
       $\varlambda_1^u$ & $\textbf{0.057}$ & $\textbf{0.124}$ & $\textbf{0.127}$ & $\textbf{0.092}$  \\
       $R_1^u$ & $\textbf{0.156}$ & $\textbf{0.103}$ & $\textbf{0.164}$ & $\textbf{0.174}$  \\
       $n_1^u$ & $\textbf{0.106}$ & $\textbf{0.148}$ & $\textbf{0.118}$ & $\textbf{0.176}$  \\
       \bottomrule
    \end{tabular}%
    }
    \caption{Prior sensitivity analysis of parameters/quantities associated with UDGs for all four models fitted to the image V12-ACS. Values greater than $0.05$ serves as a general rule for indicating the parameters/quantities are sensitive to the prior. Bold numbers indicate diagnostic statistics being greater than $0.05$.}
    \label{tab:ps_v12}
\end{table}

For all $\lambda_c$ and the majority of quantities associated with the UDG locations, the results are not sensitive to the prior choice. V12-ACS, as shown in Table \ref{tab:ps_v12}, is the only image where the posterior of UDG location related quantities seem to be sensitive to the prior. It should be noted that in \citeauthor{kallioinen2023detecting}, the diagnostic statistic $0.05$ is chosen assuming the target distribution resembles a Gaussian distribution. Under this assumption, the value of $0.05$ represents a change in mean by $0.3$ standard deviations. Therefore, a diagnostic of $< 0.05$ for a more complex posterior distribution than a Gaussian should still suggest prior insensitivity. However, a diagnostic of $> 0.05$, does not necessarily imply prior \textit{sensitivity}. A more complex distribution (e.g., multi-modal) is more likely to accumulate discrepancy compared to a Gaussian-shaped distribution. However, if the discrepancy for a multi-modal posterior distribution is smaller than what one would expect as insensitive for Gaussian, then insensitivity is suggested.

In Table \ref{tab:ps_v12}, the only quantity that has a Gaussian-like posterior is the distance of UDG locations to the point $(0.5, 0.5)$. It turns out this distance is also the only quantity that is insensitive towards the prior based on the $0.05$ diagnostic threshold. Thus, we can conclude that there is sufficient information in the data to provide us with reliable estimates for the existence and locations of UDGs.

For the physical parameters of UDGs, the majority of them are quite sensitive to the prior. We can safely conclude that these parameters are sensitive to the priors since their posterior distributions resemble a Gaussian (or log-normal). The observed sensitivity of these parameters is expected since the majority of the UDGs detected here have fewer than five GCs. Such a low GC count does not provide sufficient information to the likelihood. The physical parameter that is the least sensitive to prior is $\varlambda^u$. This is expected as $\varlambda^u$ is directly related to the point pattern count data. However, $R^u$ and $n^u$ are sensitive to the prior because they require more observed GCs to inform them.

Even though the physical parameters of UDGs are sensitive to priors, it does not mean the priors we chose are inappropriate --- our priors are purposely informative and based on previous studies of GC systems around UDGs. 

\subsection{Computation of the Detection Region}

We provide the details for computing the $\vartheta$-restricted detection probability mentioned in Eq. 20 of the main paper. Recall that the $\vartheta$-restricted detection probability is defined as 
\[
p_{\mathbf{D}}(\vartheta) = \sup\{\mathbb{P}(\mathbf{X}_c \subset B \mid \mathbf{D}): |B| \leq \vartheta|S|, \ B \subset S\}.
\]
To compute the above probability, we first descritise the study region $S$ into equidistance cells:
\[
S = \bigcup_{b=1}^{\mathcal{B}}C_b.
\]
Denote the MCMC posterior sample of UDG locations as $\{\mathbf{X}_c^{(t)}\}_{t=1}^N$. We set the size of cell $C_b$ to be small enough so that there is at most one point from $\mathbf{X}_c^{(t)}, \ t \in [N]$ in any one of $C_b, \ b \in [\mathcal{B}]$. Write
\[
p_b = \frac{1}{N}\sum_{t=1}^N\boldsymbol{1}(C_b \cap \mathbf{X}_c^{(t)} \neq \emptyset), \ b \in [\mathcal{B}].
\]
Denote $\mathcal{P} = \{p_1, \dots, p_{\mathcal{B}}\}$. For any $\vartheta \in [0,1]$, the $p_{\mathbf{D}}(\vartheta)$-level detection region defined by Eq.~19 in the main paper is then
\[
B_{\mathbf{D}}(p_{\mathbf{D}}(\vartheta)) = \bigcup_{b\in \mathcal{Q}}C_{b}, \ \mathcal{Q} = \{b \in [\mathcal{B}]:p_b \geq \text{quantile}(\mathcal{P}, 1 - \vartheta)\}.
\]
$\text{quantile}(\mathcal{P}, 1 - \vartheta)$ is the $(1-\vartheta)\times100$-th quantile of $\mathcal{P}$. Essentially, the $p_{\mathbf{D}}(\vartheta)$-level detection region can be found by ranking the elements in $\mathcal{P}$ and merge the cells with $p_b$ value being the top $(1-\vartheta)\times100\%$ in $\mathcal{P}$.

After $B_{\mathbf{D}}(p_{\mathbf{D}}(\vartheta))$ is determined, the corresponding detection probability $p_{\mathbf{D}}(\vartheta)$ is
\[
p_{\mathbf{D}}(\vartheta) = \frac{1}{N}\sum_{t=1}^N\boldsymbol{1}(B_{\mathbf{D}}(p_{\mathbf{D}}(\vartheta)) \cap \mathbf{X}_c^{(t)} \neq \emptyset).
\]

\subsection{Posterior Detection Results for Known UDGs}

\begin{figure}[!ht]
    \centering
    \includegraphics[width = 0.6\columnwidth]{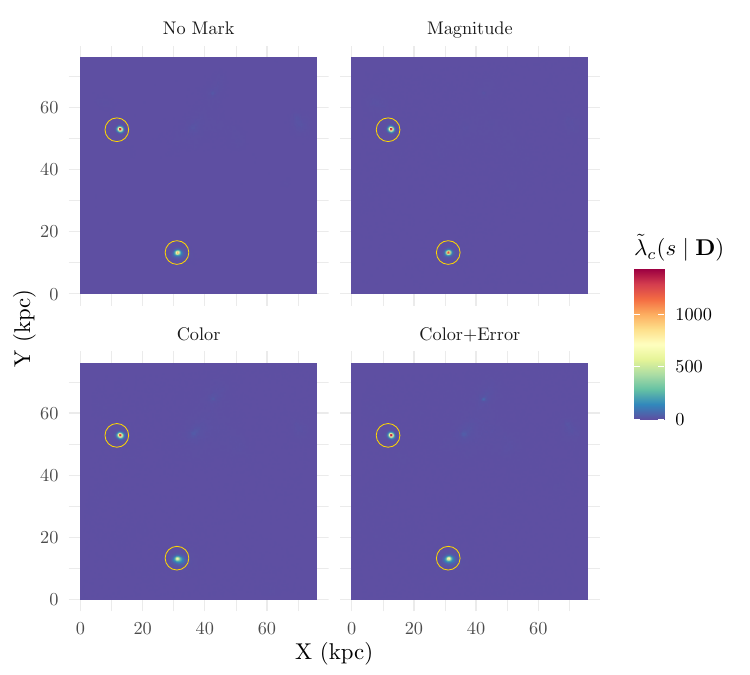}
    \caption{\footnotesize{The scaled posterior intensity estimates $\Tilde{\lambda}_c(s \mid \mathbf{D})$ of $\mathbf{X}_c$ for the field V7-ACS under Model 1 and Model 2 with magnitude, color variation, and color variation with error considered. Golden circles in the figures indicate the locations of two confirmed UDGs in the image.}}
    \label{fig:v7acs_result}
\end{figure}

\begin{figure}[!ht]
    \centering
    \includegraphics[width = 0.6\columnwidth]{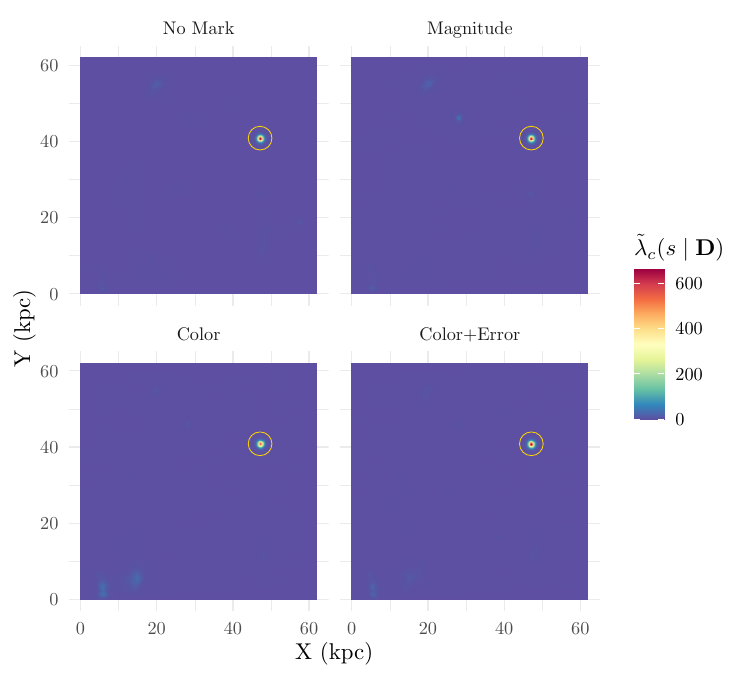}
    \caption{\footnotesize{The scaled posterior intensity estimates $\Tilde{\lambda}_c(s \mid \mathbf{D})$ of $\mathbf{X}_c$ for the field V8-WFC3 under Model 1 and Model 2 with magnitude, color variation, and color variation with error considered. Golden circle in the figures indicate the locations of the confirmed UDG in the image.}}
    \label{fig:v8wfc3_result}
\end{figure}

\begin{figure}[!ht]
    \centering
    \includegraphics[width = 0.6\columnwidth]{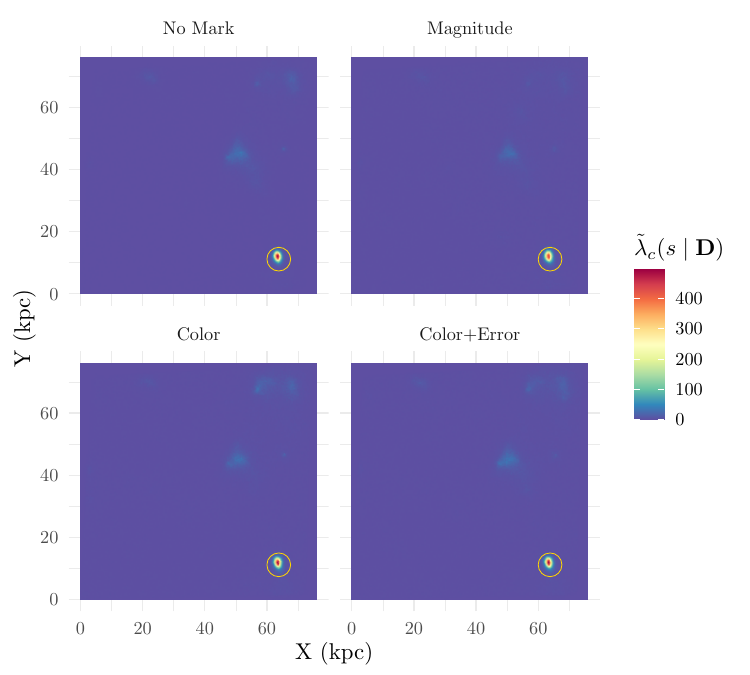}
    \caption{\footnotesize{The scaled posterior intensity estimates $\Tilde{\lambda}_c(s \mid \mathbf{D})$ of $\mathbf{X}_c$ for the field V10-ACS under Model 1 and Model 2 with magnitude, color variation, and color variation with error considered. Golden circle in the figures indicate the locations of the confirmed UDG in the image.}}
    \label{fig:v10acs_result}
\end{figure}

\begin{figure}[!ht]
    \centering
    \includegraphics[width = 0.6\columnwidth]{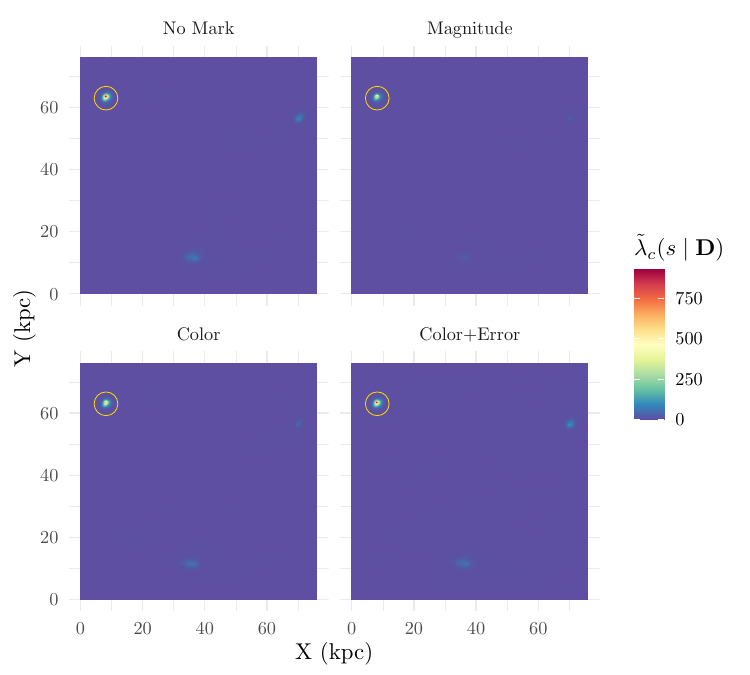}
    \caption{\footnotesize{The scaled posterior intensity estimates $\Tilde{\lambda}_c(s \mid \mathbf{D})$ of $\mathbf{X}_c$ for the field V15-ACS under Model 1 and Model 2 with magnitude, color variation, and color variation with error considered. Golden circle in the figures indicate the locations of the confirmed UDG in the image.}}
    \label{fig:v15acs_result}
\end{figure}

This section contains the posterior intensity of $\mathbf{X}_c$ for the images V7-ACS, V8-WFC3, V10-ACS, and V15-ACS. These images all have previously known GC-rich UDGs. Figure~\ref{fig:v7acs_result} to Figure~\ref{fig:v15acs_result} show the scaled posterior intensity of $\mathbf{X}_c$ for the four images. Golden circles in the figures are the previously confirmed UDGs.

\subsection{Posterior Distributions and Model Checking}

This section contains the posterior distributions for the UDG physical parameters of seven previously confirmed UDGs that we have detected from all four models, namely the mean number of GCs $\varlambda^u$, the GC system radius $R^u$, and the S\'{e}rsic index $n^u$. We conducted a simple posterior predictive check by comparing our results to results obtained using traditional astronomical techniques through maximum likelihood. We also perform a more traditional posterior predictive check by simulating GC point pattern using our full posterior distribution. We then compare the simulated data and the real data through simple summary statistics.

\begin{figure}[!ht]
    \centering
    \subfigure[No-mark model]{\includegraphics[width=0.45\columnwidth]{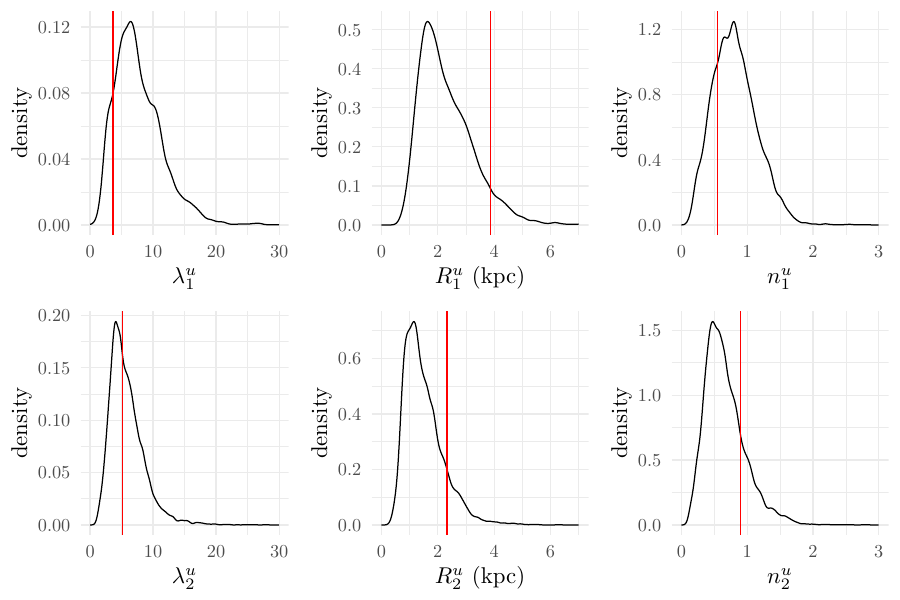}}
     \subfigure[Magnitude-marked model]{\includegraphics[width=0.45\columnwidth]{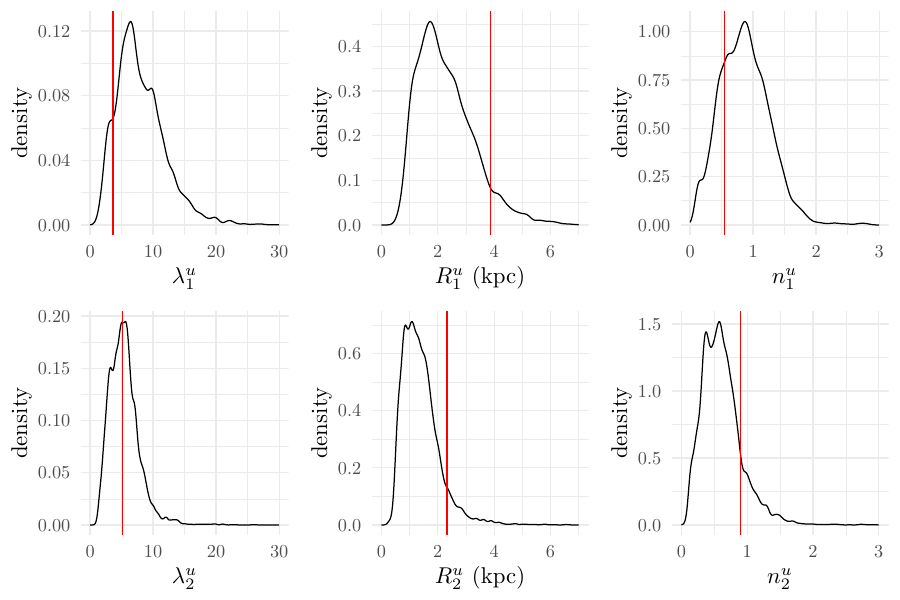}}
     \subfigure[Color-marked model]{\includegraphics[width=0.45\columnwidth]{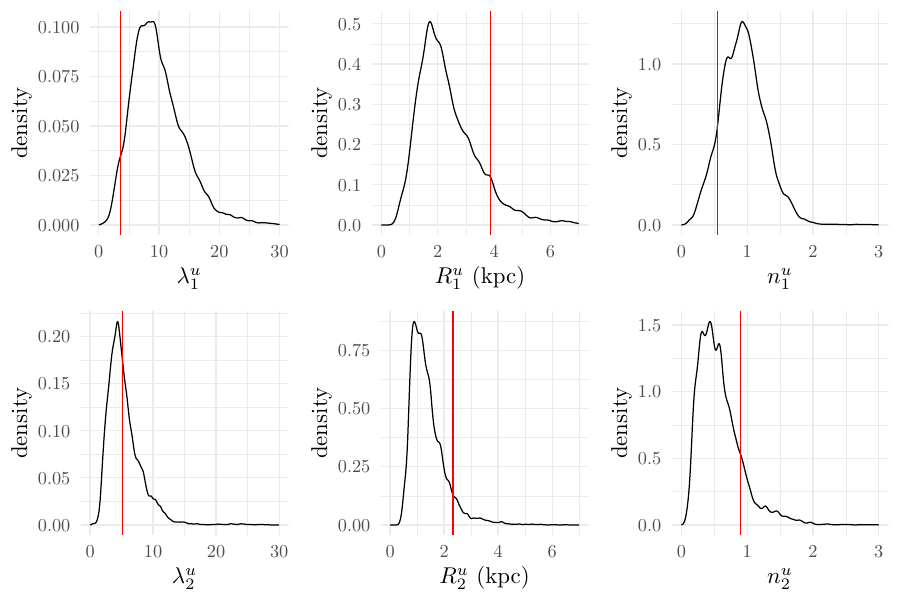}}
     \subfigure[Color-marked model with error]{\includegraphics[width=0.45\columnwidth]{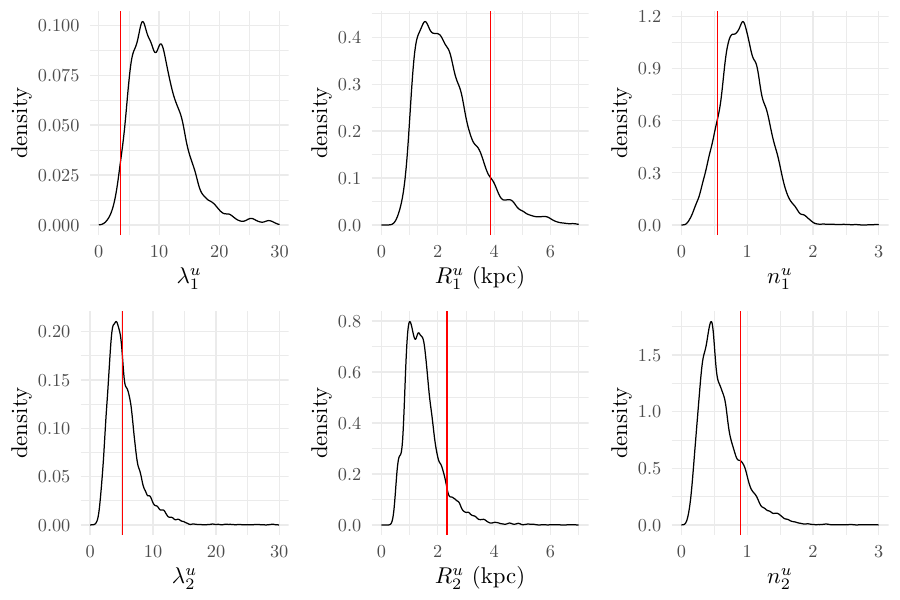}}
     \caption{\footnotesize{Posterior of physical parameters of previously confirmed UDGs that we have detected from all four models in the image V11-ACS, including the mean number of GCs $\varlambda^u$, the GC system radius $R^u$, and the S\'{e}rsic index $n^u$. The red lines are maximum likelihood estimates using traditional astronomical model fitting.}}
     \label{fig:v11acs_UDG_dens}
\end{figure}

Figure \ref{fig:v11acs_UDG_dens} shows an example of one of our simple posterior predictive checks for image V11-ACS. The black curves repsent the marginal posterior distributions for each parameter, and the red vertical lines show the MLE using traditional astronomical model fitting. The MLE results were obtained by Steven Janssens in preparation for the \textit{HST} proposal for the PIPER survey \citep{Harris2020}. We can see from Figure \ref{fig:v11acs_UDG_dens} that our results in general align with the ones from MLE.  Since the physical parameters are ultimately unknown quantities, we cannot compare our results with their true values (or true distributions). Thus, we can only compare our results with those of other methods as a sanity check. Results for other images can be accessed through the project repository and they are similar to that in Figure \ref{fig:v11acs_UDG_dens}. 

\begin{figure}[!ht]
    \centering
    \subfigure[Total GC counts]{\includegraphics[width=0.47\columnwidth]{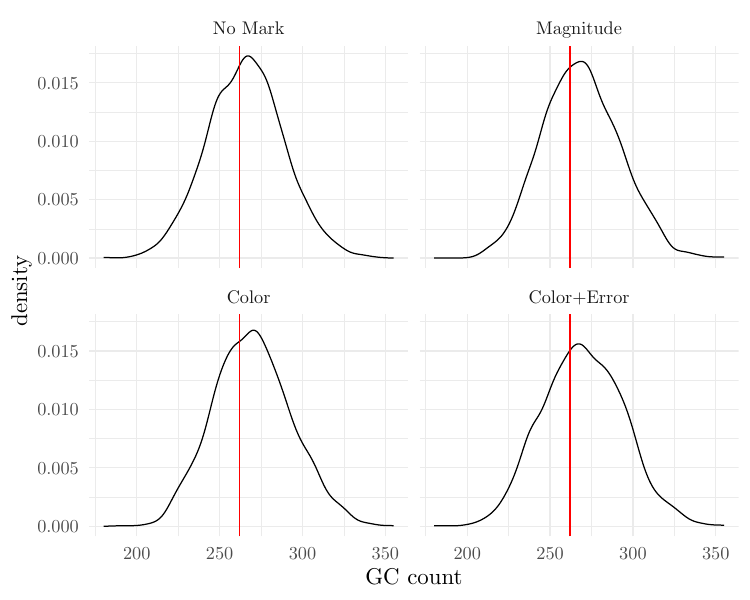}}
     \subfigure[Differences in nearest-neighbor distance distribution]{\includegraphics[width=0.47\columnwidth]{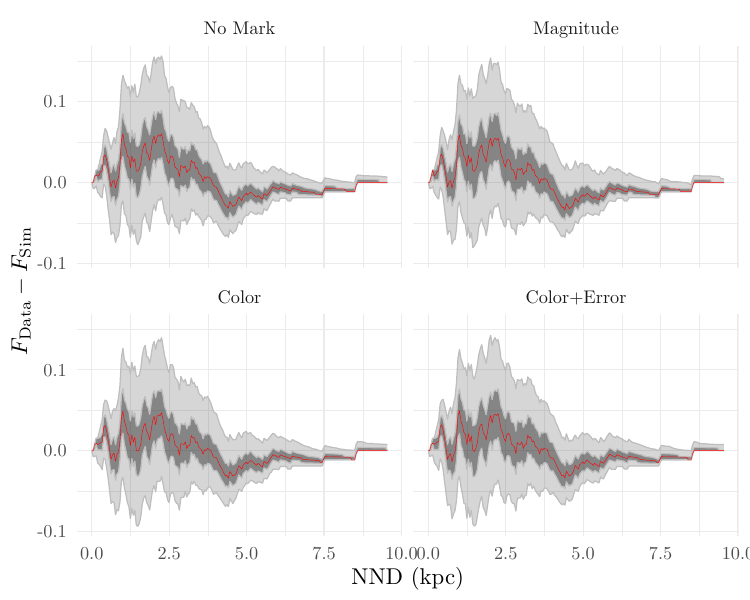}}
     \caption{\footnotesize{(a) Posterior predictive checks of total GC counts in the image from all four models fitted to V11-ACS (the red line is the true GC count from data); (b) Differences between the nearest-neighbor distance (NND) distribution of GCs obtained from GC data in V11-ACS and the posterior predictive distribution (the red line is the difference between the data and the posterior median NND distribution, the dark grey band is the $50\%$ credible interval, and the light grey band is the $95\%$ credible interval).}}
     \label{fig:v11acs_posterior_checks}
\end{figure}

We show in Figure \ref{fig:v11acs_posterior_checks} the posterior predictive check for image V11-ACS. Again, results for other images can be accessed in the provided project repository and they are similar to ones in Figure \ref{fig:v11acs_posterior_checks}. We compare the posterior predictive distribution of the total GC counts in each image to that from the real data. We also compare the posterior predictive distribution of the nearest-neighbor distance (NND) distribution of the simulated GC point patterns to that from the real data. All of our models produce posterior predictive distributions of total GC counts that agree well with the actual GC counts from data. For the NND distribution, our posterior predictive distributions also align very well with the NND distribution from the data.

\subsection{Simulation}\label{supp_sec: simulation}

We conduct a simple simulation to confirm the validity of our inference algorithm. We simulate the following data in a $[0,76]^2$~kpc square: (1) a giant elliptical galaxy centered at $(60.8, 38)$~kpc with $150$ GCs, $R=8.36$~kpc, S\'{e}rsic index $n = 1$ , 
rotation angle $\varphi=\pi/6$, and $\rho = 1.3$, (2) two UDGs with $10$ and $4$ GCs centered at $(15.2, 15.2)$~kpc and $(30.4, 53.2)$~kpc respectively, and (3) $100$ GCs in the IGM. For each of the UDGs, the corresponding quantities are used: GC system radii of $2.28$ and $0.38$~kpc, S\'{e}rsic indices of $0.5$ and $0.9$, rotation angles of $\pi/4$ and $-\pi/4$, semi-axis ratios of $1.2$ and $0.8$. 

We simulate the GC magnitude based on the canonical GCLF after correcting for the distance to the Perseus cluster. Under the F814W filter, we assume the GCLF has $\mu_{\mathrm{F814W}} = 26.2$~mag, and $\sigma = 1$~mag for GCs in the IGM and the giant elliptical galaxy respectively. The first UDG (with $10$ GCs) is given a brighter GCLF with $\mu_{\mathrm{F814W}} = 25.2$~mag, and $\sigma = 1$~mag. The second UDG (with $4$ GCs) is given a lower color variation by setting $\sigma = 0.6$~mag with $\mu_{\mathrm{F814W}} = 26.2$~mag. Under the F475W filter, we set $\mu_{\mathrm{F475W}} = \mu_{\mathrm{F814W}}+1.5$~mag, whilst maintaining the same $\sigma$ for all GC sub-populations. This choice is to ensure that the GC color mean is centered at the typical value of $1.5$~mag. The measurement error of the GC magnitude is then simulated as an exponential function of the GC magnitude \citep[i.e., following typical observations, see for example][Equation 1]{Harris_2023}. 

\begin{figure}[t]
    \centering
    \includegraphics[width = 0.5\textwidth]{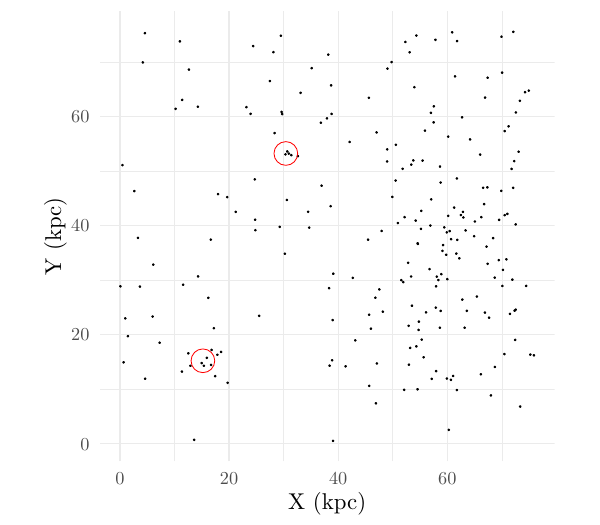}
    \caption{\footnotesize{Simulated GC point pattern with one giant elliptical galaxy (in the right part of the observation window) and two UDGs (indicated by red circles).}}
    \label{fig:simulated GC}
\end{figure}

Figure \ref{fig:simulated GC} shows the point pattern of the simulated GCs. The simulated giant elliptical galaxy is on the right-hand side of the field. The red circles in the figure show the locations of the two simulated UDGs.

\begin{figure}[t]
    \centering
    \includegraphics[width = 0.7\textwidth]{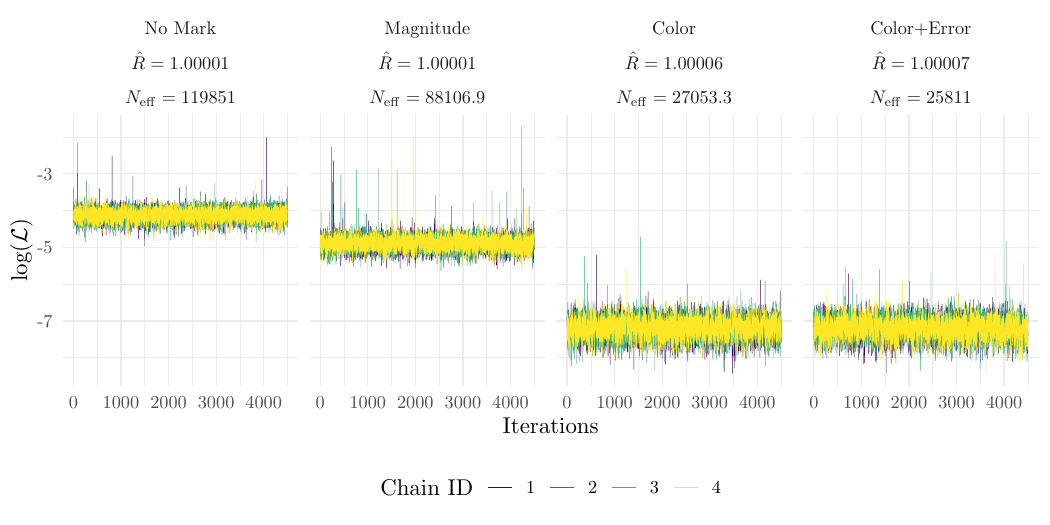}
    \caption{\footnotesize{Traceplots, PSRF, and effective sample size of quantities shown in Eq.~\ref{eqn: model1_L} and \ref{eqn: model2_L} for selected GCs in the simulated data.}}
    \label{fig:simulate traceplot}
\end{figure}

We fit all four of our models to the simulated GC data. For the prior distributions, we have set $\beta_0 \sim \text{LN}(\log(80), 0.5^2)$, $\varlambda^g \sim \text{LN}(\log(200), 0.25^2)$, and $R^g (\text{kpc}) \sim \text{LN}(\log(11.4), 0.25^2)$. These priors are intentionally set with significant bias from the true values. The (hyper-) priors for all other (hyper-) parameters are the same as ones used in our analysis of real data. 

We conduct our diagnostic procedures as before. Figure \ref{fig:simulate traceplot} shows the traceplots from the MCMC algorithms under the four models. There is sufficient evidence of convergence based on the results in the figure.

We then conducted the same estimation procedure to determine the existence of UDGs based on our posterior samples. We find $p_0^{pos} \gtrsim 0.9 > p_0^{pri}$ for all models, indicating that there are UDGs in the data.

\begin{figure}[t]
    \centering
    \includegraphics[width = 0.6\columnwidth]{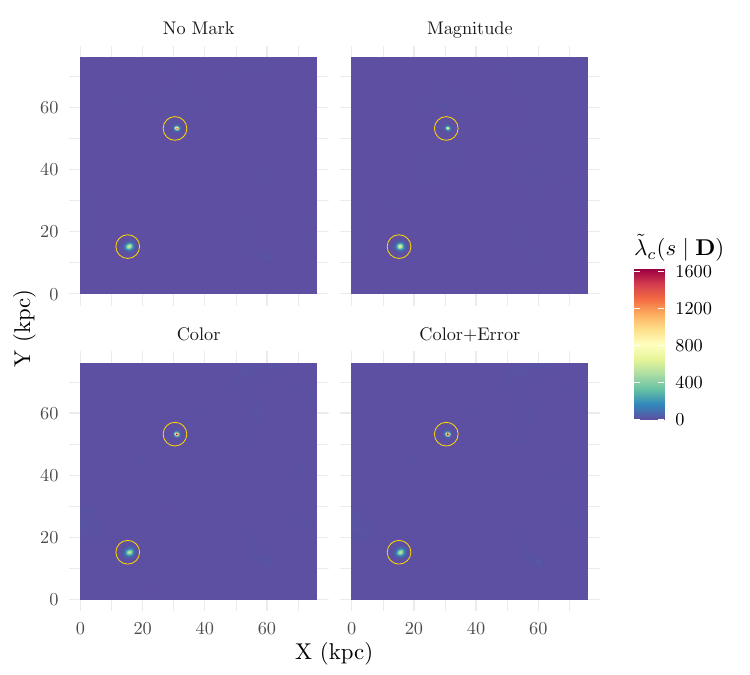}
    \caption{\footnotesize{The scaled posterior intensity estimates $\Tilde{\lambda}_c(s \mid \mathbf{D})$ of $\mathbf{X}_c$ for the simulated GC data under Model 1 and Model 2 with magnitude, color variation, and color variation with error considered. Golden circles in the figures indicate the locations of two simulated UDGs in the data.}}
    \label{fig:simulate_result}
\end{figure}

Figure \ref{fig:simulate_result} shows the posterior intensity $\Tilde{\lambda}_c(s \mid \mathbf{D})$ of $\mathbf{X}_c$ for the simulated GC data for all our models. All models successfully detected both simulated UDGs in the data. Furthermore, the change in the strength of the detection signals under Model 2 coincides with the intended simulated marks: under the magnitude-marked model, the UDG at $(15.2, 15.2)$~kpc (bottom left) has a much stronger detection signal because of its brighter GCLF; under the color-marked models, the UDG at $(30.4, 53.2)$~kpc (upper mid) has a stronger detection signal because of its smaller color variation.

\begin{figure}
    \centering
    \subfigure[No-mark model]{\includegraphics[width=0.7\columnwidth]{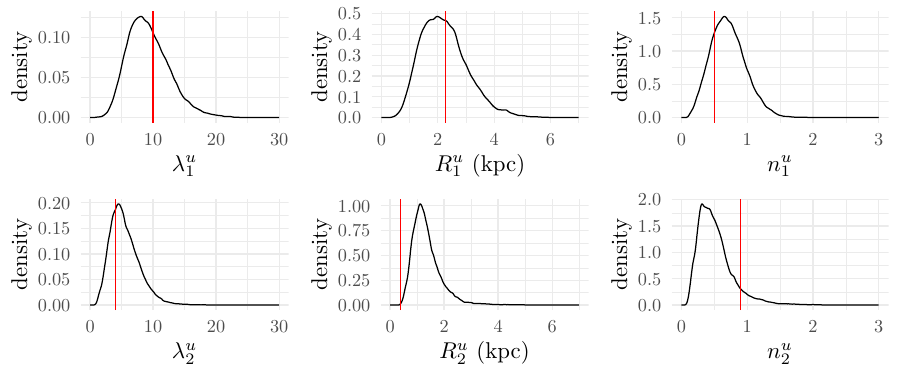}}
     \subfigure[Magnitude-marked model]{\includegraphics[width=0.85\columnwidth]{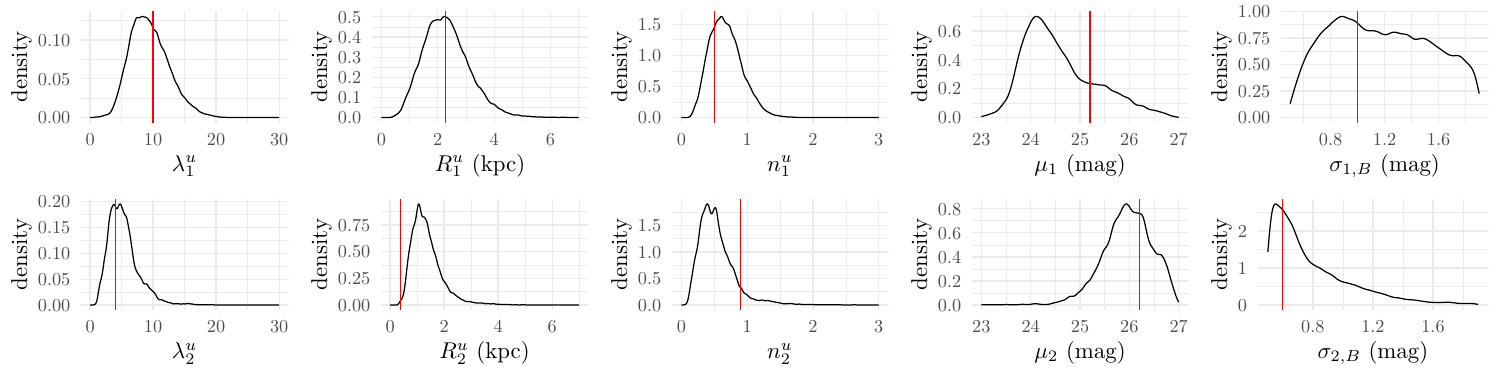}}
     \subfigure[Color-marked model]{\includegraphics[width=0.85\columnwidth]{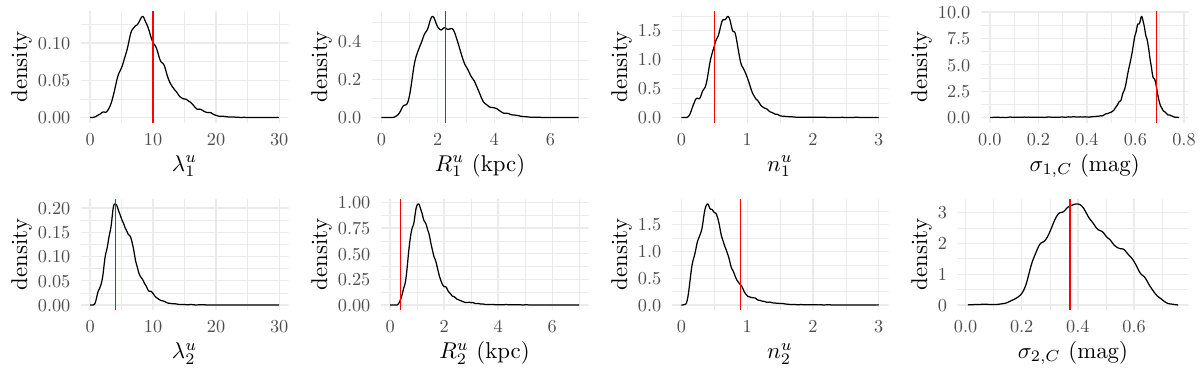}}
     \subfigure[Color-marked model with error]{\includegraphics[width=0.85\columnwidth]{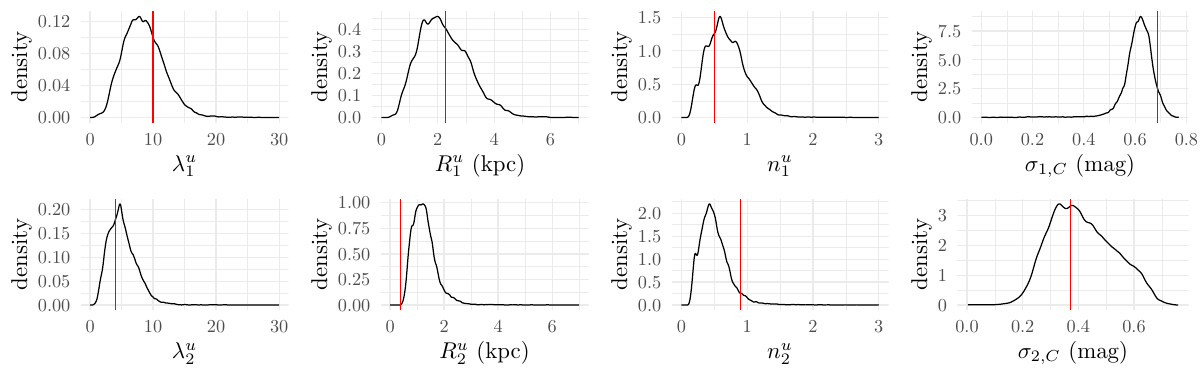}}
     \caption{\footnotesize{Posterior densities of the selected physical and mark parameters of simulated UDGs from all four models. The red lines are the true values.}}
     \label{fig:sim_UDG_dens}
\end{figure}

Figure \ref{fig:sim_UDG_dens} show the posterior densities of the selected physical and mark parameters of simulated UDGs from all four models. The results indicate that our inference algorithm managed to recover the true values of these parameters.